\documentclass[journal,twoside]{IEEEtran}
\usepackage{amsfonts}
\usepackage{amssymb}
\usepackage{psfig}
\usepackage{subfigure}
\usepackage[centertags]{amsmath}
 \newcommand{\RE}{Re}
 \newcommand{\IM}{Im}

 \newcommand{\h}{{\mathcal{H}}}

 \newcommand{\A}{{\mathcal{A}}}
 \newcommand{\V}{\mathcal{V}}
 
  \newcommand{\T}{{\mathcal{T}}}
 \newcommand{\R}{{\mathcal{R}}}
\newcommand{\J}{{\mathbf{j}}}

 \newcommand{\Z}{{\mathbb{Z}}}
 
 \newcommand{\BOP}{{\mathbf{B}}}

 \newcommand{\Real}{{\mathbb{R}}}
 \newcommand{\cp}{{\mathbb{C}}}

 \newcommand{\Xe}{{\mathcal{E}}}
 \newcommand{\EssD}{{\mathcal{D}}}
 \newcommand{\Lom}{{{\mathcal{L}}}}

\newcommand{\tr}[1]{{\mathrm{tr}}\left(#1\right)}
\newcommand{\hn}[1]{{H}\left(#1\right)}
\newcommand{\Hu}[1]{{H}\left(#1\right)}

 \newcommand{\bV}{{\mathbf {V}}}
\newcommand{\bH}{{\mathbf {H}}}
\newcommand{\bS}{{\mathbf {S}}}
\newcommand{\bW}{{\mathbf {W}}}
\newcommand{\E}[1]{\mathrm{E}\left(#1\right)}

 \newtheorem{thm}{\bf Theorem}

\newtheorem{cons}{ Construction}[section]

 \newtheorem{cor}[thm]{{ Corollary}}

\newtheorem{lem}{{ Lemma}}
 \newtheorem{prop}[thm]{{\it Proposition}}

\newtheorem{defn}{{ Definition}}

 \newtheorem{rem}[thm]{Remark}
 \newtheorem{pr}{Property}[section]

\newtheorem{eg}{Example}[section]
\begin{document}
\title{Single-Symbol Maximum Likelihood Decodable Linear STBCs}
\author{Md. Zafar Ali Khan,~\IEEEmembership{Member,~IEEE,}
        B. Sundar Rajan,~\IEEEmembership{Senior Member,~IEEE,}%
\thanks{This work was supported through grants to
B.S.~Rajan; partly by the
IISc-DRDO program on Advanced Research in
Mathematical Engineering, and partly
by the Council of Scientific \& Industrial
Research (CSIR, India) Research
Grant (22(0365)/04/EMR-II).}% <-this % stops a space
\thanks{Md. Zafar Ali Khan is with the International Institute of Information Technology, Hyderabad, India and B. Sundar Rajan is with Indian Institute of Science, Bangalore, India.}}
% note the % following the last \IEEEmembership and also the first \thanks - 
% instance, "A\textbf{} \textbf{}B" will typeset as "A B" not "AB". If you want
% "AB" then you have to do: "A\textbf{}\textbf{}B"
\markboth{IEEE Transactions on Information Theory,~Vol.~XX, No.~XX,~XXX~XXXX}{Khan and Rajan: Single-Symbol Maximum Likelihood Decodable Linear STBCs}
% *** Note that you probably will NOT want to include the author's name in ***
% *** the headers of peer review papers.                                   ***

% If you want to put a publisher's ID mark on the page
% (can leave text blank if you just want to see how the
% text height on the first page will be reduced by IEEE)
%\pubid{0000--0000/00\$00.00~\copyright~2002 IEEE}

%\specialpapernotice{(Invited Paper)}

\maketitle

\begin{abstract}
Space-Time block codes (STBC) from Orthogonal Designs (OD)
 and Co-ordinate Interleaved
Orthogonal Designs (CIOD) have been attracting wider attention due
to their amenability for fast (single-symbol) ML decoding, and
full-rate with full-rank over quasi-static fading channels. However,
these codes are instances of single-symbol decodable codes and it is
natural to ask, if there exist codes other than STBCs form ODs and
CIODs that allow single-symbol decoding?

In this paper, the above question  is answered in the affirmative by
  characterizing all linear
STBCs, that allow single-symbol ML decoding (not necessarily
full-diversity) over quasi-static fading channels-calling them
single-symbol decodable designs (SDD). The class SDD includes ODs
and CIODs as  proper subclasses. Further, among the SDD, a class of
those that offer full-diversity, called Full-rank SDD (FSDD) are
characterized and classified.

We then concentrate on square designs and  derive the maximal rate
for square FSDDs using a constructional proof. It follows that (i)
except for $N=2$, square Complex ODs are not maximal rate and (ii)
square FSDD exist only for 2 and 4 transmit antennas. For non-square
designs, generalized co-ordinate-interleaved orthogonal  designs (a
superset of CIODs) are presented and analyzed.

Finally, for rapid-fading channels an equivalent matrix channel
representation is developed, which allows the results of
quasi-static fading channels to be applied to rapid-fading channels.
Using this representation we show that for rapid-fading channels the
rate of single-symbol decodable STBCs are independent of the number
of transmit antennas and inversely proportional to the block-length
of the code. Significantly, the CIOD for two transmit antennas is
the only STBC that is single-symbol decodable over both quasi-static
and rapid-fading channels.
\end{abstract}


% Will be ignored for other modes.
\IEEEpeerreviewmaketitle

\section{Introduction}
% \PARstart{A}{demo} file is ....
% \PARstart{A}{}demo file is ....
% \PARstart{T}{his demo} file is ....
\PARstart{S}{ince}  the publication of capacity gains of MIMO systems \cite{FoG,Tel} coding for   MIMO systems
 has been an active area of research and such codes have been
 christened
Space-Time Codes (STC). The primary difference between coded
modulation (used for SISO, SIMO) and space-time codes is that in
coded modulation the coding is in time only while in space-time
codes the coding is in both space and time and hence the name.
Space-time Codes (STC) can be thought of as a signal design
problem at the transmitter to realize the capacity benefits of
MIMO systems \cite{FoG,Tel}, though, several developments towards
STC
  were presented in \cite{SeW,GFBK,Wee,Win1,Win2}  which combine transmit and
  receive diversity, much prior to the results on capacity.  Formally, a
   thorough treatment of STCs was first presented in \cite{TSC} in the form of
trellis codes (Space-Time Trellis Codes (STTC)) along with
appropriate design and performance criteria,

  The decoding
complexity of STTC is exponential in bandwidth efficiency and
required diversity order. Starting from Alamouti \cite{Ala},
several authors have studied
 Space-Time Block Codes (STBCs) obtained from Orthogonal Designs
 (ODs) and their variations that  offer fast decoding (single-symbol
 decoding or double-symbol decoding) over quasi-static fading channels
 \cite{TJC}-\cite{Gan}, \cite{Jaf}-\cite{SuX4}.
 But the STBCs from
  ODs are a class of codes that are amenable to single-symbol
  decoding. Due to the importance of single-symbol decodable
  codes,  need was felt for rigorous characterization of single-symbol
  decodable linear STBCs.

Following the spirit of \cite{TiH1}, by a linear STBC\footnote{
Also referred to as a Linear Dispersion code \cite{HaH}} we mean
those covered by the following definition.
\begin{defn}[ Linear STBC]\label{c0def1}
 A linear design, $S$, is a $L \times N$ matrix
whose entries are complex linear combinations of $K$ complex
indeterminates $x_k=x_{kI}+\J x_{kQ}$, $k=0,\cdots,K-1$ and their
complex conjugates. The STBC obtained by letting each
indeterminate to take all possible values from a complex
constellation $\A$ is called a linear STBC over $\A$. Notice that
$S$ is basically a ``design''and by the STBC $(S,\A)$ we mean the
STBC obtained using the design $S$ with the indeterminates taking
values from the signal constellation $\A$. The rate of the
code/design\footnote{Note that if the signal set is of size $2^b$
the throughput rate  $R$ in bits per second per Hertz is related
to the rate of the design ${\cal R}$ as $R= {\cal R}b$.} is given
by $K/L$ symbols/channel use. Every linear design $S$ can be
expressed as
\begin{equation}\label{c0eq13}
  S=\sum_{k=0}^{K-1}x_{kI}A_{2k}+x_{kQ}A_{2k+1}
\end{equation}
where $\{A_k\}_{k=0}^{2K-1}$ is a set of complex matrices called
weight matrices of $S$. When the signal set $\A$ is understood
from the context or with the understanding that an appropriate
signal set $\A$ will be specified subsequently,  we will use the
terms Design and STBC interchangeably.
\end{defn}

Throughout the  paper, we consider only those linear STBCs that
are obtained from designs. Linear STBCs  can be decoded using
simple
 linear processing at the receiver
with algorithms like sphere-decoding \cite{DMB2,DMB3} which have
polynomial complexity in, $N$, the number of transmit antennas.
But STBCs from ODs  stand out because of their amenability to very
simple (linear complexity in $N$) decoding. This is because the ML
metric can be written as a sum of several square terms, each
depending on at-most one variable for OD. However, the rates of
ODs is restrictive; resulting in search of other codes that allow
simple decoding similar to ODs. We call such codes ``single-symbol
decodable''.
 Formally
 \begin{defn}[Single-symbol Decodable (SD)
 STBC]\label{c0def2}
A Single-symbol Decodable (SD) STBC of rate $K/L$ in $K$ complex
indeterminates $x_k=x_{kI}+\J x_{kQ}$, $k=0,\cdots,K-1$ is a
linear STBC such that
 the ML decoding metric can be written as a   square of several terms
  each depending on at most one indeterminate.
\end{defn}
Examples of SD STBCs are STBCs from Orthogonal Designs of
\cite{TJC}.

 In this paper, we first
characterize all linear STBCs that admit single-symbol ML decoding, (not necessarily full-rank)
  over quasi-static fading channels, the class of Single-symbol Decodable Designs (SDD).
Further, we characterize a class of full-rank SDDs called
Full-Rank SDD (FSDD).

Fig. \ref{c1fig1} shows the various classes of SD STBCs identified
in this paper. Observe that the class of FSDD consists of only

\begin{itemize}
\item  an extension of Generalized Linear Complex Orthogonal Design
(GLCOD\footnote{GLCOD is the same as the Generalized Linear Processing
Complex Orthogonal Design of \cite{TJC}-the word ``Processing'' has nothing
to be with the linear processing operations in the receiver and means basically
 that the entries are linear combinations of the variables of the design. Since
 we feel that it is better to drop this word to avoid possible confusion we call
  it GLCOD. GLCOD is formally defined in Definition \ref{c0def3}})  which we have called Unrestricted
Full-rank Single-symbol Decodable Designs (UFSDD) and \item  a class
of non-UFSDDs called Restricted Full-rank Single-symbol Decodable
Designs (RFSDD)\footnote{The word ``Restricted'' reflects the fact
that the STBCs obtained from these designs can achieve full
diversity for those complex constellations that satisfy a (trivial)
restriction. Likewise, ``Unrestricted'' reflects the fact that the
STBCs obtained from these designs achieve full diversity for all
complex constellations.}.
\end{itemize}

The rest of the material of this paper is organized as follows:
 In section \ref{c5ch} the channel model and the design criteria for
  both quasi-static and rapid-fading channels are reviewed. A
brief presentation of basic, well known results concerning GLCODs
 is given in Section
\ref{glpcods}. In Section  \ref{ssdd} we characterize the class
SDD of all SD (not necessarily full-rank) designs and within the
class of SDD the class FSDD consisting of  full-diversity SDD is
characterized. % in Section \ref{frssdd}.
 Section \ref{sqssdd} deals exclusively with the maximal rate of square designs  and
construction of such maximal rate designs.

In  Section \ref{chap3}  we generalize the construction of  square
RFSDDs given in Subsection \ref{frssdd}, and give a formal
definition for Co-ordinate Interleaved Orthogonal Designs (CIOD)
and its generalization, Generalized Co-ordinate Interleaved
Orthogonal Designs (GCIOD). This generalization is basically a
construction of RFSDD; both square and non-square and results in
construction of various high rate RFSDDs. The signal set expansion
due to co-ordinate interleaving is then highlighted and the coding
gain of GCIOD is shown to be equal to what is defined as the
generalized co-ordinate product distance (GCPD) for a signal set.
A special case of GCPD, the co-ordinate product distance (CPD) is
derived for lattice constellations. We then show that, for lattice
constellations, GCIODs have higher coding gain as compared to
GLCODs. Simulation results are also included for completeness. The
maximum mutual information (MMI) of GCIODs is then derived and
compared with that of GLCODs to show that, except for $N=2$, CIODs
have higher MMI. In short, this section shows that, except for
$N=2$ (the Alamouti code), CIODs are better than GLCODs in terms
of rate, coding gain and MMI.

In section \ref{fastsdd}, we study STBCs for use in rapid-fading
channels by giving a matrix representation of the multi-antenna
rapid-fading channels. The emphasis is on finding STBCs that allow
single-symbol decoding for both quasi-static and rapid-fading
channels as BER performance such STBCs will be invariant to any
channel variations. Therefore, we  characterize all linear STBCs
that allow single-symbol ML decoding when used in rapid-fading
channels.
 Then, among these we identify those with full-diversity, i.e.,
those with diversity $L$ when the STBC is of size $ L \times N, (L
\ge N)$, where $N$ is the number of transmit antennas and $L$ is the
length of the code. The maximum rate for such a full-diversity, SD
code is shown to be $2/L$ from which it follows that rate-one is
possible only for 2 Tx. antennas. The co-ordinate interleaved
orthogonal design (CIOD) for 2 Tx (introduced in Section \ref{ssdd})
is shown to be one such rate-one, full-diversity and SD code. (It
turns out that Alamouti code is not SD for rapid-fading channels.)
 Finally, Section \ref{conc}
consists of some concluding remarks and a couple of directions for
further research.
%%%%%%%%%%%%%%%%%%%%%%%%%%%%%%%%%%%%%%%%%%%%%%%%%%%%%5
\section{Channel Model}\label{c5ch} In this section we present
the channel model and review the design criteria for both
quasi-static and rapid-fading channels.
 Let the number of transmit antennas be $N$ and the
number of receive antennas be $M$. At each time slot $t$, complex
signal points, $s_{it}, ~i=0,1,\cdots, N-1$ are transmitted from
the $N$ transmit antennas simultaneously. Let
$h_{ijt}=\alpha_{ijt}e^{\J \theta_{ijt}}$ denote the path gain
from the transmit antenna $i$ to the receive antenna $j$ at time
$t$, where $\J=\sqrt{-1}$. The received signal $v_{jt}$ at the
antenna $j$ at time $t$,  is given by
\begin{equation}\label{c5chmod1}
  v_{jt}=\sum_{i=0}^{N-1} h_{ijt} s_{it} + n_{jt}, ~~
\end{equation}
$j=0,\cdots,M-1; ~~t=0,\cdots,L-1$.
  Assuming that perfect channel state information (CSI) is available at
the receiver, the decision rule for ML decoding is to minimize the
metric
\begin{equation}\label{c5chmod2}
  \sum_{t=0}^{L-1}\sum_{j=0}^{M-1}\left|v_{jt}-\sum_{i=0}^{N-1}
  h_{ijt}
  s_{it}\right|^2
\end{equation}
over all codewords. This results in exponential decoding
complexity, because of the joint decision on all the symbols
$s_{it}$ in the matrix ${\mathbf S}$. If the throughput rate of
such a scheme is $R$ in bits/sec/Hz, then $2^{RL}$ metric
calculations are required; one for each possible transmission
matrix ${\mathbf S}$. Even for modest antenna configurations and
rates this could be very large resulting in search for codes that
admit a simple decoding while providing full diversity gain.

\subsection{Quasi-Static Fading Channels} For quasi-static fading
channels  $h_{ijt}=h_{ij}$ and (\ref{c5chmod1}) can be written in
matrix notation as,
\begin{equation}\label{chmod1}
  v_{jt}=\sum_{i=0}^{N-1} h_{ij} s_{it} + n_{jt}, ~~   j=0,\cdots,M-1; ~~t=0,\cdots,L-1.
\end{equation}
 In matrix notation,
\begin{equation}\label{c0eq6}
  \bV=\bS \bH+\bW
\end{equation}
\noindent where ${\mathbf V} \in \cp^{L \times M}$ ($\cp$ denotes
the complex field) is the received signal matrix, ${\mathbf S} \in
\cp^{L \times N}$ is the transmission matrix (codeword matrix),
${\mathbf H} \in \cp^{N \times M}$ denotes the channel matrix  and
${\mathbf W}\in \cp^{L\times M} $ has entries that are Gaussian
distributed with zero mean and unit variance and also are
temporally and spatially white. In ${\bV, \bS}$ and ${\mathbf W}$
time runs vertically and space runs horizontally. The channel
matrix ${\mathbf H}$ and the transmitted codeword ${\mathbf S}$
are assumed to have unit variance entries.  The ML metric can then
be written as
\begin{equation}\label{c1eq3}
M({\bS})= \tr{({\bV}-{\bS \bH})^H({\mathbf V}-{\bS \bH})}.
\end{equation}
This ML metric (\ref{c1eq3}) results in exponential decoding
complexity with the rate of transmission in bits/sec/Hz.
\subsubsection{{\it Design Criteria for STC over
quasi-static fading channels}} The design criteria for STC over
quasi-static fading channels are \cite{TSC}:
\begin{itemize}
  \item {\it Rank Criterion:} In order to achieve
  diversity of $rM$, the matrix $\BOP({\bS} , { \hat{\bS}})\triangleq \bS-\hat{\bS}$ has
   to be full rank for any two distinct codewords ${\bS}, {
 ~ \hat{\bS}}$. If  $\BOP({\bS} , { \hat{\bS}})$ has  rank $N$,
 then the STC achieves full-diversity.
  \item {\it Determinant Criterion:}After ensuring full diversity
  the next criteria is to maximize the coding gain given by,
   \begin{equation}\label{pr3}
\Lambda({\bS},\hat{{\bS}})=\min_{\bS,\hat{\bS}}|({\bS}-\hat{{\bS}})({\bS}-\hat{{\bS}})^H|_+^{1/r}
\end{equation}
where $|{A}|_+$ represents the product of the non-zero eigen
values of the matrix ${A}$.
\end{itemize}
\subsubsection{ Design Criteria for STC over Rapid-Fading Channels:} We recall that the design
criteria for rapid-fading channels are \cite{TSC}:
\begin{itemize}
  \item The Distance Criterion : In order to achieve the diversity $rM$
  in rapid-fading channels, for any two distinct codeword matrices $\mathbf{S}$ and
  $\mathbf{\hat{S}}$, the strings $s_{0t},s_{1t},\cdots$, $s_{(N-1)t}$ and $\hat{s}_{0t},\hat{s}_{1t},\cdots,\hat{s}_{(N-1)t}$
  must differ at least for $r$ values of $0 \le t \le L-1$. (Essentially, the distance criterion implies that if a codeword is viewed
as a $L$ length vector with each row of the transmission matrix
viewed as a single element of $\cp^N$, then the diversity gain is
equal to the Hamming distance of this $L$ length codeword over
$\cp^N$.
  \item The Product Criterion : Let $\V(\mathbf{S},\mathbf{\hat{S}})$ be the indices of the
  non-zero rows of $\mathbf{S}-\mathbf{\hat{S}}$ and let$
  |\mathbf{s}_t-\mathbf{\hat{s}}_t|^2=\sum_{i=0}^{N-1}|s_{it}-\hat{s}_{it}|^2
  $, where $\mathbf{s}_t $ is the $t$-th row of $\mathbf{S}$, $0\le t \le L-1$.
  Then the coding gain is given by
  \[
  \min_{\mathbf{s}\ne \mathbf{\hat{s}}}\prod_{t \in \V(\mathbf{s},\mathbf{\hat{s}})}
  |\mathbf{s_t}-\mathbf{\hat{s}_t}|^2.
  \]
  The product criterion is to maximize the coding gain.
\end{itemize}

%%%%%%%%%%%% Section 2 begins %%%%%%%%%%%%%%%%%%%%%%%%%%5555
\section{Generalized Linear Complex Orthogonal Designs (GLCOD)}\label{glpcods}
The class of GLCOD was first discovered and studied in the context
of single-symbol decodable designs by coding theorists in
\cite{TJC,TiH1,SuX2,Lia,KhR6}. It is therefore proper to recollect
the main results concerning GLCODs before the characterization of
SSD. In this section we review the definition of GLCOD and
summarize important results on square as well as non-square GLCODs
from \cite{TJC,TiH1,SuX2,Lia,KhR6}.

\begin{defn}[GLCOD]\label{c0def3}A Generalized Linear Complex Orthogonal Design (GLCOD)  in $k$ complex
indeterminates $x_1,x_2,\cdots,x_k$ of size $N$ and rate ${\cal
R}=k/p$, $p \ge N$ is a $p \times N$ matrix $\Theta$, such that
\begin{itemize}
  \item the entries of $\Theta$ are  complex linear combinations of   $0, \pm x_{i}, ~~i=1,\cdots, k$  and  their conjugates.
  \item $\Theta^{\cal H} \Theta = {\mathbf D}$, where $\mathbf D$ is a diagonal matrix whose
  entries are a linear combination of $|x_{i}|^2, i=1,\cdots,
  k$ with all strictly positive real coefficients.
\end{itemize}
If $k$=$N$=$p$ then  $\Theta$ is called a Linear Complex
Orthogonal Design (LCOD). Furthermore, when the entries are only
from $\{0, \pm x_1,\pm x_2,\cdots,\pm x_k \}$, their conjugates
and multiples of $\mathbf j$ then $\Theta$ is called a Complex
Orthogonal Design (COD). STBCs from ODs are obtained by replacing
$x_i$ by $s_i$ and allowing $s_i$ to take all values from a signal
set ${\A}$. A GLCOD is said to be of minimal-delay if $N=p$.
\end{defn}
{\bf Actually, according to  \cite{TJC} it is required that
$\mathbf D=\sum_{i=1}^k |x_i|^2I$, which is a special case of the
requirement that $\mathbf D$ is a diagonal matrix with the
conditions in the above definition. In other words, we have
presented a generalized version of the definition of GLCOD of
\cite{TJC}.} Also we say that a GLCOD  satisfies {\bf
Equal-Weights condition} if $\mathbf D=\sum_{i=1}^k |x_i|^2I$.

The Alamouti scheme \cite{Ala}, which is of minimal-delay,
full-rank and rate-one is basically the STBC arising from the size
2 COD.

 Consider a square GLCOD\footnote{A rate-1, square GLCOD is referred to
  as complex linear processing orthogonal design (CLPOD) in \cite{TJC}.},
$S=\sum_{k=0}^{K-1}x_{kI}$ $A_{2k}+ x_{kQ}A_{2k+1}$. The weight
matrices satisfy,
\begin{eqnarray}\label{cond2g}
A_{k}^{\cal H} A_{k}=\hat{\EssD}_k, ~~~~ k=0,\cdots,2K-1\\
A_{l}^{\cal H} A_{k}+A_{k}^{\cal H} A_{l}=0,  ~~~~ 0\le k \ne l
\le 2K-1 \label{cond2gl}.
\end{eqnarray}
where $\hat{\EssD}_k$ is a diagonal matrix of full-rank for all
$k$. Define
 $B_k=A_k\hat{\EssD}_k^{-1/2}$.
Then the matrices $B_k$ satisfy (using the results shown in
\cite{KhR6})
\begin{eqnarray}\label{cond3g}
B_{k}^{\cal H} B_{k}=I_N,  ~~~~ k=0,\cdots,2K-1\\
B_{l}^{\cal H} B_{k}+B_{k}^{\cal H} B_{l}=0,  ~~~~ 0\le k \ne l
\le 2K-1 \label{cond3gl}
\end{eqnarray}
and again defining
\begin{equation}\label{cond3ga}
C_k={B}_0^{\cal H}B_k, ~~~~ k=0,\cdots, 2K-1,
\end{equation}
we end up with  $C_0=I_N$ and
\begin{eqnarray}\label{condg4}
C_{k}^{\cal H} =-C_{k},  ~~~~ k=1,\cdots,2K-1\\
C_{l}^{\cal H} C_{k}+C_{k}^{\cal H} C_{l}=0,  ~~~~ 1\le k \ne l
\le 2K-1 \label{cond4gl}.
\end{eqnarray}
 The above normalized set of matrices $\{C_1,\cdots,C_{2K-1}\}$
  constitute a {\it Hurwitz family of order $N$ \cite{Jos}}.
  Let $\Hu{N}-1$ denote the number of matrices in a Hurwitz family
  of order $N$, then the Hurwitz Theorem can be stated as
  \begin{thm}[Hurwitz \cite{Jos}]\label{hur}
  If $N=2^ab$, $b$ odd and $a,b>0$ then
 \[
 \Hu{N} \le 2a + 2.
 \]
 \end{thm}
\noindent Observe that $\Hu{N}=2K$.  An immediate consequence of
the Hurwitz Theorem are the following results:
\begin{thm}[Tarokh, Jafarkhani and Calderbank \cite{TJC}]
A square GLCOD of rate-1 exists iff $N=2$.
\end{thm}
\begin{thm}[Trikkonen and Hottinen \cite{TiH1}]
The maximal rate, $\R$ of a square GLCOD  of size $N=2^ab, b$ odd,
satisfying equal weight condition is
\[
\R=\frac{a+1}{N}.
\]
\end{thm}
This result was generalized to all square GLCODs in \cite{KhR6}
using the theorem:
\begin{thm}[Khan and Rajan \cite{KhR6}]\label{c1a2prop1ca} With the Equal-Weights condition removed from the definition of GLCODs, an $N\times N$  square (GLCOD), $\Xe_c$ in variables
$x_0,\cdots,x_{K-1}$ exists iff there exists a GLCOD $\Lom_c$ such
that
\begin{equation}\label{c1a2eq1}
\Lom_c^{\cal H}\Lom_c=(|x_0|^2+\cdots+|x_{K-1}|^2)I.
\end{equation}
 \end{thm}
 Hence we have the
following corollary.
\begin{cor}[Khan and Rajan \cite{KhR6}]
Let $N=2^ab$ where $b$ is an odd integer and $a=4c+d$, where
$0\leq d < c$
 and $c \ge 0$. The maximal rate of size $N$, square GLROD without
 the Equal-Weights condition satisfied is $\frac{8c+2^d}{N}$ and
 of size $N$, square GLCOD without the Equal-Weights condition satisfied
  is $\frac{a+1}{N}$.
\end{cor}
%%%%%%%%%%%%%%%%%%%%
An intuitive and simple realization of such GLCODs based on
Josefiak's realization of the Hurwitz family, was presented in
\cite{SuX2} as
\begin{cons}[Su and Xia \cite{SuX2}]\label{c1cons1}
Let $G_1(x_0)=x_0I_1$, then the GLCOD of size $2^K$,
$G_{2^K}(x_0,x_1,\cdots,x_K)$, can be constructed iteratively for
$K=1,2,3,\cdots$ as
\begin{eqnarray}\label{glpc}
  G_{2^K}(x_0,x_1,\cdots,x_K)=~~~~~~~~~~~~~~~~~~~~~~~~~~~~~~~~~~~~~~~~~\nonumber\\
  \begin{bmatrix}
   G_{2^{K-1}}(x_0,x_1,\cdots,x_{K-1})  & x_KI_{2^{K-1}} \\
    -x_K^*I_{2^{K-1}} &  G_{2^{K-1}}^{\cal H}(x_0,x_1,\cdots,x_{K-1}) \
  \end{bmatrix}.
\end{eqnarray}
\end{cons}
While  square GLCODs have been completely characterized non-square
GLCODs are not well understood. The main results for non-square
GLCODs are due to Liang and Xia. The primary result is
\begin{thm}[Liang and Xia \cite{LiX}]\label{lix}
A rate 1 GLCOD exists iff $N=2$.
\end{thm}
This was further, improved later to,
\begin{thm}[Su and Xia \cite{SuX2}]\label{Sux}
The maximum rate of  GCOD (without linear processing) is upper
bounded by 3/4.
\end{thm}
Xue bin-Liang \cite{Lia} gave the construction of maximal rates
GCOD
\begin{thm}[Liang \cite{Lia}]\label{thm3ex}
The maximal rate of a GCOD for $N \in \mathbb{N}$ transmit
antennas is given by  $\R=\frac{m+1}{2m}$ where $m=\lfloor N/2
\rfloor$.
\end{thm}
 The maximal rate and the
construction of such maximal rate non-square GLCODs for $N > 2$
remains an open problem. %Finally, rate 1/2 constructions for
\section{Single-symbol Decodable Designs}\label{ssdd}
In the first part of this section we characterize all STBCs that
allow single-symbol ML decoding in quasi-static fading channel and
using this characterization define Single-symbol Decodable Designs
(SDD) in terms of the weight matrices and discuss several examples
of such designs. In the second part, we characterize the class
FSDD and classify the same.
\subsection{Characterization of SD STBCs}\label{charssdd}
Consider the matrix channel model for quasi-static fading channel
given in (\ref{c0eq6}) and the corresponding  ML decoding metric
(\ref{c1eq3}). For a linear STBC with $K$ variables,  we are
concerned about those STBCs for which the ML metric (\ref{c1eq3})
can be written as sum of several terms with each term involving
at-most one variable only and hence SD.

The following theorem characterizes \textbf{all linear STBCs}, in
terms of the weight matrices, that will allow single-symbol
decoding.
\begin{thm}\label{prop1}
For a linear STBC in $K$ variables,
$S=\sum_{k=0}^{K-1}x_{kI}A_{2k}+x_{kQ}A_{2k+1}$, the ML metric,
 $M(S)$ defined in (\ref{c1eq3}) decomposes as
 $M(S)=\sum_{k=0}^{K-1} M_k(x_k) +M_c$ where $M_c=-(K-1)\tr{V^{\cal H}V}$ is independent of all the variables and $M_k(x_k)$ is a function only of the variable $x_k$,  iff\footnote{The condition (\ref{chr1}) can also be given as
$$
 A_kA_l^{\cal H}+A_lA_k^{\cal H}=0~~ \left\{\begin{array}{c}\forall l \ne  k ,k+1 \mbox{ if k is even} \\   \forall l \ne k, k-1 \mbox{ if k is odd}\end{array}\right.
 $$
due to the identity $tr\left\{{(\mathbf{V-SH})}^{\cal H}{(\mathbf{
V-SH})}\right\}= $ $tr\left\{(\mathbf{ V-SH})(\mathbf{V-SH})^{\cal
H}\right\}$ when $S$ is a square matrix. }

\begin{equation}\label{chr1}
 A_k^{\cal H}A_l+A_l^{\cal H}A_k=0~~ \left\{\begin{array}{c}\forall l \ne  k ,k+1 \mbox{ if k is even} \\   \forall l \ne k, k-1 \mbox{ if k is odd}\end{array}\right..
 \end{equation}
 \end{thm}
 \begin{proof}
From (\ref{c1eq3}) we have
\begin{eqnarray*}
 M(S) =\tr{{\mathbf V}^{\cal H}{\mathbf V}} -\tr{(S{\mathbf  H})^{\cal H}{\mathbf V}}-\tr{{\mathbf V}^{\cal H}{ S {\mathbf H}})}
   \\
+\tr{{ S}^{\cal H}{ S}{\mathbf H}{\mathbf H}^{\cal H}}.
\end{eqnarray*}
Observe that  $\tr{{\mathbf V}^{\cal H}{\mathbf V}}$ is
independent of ${ S}$. The next two terms in $M(S)$ are functions
of ${ S,S^{\cal H}}$ and hence linear in $x_{kI}, x_{kQ}$. In the
last term,
\begin{eqnarray}\label{chr1a}
  S^{\cal H}S&=&\sum_{k=0}^{K-1}(A_{2k}^{\cal H}A_{2k} x_{kI}^2+A_{2k+1}^{\cal H}A_{2k+1}
  x_{kQ}^2)
\nonumber\\&&+
  \sum_{k=0}^{K-1}\sum_{l=k+1}^{K-1}(A_{2k}^{\cal H}A_{2l}+A_{2l}^{\cal H}A_{2k}
 ) x_{kI} x_{lI}\nonumber\\
 &&+\sum_{k=0}^{K-1}\sum_{l=k+1}^{K-1}(A_{2k+1}^{\cal H}A_{2l+1}+A_{2l+1}^{\cal H}
 A_{2k+1}) x_{kQ} x_{lQ}\nonumber\\&&+
 \sum_{k=0}^{K-1}\sum_{l=0}^{K-1}(A_{2k}^{\cal H}A_{2l+1}+A_{2l+1}^{\cal H}
 A_{2k}) x_{kI} x_{lQ}.\
\end{eqnarray}
(a) Proof for the  ``if part'': If (\ref{chr1}) is satisfied then
(\ref{chr1a}) reduces to
\begin{eqnarray}\label{ex5}
{S}^{\cal H}{S}&=&\sum_{k=0}^{K-1} \left(
A_{2k}^{\cal H}A_{2k}x_{kI}^2+A_{2k+1}^{\cal H}A_{2k+1}x_{kQ}^2
\right.\nonumber\\&&\left.
+\left(A_{2k}^{\cal H}A_{2k+1}+A_{2k+1}^{\cal H}A_{2k}\right)x_{kI}x_{kQ}\right)\nonumber\\
&=&\sum_{k=0}^{K-1}T_k^{\cal H}T_k, \mbox{ where }\\
T_k&=&A_{2k}x_{kI}+A_{2k+1}x_{kQ}
\end{eqnarray}
 and using linearity of the trace operator, $M(S)$ can be written as
\begin{eqnarray}\label{met2}
 M(S) &=&\tr{{\mathbf V}^{\cal H}{\mathbf V}}
 -\sum_{k=0}^{K-1}\left\{\tr{(T_k{\mathbf  H})^{\cal H}{\mathbf V}}\right.\nonumber\\&& \left.
- \tr{{\mathbf V}^{\cal H}{ T_k {\mathbf H}}}
   +\tr{{T_k}^{\cal H}{T_k}{\mathbf H}{\mathbf H}^{\cal H}}\right\}\nonumber\\
    &=&\sum_k \underbrace{\left\|{\mathbf V}-(A_{2k}x_{kI}+A_{2k+1}x_{kQ}){\mathbf H}\right\|^2}_{M_k(x_k)}+M_c
\end{eqnarray}
 where $M_c=-(K-1)\tr{V^{\cal H}V}$ and $\|.\|$ denotes the Frobenius norm.

(b) Proof for the ``only if part'': If (\ref{chr1}) is not
satisfied for any $A_{k_1},A_{l_1}, k_1 \ne l_1$ then
\begin{eqnarray}\label{crossterms}
 M(S)
    &=&\sum_k \left|\left|{\mathbf V}-(A_{2k}x_{kI}+A_{2k+1}x_{kQ}){\mathbf
    H}\right|\right|^2
    \nonumber\\&&
+ \tr{(A_{k_1}^{\cal H}A_{l_1}+A_{l_1}^{\cal H}A_{k_1}){\mathbf H}^{\cal H}{\mathbf H}}y +M_c
\end{eqnarray}
where
\begin{eqnarray*}
  y=\left\{\begin{array}{ll}
x_{(k_1/2)I}x_{(l_1/2)I} & \mbox{ if both $k_1,l_1$ are
even}\\
x_{((k_1-1)/2)Q}x_{((l_1-1)/2)Q} & \mbox{ if both $k_1,l_1$ are
odd}
\\
x_{((k_1-1)/2)Q}x_{(l_1/2)I} & \mbox{ if  $k_1$ odd, $l_1$ even}.
\end{array}\right.
\end{eqnarray*}
Now, from the above it is clear that $M({{S}})$ can not be
decomposed into terms involving only one variable.
\end{proof}

It is important to observe that (\ref{chr1}) implies that it is
not necessary for the weight matrices associated with the in-phase
and quadrature-phase of a single variable (say $k$-th) to satisfy
the condition $A_{2k+1}^{\cal H}A_{2k}+A_{2k}^{\cal H}A_{2k+1}=0$.
Since $A_{2k+1}^{\cal H}A_{2k}+A_{2k}^{\cal H}A_{2k+1}$ is indeed
the coefficient of $x_{kI}x_{kQ}$ in $S^{\cal H}S$, this implies
that  terms of the form $x_{kI}x_{kQ}$ can appear in $S^{\cal H}S$
without violating single-symbol decodability. An example of such a
STBC is given in Example \ref{c1ex1}.

\begin{eg}\label{c1ex1}
Consider
\begin{equation}\label{anw21}
  S(x_0,x_1)=\left[\begin{array}{cc} x_{0I}+\J x_{1I} & x_{0Q}+\J x_{1Q}\\
    x_{0Q}+\J x_{1Q} &x_{0I}+\J x_{1I}
  \end{array}\right].
\end{equation}
The corresponding weight matrices are given by
\begin{eqnarray*}
&&  A_0=\left[\begin{array}{cc}
    1 & 0 \\
    0 & 1 \
  \end{array}\right],
A_1=\left[\begin{array}{cc}
    0 & 1 \\
  1 & 0 \
  \end{array}\right],\\
&&
A_2=\left[\begin{array}{cc}
    \J & 0 \\
    0 & \J \
  \end{array}\right],
A_3=\left[\begin{array}{cc}
    0 & \J \\
    \J & 0 \
  \end{array}\right]
\end{eqnarray*}
and it is easily verified that (\ref{chr1}) is
satisfied
 and $A_{2k+1}^{\cal H}A_{2k}+A_{2k}^{\cal H}A_{2k+1} \neq 0$ for $k=0$ as well as $k=1$. Explicitly,
\begin{eqnarray}
A_0^{\cal H}A_1+A_1^{\cal H}A_0 & \neq & 0 \\
A_2^{\cal H}A_3+A_3^{\cal H}A_2 & \neq & 0 \\
A_0^{\cal H}A_2+A_2^{\cal H}A_0 & = & 0 \\
A_0^{\cal H}A_3+A_3^{\cal H}A_0 & = & 0 \\
A_1^{\cal H}A_2+A_2^{\cal H}A_1 & = & 0 \\
A_1^{\cal H}A_2+A_2^{\cal H}A_1 & = & 0.
\end{eqnarray}
\end{eg}
\begin{rem}
However note that for the SD STBC in Example \ref{c1ex1},
\begin{eqnarray*}\label{anw22}
 &&\hspace*{-9mm} \det\left\{(S-\hat{S})^{\cal H}(S-\hat{S})\right\}=
  \left[(\triangle x_{0I}- \triangle x_{0Q})^2
\right.+\\
&&\hspace*{-9mm}\left.(\triangle x_{1I}- \triangle x_{1Q})^2\right]
\left[(\triangle x_{0I}+\triangle x_{0Q})^2+(\triangle
x_{1I}+\triangle x_{1Q})^2\right]
\end{eqnarray*}
where $x_i-\hat{x}_i=\triangle x_{iI}+\J \triangle x_{iQ}$. If we
set $\triangle x_{1I}=\triangle x_{1Q}=0$ we have
\begin{equation}\label{anw23}
  \det\left\{(S-\hat{S})^{\cal H}(S-\hat{S})\right\}=\left[(\triangle^2 x_{0I}- \triangle^2
  x_{0Q})^2\right]
\end{equation}
which is maximized (without rotation of the signal set) when either $\triangle^2 x_{0I}=0$ or
$\triangle^2 x_{0Q}=0$, i.e. the $k$-th indeterminate should take
values from a constellation that is parallel to the ``real axis''
or the ``imaginary axis''.  Such codes are closely related to Quasi-Orthogonal Designs (QOD) and the  maximization of the corresponding coding gain with signal set rotation has been considered in \cite{CYT1,HWX}.
\end{rem}
%As shown in the above example, it is easily seen that whenever the weight matrices corresponding to one indeterminate (say $k$-th) of SD design does not satisfy $A_{2k+1}^{\cal H}A_{2k}+A_{2k}^{\cal H}A_{2k+1}=0$, then whenever all the indeterminates of the design take value from the same constellation, the coding gain is maximized when the constellation is a one dimensional constellation parallel to the $x$-axis or the $y$-axis\footnote{ discussed in Chapter \ref{chap5} where we show that such codes are constructible from QODs.}.

{\bf Henceforth, we consider only those STBCs
$S=\sum_{k=0}^{K-1}x_{kI}A_{2k}+x_{kQ}A_{2k+1}$, which have the
property that the weight matrices of the in-phase and quadrature
components of any variable are orthogonal, that is
\begin{equation}\label{ch}
A_{2k}^{\cal H}A_{2k+1}+A_{2k+1}^{\cal H}A_{2k}=0,~~~~~ 0 \le k
\le K-1
\end{equation}
since all known STBCs satisfy (\ref{ch}) and we are able to tract
and obtain several  results concerning full-rankness, coding gain
and existence results with this restriction. }

%for the following reasons: (i) it is customary and also convenient to assume that all indeterminates take values from one and the same complex constellations and and (ii)
\noindent Theorem \ref{prop1} for this case specializes to:
\begin{thm}\label{prop1a}
For a linear STBC in $K$ complex variables,
$S=\sum_{k=0}^{K-1}x_{kI}A_{2k}+x_{kQ}A_{2k+1}$ satisfying the
necessary condition $A_{2k}^{\cal H}A_{2k+1}+A_{2k+1}^{\cal
H}A_{2k}=0, 0 \le k \le K-1$, the ML metric,
 $M(S)$ defined in (\ref{c1eq3}) decomposes as
 $M(S)=\sum_{k=0}^{K-1} M_k(x_k) +M_c$ where $M_c=-(K-1)\tr{V^{\cal H}V}$,
 iff
 \begin{equation}\label{chr}
A_k^{\cal H}A_l+A_l^{\cal H}A_k=0,~~~~ 0 \le k \ne l \le 2K-1.
\end{equation}
 \end{thm}
We also have
\begin{prop}\label{cor}
For a linear STBC in $K$ complex variables,
$S=\sum_{k=0}^{K-1}x_{kI}$ $A_{2k}+x_{kQ}A_{2k+1}$ satisfying the
necessary condition $A_{2k}^{\cal H}A_{2k+1}+A_{2k+1}^{\cal
H}A_{2k}=0, 0 \le k \le K-1$, the ML metric,
 $M(S)$ defined in (\ref{c1eq3}) decomposes as
 $M(S)=\sum_{k=0}^{K-1}$ $ M_k(x_k) +M_c$ where $M_c=-(K-1)\tr{V^{\cal H}V}$,
 iff
 \begin{equation}\label{cor1a}
\tr{A_k\mathbf{H}\mathbf{H}^{\cal H}A_l^{\cal
H}+A_l\mathbf{H}\mathbf{H}^{\cal H}A_k^{\cal H}}=0,  ~~~~ 0 \le k
\ne l \le 2K-1.
\end{equation}
If, in addition, $S$ is square ($N=L$), then (\ref{cor1a}) is
satisfied if and only if
\begin{equation}\label{cor2a}
A_kA_l^{\cal H}+A_lA_k^{\cal H}=0, ~~~~  0 \le k \ne l \le 2K-1.
\end{equation}
\end{prop}

\begin{proof}
Using the identity, $$\tr{{(\mathbf{V-SH})}^{\cal H}{(\mathbf{
V-SH})}}= \tr{(\mathbf{ V-SH})(\mathbf{V-SH})^{\cal H}},$$
(\ref{c1eq3}) can be written as
\begin{eqnarray*}
 M(S) &=&\tr{{\mathbf V}{\mathbf V}^{\cal H}} -\tr{(S{\mathbf  H}){\mathbf V}^{\cal H}} -
 \tr{{\mathbf V}( S {\mathbf H})^{\cal H}}
   \\&&+\tr{{ S}{\mathbf H}{\mathbf H}^{\cal H}{ S}^{\cal H}}.
\end{eqnarray*}
Observe that  $\tr{{\mathbf V}{\mathbf V}^{\cal H}}$ is
independent of ${ S}$. The next two terms in $M(S)$ are functions
of ${ S,S^{\cal H}}$ and hence linear in $x_{kI}, x_{kQ}$. In the
last term,
\begin{eqnarray}\label{chr1a11}
  S \bH\bH^\h S^{\cal H}&=&\sum_{k=0}^{K-1}(B_{2k} B_{2k}^{\cal H} x_{kI}^2+ B_{2k+1} B_{2k+1}^{\cal H} x_{kQ}^2)
  \nonumber\\
 + &&\hspace*{-6mm}  \sum_{k=0}^{K-1}\sum_{l=k+1}^{K-1}(B_{2k}  B_{2l}^{\cal
H}+A_{2l} \bH\bH^\h A_{2k}^{\cal H}
 ) x_{kI} x_{lI}\nonumber\\
 +&&\hspace*{-6mm}\sum_{k=0}^{K-1}\sum_{l=k+1}^{K-1}(B_{2k+1} B_{2l+1}^{\cal H}+B_{2l+1}
  B_{2k+1}^{\cal H}) x_{kQ} x_{lQ}\nonumber\\
+&&\hspace*{-6mm}
 \sum_{k=0}^{K-1}\sum_{l=0}^{K-1}(B_{2k} B_{2l+1}^{\cal H}+B_{2l+1}
  B_{2k}^{\cal H}) x_{kI} x_{lQ}\
\end{eqnarray}
where $B_{k}=A_{k} \bH$
(a) Proof for the  ``if part'': If (\ref{cor1a}) is satisfied then
(\ref{chr1a11}) reduces to
\begin{eqnarray}\label{ex511}
{S}^{\cal H}{S}&=&\sum_{k=0}^{K-1} \left(
A_{2k} \bH\bH^\h A_{2k}^{\cal H}x_{kI}^2+A_{2k+1} \bH\bH^\h A_{2k+1}^{\cal H}x_{kQ}^2\right) \nonumber\\
&=&\sum_{k=0}^{K-1}T_k T_k^{\cal H}, \mbox{ where }\\
T_k&=&\left(A_{2k}x_{kI}+A_{2k+1}x_{kQ}\right)\bH
\end{eqnarray}
 and using linearity of the trace operator, $M(S)$ can be written as
\begin{eqnarray}\label{met211}
 M(S) &=&\tr{{\mathbf V}{\mathbf V}^\h}
 -\sum_{k=0}^{K-1}\left\{\tr{T_k{\mathbf V}^{\cal H}}\right.%\\
- \tr{{\mathbf V}{ T_k }^{\cal H}}\nonumber\\&&
   +\tr{{ T_k}{ T_k}^{\cal H}}\nonumber\\
    &=&\sum_k \underbrace{\left\|{\mathbf V}-(A_{2k}x_{kI}+A_{2k+1}x_{kQ}){\mathbf H}\right\|^2}_{M_k(x_k)}+M_c
\end{eqnarray}
 where $M_c=-(K-1)\tr{V^{\cal H}V}$ and $\|.\|$ denotes the Frobenius norm.

(b) Proof for the ``only if part'': If (\ref{cor1a}) is not
satisfied for any $A_{k_1},A_{l_1}, k_1 \ne l_1$ then
\begin{eqnarray*}
 M(S)
    &=&\sum_k \left|\left|{\mathbf V}-(A_{2k}x_{kI}+A_{2k+1}x_{kQ}){\mathbf
    H}\right|\right|^2
    \nonumber\\ &+& \tr{(A_{k_1} {\bH}{\bH}^{\h}A_{l_1}^\h+A_{l_1}{\bH}{\bH}^{\h} A_{k_1}^\h)}y +M_c 
\end{eqnarray*}
  where 
\[
y=\left\{\begin{array}{ll}
x_{(k_1/2)I}x_{(l_1/2)I} & \mbox{ if both $k_1,l_1$ are
even}\\
x_{((k_1-1)/2)Q}x_{((l_1-1)/2)Q} & \mbox{ if both $k_1,l_1$ are
odd}
\\
x_{((k_1-1)/2)Q}x_{(l_1/2)I} & \mbox{ if  $k_1$ odd, $l_1$ even}.
\end{array}\right.
\]
Now, from the above it is clear that $M({{S}})$ can not be
decomposed into terms involving only one variable.

 For square $S$, (\ref{cor1a}) can be written as
\begin{equation}\label{cor1aa}
\tr{\mathbf{H}\mathbf{H}^{\cal H}\left\{A_kA_l^{\cal
H}+A_lA_k^{\cal H}\right\}}=0,  ~~~~ 0 \le k \ne l \le 2K-1
\end{equation}
 which is satisfied iff $A_kA_l^{\cal H}+A_lA_k^{\cal H}=0,  ~~~~ 0 \le k \ne l \le 2K-1$.
\end{proof}

%%%%%%%%%%%%%%%%%%%%%%insertion of 110603 ends here %%%%%%%%
%\begin{proof}
%Using the identity, $tr{{(\mathbf{V-SH})}^{\cal H}{(\mathbf{ V-SH})}}= $
%$tr{(\mathbf{ V-SH})(\mathbf{V-SH})^{\cal H}}$ and proceeding as in the
%proof of Theorem \ref{prop1} we have (\ref{cor1a}). For square
%$S$, (\ref{cor1a}) can be written as
%\begin{equation}\label{cor1aa}
%\tr{\mathbf{H}\mathbf{H}^{\cal H}\left\{A_kA_l^{\cal H}+A_lA_k^{\cal H}\right\}}=0,  ~~~~ 0
%\le k \ne l \le 2K-1
%\end{equation}
% which is satisfied iff $
%A_kA_l^{\cal H}+A_lA_k^{\cal H}=0,  ~~~~ 0 \le k \ne l \le 2K-1$.
%\end{proof}
Examples of SD STBCs are those from OD, in-particular the Alamouti
code. The following example gives two STBCs that are not
obtainable as STBCs from ODs.
\begin{eg}\label{ssdds1}
For $N=K=2$ consider
\begin{equation}\label{nw21}
  S=\left[\begin{array}{cc} x_{0I}+\J x_{1Q} & 0\\ 0 &
 x_{1I}+\J x_{0Q} \end{array}\right].
\end{equation}
The corresponding weight matrices are given by
\begin{eqnarray*}
  A_0=\left[\begin{array}{cc}
    1 & 0 \\
    0 & 0 \
  \end{array}\right],A_1=\left[\begin{array}{cc}
    0 & 0 \\
    0 & \J \
  \end{array}\right],\\
 A_2=\left[\begin{array}{cc}
    0 & 0 \\
    0 & 1 \
  \end{array}\right],A_3=\left[\begin{array}{cc}
    \J & 0 \\
    0 & 0 \
  \end{array}\right].
\end{eqnarray*}
Similarly, for $N=K=4$ consider the design given in (\ref{exeq1}).
\begin{table*}[t]
\begin{equation}\label{exeq1}
S= \left[\begin{array}{cccc}
x_{0I}+\J x_{2Q} & x_{1I}+\J x_{3Q} & 0 &0\\
-x_{1I}+\J x_{3Q} & x_{0I}-\J x_{2Q} & 0 &0\\
 0 &0 & x_{2I}+\J x_{0Q} & x_{3I}+\J x_{1Q} \\
 0 &0 & -x_{3I}+\J x_{1Q} &  x_{2I}-\J x_{0Q}
\end{array}\right].
\end{equation}
\end{table*}
The corresponding weight matrices are
$$
A_0=\left[\begin{array}{cccc}
1 & 0 & 0 &0\\
0 & 1 & 0 &0\\
 0 &0 & 0 & 0 \\
 0 &0 & 0 & 0
\end{array}\right],
 A_1=\left[\begin{array}{cccc}
0 & 0 & 0 &0\\
0 & 0 & 0 &0\\
 0 &0 & \J & 0 \\
 0 &0 & 0 & -\J
\end{array}\right],
$$
$$
 A_2=\left[\begin{array}{cccc}
 0 & 1 & 0 &0\\
-1 & 0 & 0 &0\\
 0 &0 & 0 & 0 \\
 0 &0 & 0 & 0
\end{array}\right],
A_3=\left[\begin{array}{cccc}
0 & 0 & 0 &0\\
0 & 0 & 0 &0\\
 0 &0 & 0 & \J \\
 0 &0 & \J & 0
\end{array}\right],
$$
$$
A_4=\left[\begin{array}{cccc}
 0 & 0 & 0 &0\\
 0 & 0 & 0 &0\\
 0 &0 & 1 & 0 \\
 0 &0 & 0 & 1
\end{array}\right],
 A_5=\left[\begin{array}{cccc}
\J & 0 & 0 &0\\
0 & -\J & 0 &0\\
 0 &0 & 0 & 0 \\
 0 &0 & 0 & 0
\end{array}\right],
$$
%$$

%$$
$$
A_6=\left[\begin{array}{cccc}
 0 & 0 & 0 &0\\
 0 & 0 & 0 &0\\
 0 &0 & 0 & 1 \\
 0 &0 & -1 & 0
\end{array}\right],
 A_7=\left[\begin{array}{cccc}
0 & \J & 0 &0\\
\J & 0 & 0 &0\\
 0 &0 & 0 & 0 \\
 0 &0 & 0 & 0
\end{array}\right].
$$
\end{eg}
It is easily seen that the two codes of the above example are not
covered by GLCODs and satisfy the requirements of Theorem
\ref{prop1a} and hence are SD. These two STBCs are instances of
the so called  Co-ordinate Interleaved Orthogonal Designs (CIOD),
which is discussed in detail  in Section \ref{chap3} and a formal
definition of which is Definition \ref{c2def1}. These codes apart
from being SD can give STBCs with full-rank also when the
indeterminates take values from appropriate signal sets- an aspect
which is discussed in detail in Subsection \ref{frssdd} and in
Section \ref{chap3}.

%%%%%%%%%%%%%%% Section 4 begins %%%%%%%%%%%%%%%%%
\subsection{Full-rank SDD}\label{frssdd} In this subsection we
identify all full-rank designs with in the class of SDD that
satisfy (\ref{chr}), calling them the class of Full-rank
Single-symbol Decodable Designs (FSDD), characterize the class of
FSDD and classify the same. Towards this end, we have for square
($N=L$) SDD
\begin{prop}\label{prop1b}
A square  SDD $S=\sum_{k=0}^{K-1}x_{kI}A_{2k}+x_{kQ}A_{2k+1}$,
exists if and only if there exists a square SDD,
 $\hat{S}=\sum_{k=0}^{K-1}x_{kI}\hat{A}_{2k}+x_{kQ}\hat{A}_{2k+1}$  such that
 $$\hat{A}_k^{\cal H}\hat{A}_l+\hat{A}_l^{\cal H}\hat{A}_k=0, ~~~~ k \ne
l , ~~ \mbox{     and      } ~~ \hat{A}_k^{\cal
H}\hat{A}_k=\EssD_k, \forall k,$$ where $\EssD_k$ is a diagonal
matrix.
 \end{prop}
 \begin{proof}
 Using  (\ref{chr}) and (\ref{cor2a}) repeatedly we get
\begin{eqnarray*}
A_k^{\cal H}A_kA_l^{\cal H}A_l&=&A_k^{\cal H}(-A_lA_k^{\cal
H})A_l=(A_l^{\cal H}A_k)A_k^{\cal H}A_l\\
&=&A_l^{\cal H}A_k(-A_l^{\cal
H}A_k)=A_l^{\cal H}(A_lA_k^{\cal H})A_k,
\end{eqnarray*} 
which implies that the set of  matrices
$\{A_k^{\cal H}A_k\}_{k=0}^{2K-1}$ forms a commuting family of
Hermitian matrices and hence can be simultaneously diagonalized by
a unitary matrix, $U$. Define $\hat{A}_k=A_kU^{\cal H}$, then
$\hat{S}=\sum_{k=0}^{K-1}x_{kI}\hat{A}_{2k}+x_{kQ}\hat{A}_{2k+1}$
is a linear STBC such that $\hat{A}_k^{\cal
H}\hat{A}_l+\hat{A}_l^{\cal H}\hat{A}_k=0, \forall k \ne l,
\hat{A}_k^{\cal H}\hat{A}_k=\EssD_k, \forall k$, where $\EssD_k$
is a diagonal matrix. For the converse, given $\hat{S}$,
$S=\hat{S}U$ where $U$ is a unitary matrix.
\end{proof}
Therefore for square SDD, we may, without any loss of generality,
assume that  $S^{\cal H}S$ is diagonal. To characterize non-square
SDD, we use the following
\begin{pr}[Observation 7.1.3 of \cite{HoJ}]\label{corp1a}
Any non-negative linear combination of positive semi-definite
matrices is positive semi-definite.
\end{pr}
\noindent Property \ref{corp1a} when applied to a SDD yields
\begin{pr}\label{corp1}
For a SDD, $S=\sum_{k=0}^{K-1}x_{kI}A_{2k}+x_{kQ}A_{2k+1}$, the
matrix $S^{\cal H}S$ is positive semi-definite and  $A_k^{\cal
H}A_k, ~~ \forall k$ are positive semi-definite.
\end{pr}
 Using property \ref{corp1}, we have the following necessary condition for a SDD to have full-diversity.
\begin{prop}\label{prop2}
If an SDD, $S=\sum_{k=0}^{K-1}x_{kI}A_{2k}+x_{kQ}A_{2k+1}$, whose
weight matrices $A_k$ satisfy
\begin{equation}\label{chr2}
A_k^{\cal H}A_l+A_l^{\cal H}A_k=0,  ~~~~ \forall k \ne l
\end{equation}
achieves full-diversity then $A_{2k}^{\cal H}A_{2k}+A_{2k+1}^{\cal
H}A_{2k+1}$ is full-rank for all $k=0,1,\cdots,K-1$. In addition
if $S$ is square then the requirement specializes to
$\EssD_{2k}+\EssD_{2k+1}$ being full-rank for all
$k=0,1,\cdots,K-1$, where the diagonal matrices $\EssD_{i}$ are
those given in Proposition \ref{prop1b}.
\end{prop}
\begin{proof}
The proof is by contradiction and in two parts corresponding to
whether $S$ is square or non-square.

\noindent {\bf Part 1:} Let $S$ be a square SDD then by
Proposition \ref{prop1b}, without loss of generality, $A_k^{\cal
H}A_k=\EssD_k, \forall k$.
 Suppose
$\EssD_{2k}+\EssD_{2k+1}$, for some $ k \in [0, K-1]$, is not
full-rank. Then $ S^{\cal H}S=\sum_{k=0}^{K-1}
\EssD_{2k}x_{kI}^2+\EssD_{2k+1}x_{kQ}^2. $ Now for any two
transmission matrices $S,~ \hat{S}$ that differ only in $x_k$, the
difference matrix $B(S,\hat{S})=S-\hat{S}$, will not be full-rank
as $ B^{\cal
H}(S,\hat{S})B(S,\hat{S})=\EssD_{2k}(x_{kI}-\hat{x}_{kI})^2+\EssD_{2k+1}(x_{kQ}-\hat{x}_{kQ})^2
$ is not full-rank.

\noindent {\bf Part 2:} The proof for non-square SDD, $S$, is
similar to the above except that $ B^{\cal H}(S,\hat{S})$ $
B(S,\hat{S})=A_{2k}^{\cal
H}A_{2k}(x_{kI}-\hat{x}_{kI})^2+A_{2k+1}^{\cal
H}A_{2k+1}(x_{kQ}-\hat{x}_{kQ})^2 $ where $A_{k}^{\cal H}A_k$ are
positive semi-definite. Since a non-negative linear combination of
positive semi-definite matrices is positive semi-definite, for
full-diversity it is necessary that $A_{2k}^{\cal
H}A_{2k}+A_{2k+1}^{\cal H}A_{2k+1}$ is full-rank for all
$k=0,1,\cdots,K-1$.
\end{proof}

Towards obtaining a sufficient condition for full-diversity, we
first introduce
\begin{defn}[Co-ordinate Product Distance (CPD)]\label{c1def3}\label{cpddefinition}
The Co-ordinate Product Distance (CPD) between any two signal
points
 $u=u_I+{\mathbf j}u_Q$ and $v=v_I+{\mathbf j}v_Q$, $u \ne v$, in the signal set ${\cal A}$ is defined as
\begin{equation}
CPD(u,v)=|u_I - v_I||u_Q - v_Q|
\end{equation}
and the minimum of this value among all possible pairs is defined
as the CPD of ${\cal A}$.
\end{defn}
\begin{rem}\label{c2rem2}
The idea of rotating QAM constellation was first presented in \cite{BoB} and the term ``co-ordinate interleaving'' as also ``Co-ordinate Product Distance'' was first introduced by Jelicic and Roy in \cite{JeR1,JeR2} in the context of TCM for fading channels. This concept of rotation of QAM constellation was extended to multi-dimensional QAM constellations in \cite{BVRB,BoV} at the cost of the decoding complexity. However, for the two-dimensional case there is no increase in the decoding complexity as shown in \cite{Sli1,Sli2}.
\end{rem}

\begin{thm}\label{prop3}
A SSD, $S=\sum_{k=0}^{K-1}x_{kI}A_{2k}$+$x_{kQ}A_{2k+1}$ where
$x_k$ take values from a signal set $\A, \forall k$, satisfying
the necessary
condition of Proposition \ref{prop2} achieves full-diversity iff\\
(i) either $A_k^{\cal H}A_k$ is of full-rank for all $k$ {\bf or}
(ii) the $CPD \mbox{ of } \A \ne 0$.

%\begin{enumerate}
%  \item \label{sufcond1} either $A_k^{\cal H}A_k$ is of full-rank for all $k$ {\bf or}
%  \item \label{sufcond2} the $CPD \mbox{ of } \A \ne 0$.
%\end{enumerate}
\end{thm}
\begin{proof}
Let $S$ be a square SDD satisfying the necessary condition given
in Theorem \ref{prop2}.  We have $ B^{\cal
H}(S,\hat{S})B(S,\hat{S})$=$\sum_{k=0}^{K-1} \EssD_{2k+1}$
$(x_{kI}-\hat{x}_{kI})^2+\EssD_{2k+1}(x_{kQ}-\hat{x}_{kQ})^2. $
Observe  that under both these conditions the difference matrix
$B(S,\hat{S})$ is full-rank for any two distinct $S,~\hat{S}$.
Conversely, if the above conditions are not satisfied then for
exist distinct $S,~\hat{S}$ such that $B(S,\hat{S})$ is not
full-rank.
 The proof is similar when $S$ is a non-square design.
\end{proof}
%\begin{rem}
%From Proposition \ref{prop2} and Theorem \ref{prop3} a FSDD is a SDD such that
% $A_{2k}^{\cal H}A_{2k}+A_{2k+1}^{\cal H}A_{2k+1} $ is full-rank for all $k$ and
% $A_{k}^{\cal H}A_{l}+A_{l}^{\cal H}A_{k} =0; ~~~~ 0 \leq k \ne l \leq 2K-1.$
%\end{rem}

Examples of FSDD are the GLCODs and the STBCs of Example
\ref{ssdds1}.

Note that the sufficient condition (i) of Theorem \ref{prop3} is
an additional condition on the weight matrices whereas the
sufficient condition (ii) is a restriction on the signal set $\A$
and not on the weight matrices $A_k$. Also, notice that the FSDD
that satisfy the sufficient condition (i) are precisely an
extension of GLCODs; GLCODs have an additional constraint that
$A_k^{\cal H}A_k$ be diagonal.

An important consequence of Theorem \ref{prop3} is that there can
exist designs that are not covered by GLCODs offering
full-diversity and single-symbol decoding provided the associated
signal set has non-zero CPD. \textbf{It is important to note that
whenever we have a signal set with CPD equal to zero, by
appropriately rotating it we can end with a signal set with
non-zero CPD. Indeed, only for a finite set of angles of rotation
we will again end up with CPD equal to zero. So, the requirement
of non-zero CPD for a signal set is not at all restrictive in real
sense.} In Section \ref{chap3} we find optimum angle(s) of
rotation for lattice constellations that maximize the CPD.

For the case of square designs of size $N$ with rate-one it is
shown in  Section \ref{sqssdd} that FSDD exist for $N=2,4$ and
these are precisely the STBCs of Example \ref{ssdds1}  and the
Alamouti code.

For a SDD, when $A_k^{\cal H}A_k$ is full-rank for all $k$,
corresponding to Theorem \ref{prop3} with the condition (i) for
full-diversity satisfied, we have an extension of GLCOD in the
sense that the STBC obtained by using the design with {\bf any}
complex signal set for the indeterminates results in a FSDD. That
is, there is no restriction on the complex signal set that can be
used with such designs. So, we define,
\begin{defn}[Unrestricted FSDD (UFSDD)]\label{c1def4}\label{elpcod} A FSDD is called an Unrestricted Full-rank Single-symbol Decodable Design (UFSDD) if $A_k^{\cal H}A_k$ is of full-rank for all $k=0,\cdots,2K-1$.
\end{defn}
\begin{rem}\label{c1rema2}
Observe that for a square UFSDD $S$, $A_k^{\cal H}A_k=\EssD_k$ is
diagonal and hence UFSDD reduces to square GLCOD. For non-square
designs, GLCOD is a
 subset of UFSDD. Also the above extension of the definition of
GLCODs was hinted in \cite{SuX2} where they observe that
$A_k^{\cal H}A_k$ can be positive definite. However it is clear
from our characterization that such a generalization does not
result in any gain for square designs. For non-square designs
existence of UFSDDs that are not GLCODs or unitarily equivalent to
GLCODs is an open problem.
\end{rem}
 The FSDD that are not
UFSDDs are such that $A_{2k}^{\cal H}A_{2k}$ and/or
$A_{2k+1}^{\cal H}A_{2k+1}$ is not full-rank for at least one $k$.
(The CIOD codes  of Example \ref{ssdds1} are such that
$\EssD_{2k}+\EssD_{2k+1}$
    is full-rank $\forall k$ and $\EssD_{k}$  is not full-rank for all $k$.)
We call such FSDD codes Restricted Full-rank Single-symbol
Decodable Designs
 ({\bf RFSDD}), since any full-rank design within this class can be
there only with a restriction on the complex constellation from
which the indeterminates take values, the restriction being that
the CPD of the signal set should not be zero. Formally,
\begin{defn}[Restricted FSDD (RFSDD)]\label{c1def5} A Restricted Full-rank Single-symbol Decodable Designs (RFSDD) is a FSDD such that $A_k^{\cal H}A_k$ is not full-rank
for at least one $k$ where $k=0,\cdots,2K-1$ and the signal set,
from which the indeterminates take values from, has non-zero CPD.
\end{defn}
Observe that the CIODs are a subset of RFSDD. Figure \ref{c1fig1}
shows all the classes discussed so far, viz., SDD, FSDD, RFSDD,
UFSDD. In  Section \ref{sqssdd} we focus on the square RFSDDs as
square UFSDD have been discussed in Section \ref{glpcods}.
%%%%%%%%%%%%%%% Section 5 begins %%%%%%%%%%%%%%%%%%%%%%%5
\section{Existence of Square RFSDDs}\label{sqssdd}
The main result in this section is that {\bf there exists square
RFSDDs with the maximal rate $\frac{2a}{2^a}$ for $N=2^a$ antennas
whereas only rates up to $\frac{a+1}{2^a}$ is possible with square
GLCODs with the same number of antennas}. The other results are:
(i) rate-one square RFSDD of size $N$ exist, iff $N=2, 4$ and (ii)
a construction of RFSDDs with maximum rate from GLCODs.

Let $S=\sum_{k=0}^{K-1}x_{kI}A_{2k}+x_{kQ}A_{2k+1}$ be a square
RFSDD. We have,
\begin{eqnarray}\label{cond1}
A_{k}^{\cal H} A_{k}=\EssD_{k}, ~~~~~~k=0,\cdots,2K-1\\%>>>>>>>>>>>
A_{l}^{\cal H} A_{k}+A_{k}^{\cal H} A_{l}=0, ~~~~~~0\le k \ne l
\le 2K-1 \label{cond1l}
\end{eqnarray}
where  $\EssD_{k},k=0,\cdots,2K-1$ are diagonal matrices with
non-negative entries such that
  $\EssD_{2k}+\EssD_{2k+1}$ is full-rank $ \forall k$.
First we show that for a rate-one RFSDD, $N=2,4$ or $8$.
\begin{thm}\label{thmex1}
If $S$ is a size $N$ square RFSDD of rate-one, then $N=2,4$ or
$8$.
\end{thm}
\begin{proof}Let $B_k=A_{2k}+A_{2k+1}, ~~~~  k=0,\cdots,K-1,$ then
\begin{eqnarray}\label{cond2}
B_{k}^{\cal H} B_{k}=\hat{\EssD}_k=\EssD_{2k}+\EssD_{2k+1},  ~~~~ k=0,\cdots,K-1\\
B_{l}^{\cal H} B_{k}+B_{k}^{\cal H} B_{l}=0,  ~~~~ 0\le k \ne l
\le K-1 \label{cond2l}.
\end{eqnarray}
 Observe that $\hat{\EssD}_k$ is of full-rank for all $k$. Define $C_k=B_k\hat{\EssD}_k^{-1/2}$.
Then the matrices $C_k$ satisfy
\begin{eqnarray}\label{cond3}
C_{k}^{\cal H} C_{k}=I_N,  ~~~~ k=0,\cdots,K-1\\
C_{l}^{\cal H} C_{k}+C_{k}^{\cal H} C_{l}=0,  ~~~~ 0\le k \ne l
\le K-1 \label{cond3l}.
\end{eqnarray}
Define
\begin{equation}\label{cond3a}
\hat{C}_k={C}_0^{\cal H}C_k, ~~~~ k=0,\cdots, K-1,
\end{equation}
then $\hat{C}_0=I_N$ and
\begin{eqnarray}\label{cond4}
\hat{C}_{k}^{\cal H} =-\hat{C}_{k},  ~~~~ k=1,\cdots,K-1\\
\hat{C}_{l}^{\cal H} \hat{C}_{k}+\hat{C}_{k}^{\cal H}
\hat{C}_{l}=0,  ~~~~ 1\le k \ne l \le K-1 \label{cond4l}.
\end{eqnarray}
 The normalized set of matrices $\{\hat{C}_1,\cdots,\hat{C}_{K-1}\}$
  constitute a {\it Hurwitz family of order $N$ \cite{Jos}}
 and for $N=2^ab$, $b$ odd and $a,b>0$ the number of such matrices $K-1$ is bounded  by \cite{Jos}
 \[
 K \le 2a + 2.
 \]
 For rate-one,  RFSDD $(K=N)$, the inequality can be satisfied only for $N=2,4$ or $8$.
 \end{proof}
Therefore the search for rate-one, square RFSDDs can be restricted
to $N=2,4,8$. The rate 1, RFSDDs for $N=2,4$ have been presented
in Example \ref{ssdds1}.  We will now prove that a  rate-one,
square RFSDD for $N=8$ does not exist. Towards  this end we first
derive the maximal rates of square RFSDDs.
\begin{thm}\label{thmex2}
 The maximal rate, ${\R}$, achievable by a square RFSDD with
$N=2^ab, b$ odd (where $a,b > 0$) transmit antennas is
\begin{equation}\label{eqnthm2}
{\R}=\frac{2a}{2^ab}
\end{equation}
\end{thm}
\begin{proof}
\renewcommand{\H}{{\mathcal{H}}}
Let $ {S}=\sum_{k=0}^{K-1}x_{kI}{A}_{2k}+x_{kQ}{A}_{2k+1} $ be a
square RFSDD. Define the RFSDD
\[
{S}'=\sum_{k=0}^{K-1}x_{kI}\underbrace{{C}_0^\H{A}_{2k}\hat{\EssD}_0^{-1/2}}_{A_{2k}'}+x_{kQ}
\underbrace{{C}_0^\H{A}_{2k+1}\hat{\EssD}_0^{-1/2} }_{A_{2k+1}'}
\]
where ${C}_k$  and $\hat{\EssD}_k$ are defined in the proof of the
previous theorem. Then  the set of matrices
$\{C'_k=A_{2k}'+A_{2k+1}'\} $  is such that $C'_0=I_N$ and
$\{C'_k,k=1,\cdots,K-1\}$ is a  family of matrices of order $N$
such that
\begin{eqnarray}\label{app1a11}
&&C_k'^\H C_k'=\hat{\EssD}_0^{-1}\hat{\EssD}_k, ~~~~ 1\le k \le K-1,\\
&& C_{l}'^\H C_{k}'+C_{k}'^\H C_{l}'=0,  ~~~~ 0\le k \ne l \le
K-1,
\end{eqnarray}
where $\hat{\EssD}_0^{-1}\hat{\EssD}_k$ is diagonal and full-rank
for all $k$.
 Then we have
\begin{equation}\label{appa1}
A'_0+A'_1 =C'_0=I_N.
\end{equation}
 It is easily that the set of matrices $\{A'_k\}$ satisfy
(\ref{cond1}) and (\ref{cond1l}). Also, at least one $A'_k$ is not
full-rank. % for each pair $\{A'_{2k}, A'_{2k+1}\},k=0,\cdots,K-1$
%one of $A'_{2k}, A'_{2k+1}$ is not full-rank.
Without loss of generality we assume that $A'_{0}$ is of rank $r <
N$ (if this not so then exchange the indeterminates and/or the
in-phase and quadrature components  so that this is satisfied).
 As $A'_0$ is of rank $r$, due to
(\ref{cond1}), $n-r$ columns of $A'_0$ are zero vectors. Assume
that first $r$ columns of $A'_0$ are non-zero (If this is not the
case, we can always multiply all the weight matrices with a
Permutation matrix such that $A'_0$ is of this form) i.e.
\begin{equation}\label{appa2}
  A'_0=\left[\begin{array}{cc}
    B'_0 & 0
  \end{array}\right]
\end{equation}
where $B'_0 \in \cp^{N \times r}$. Applying (\ref{cond1l}) to
$A'_0$ and $A'_1$ and using from (\ref{appa1}) and (\ref{appa2}),
we have
\begin{eqnarray}\label{appa3}
  &  & A_0^{\prime \H}(I_N-A'_0)+(I_N-A_0^{\prime \H})A'_0=0 \\
  \Rightarrow  & & A_0^{\prime \H}+A'_0=2\EssD'_0\\
  \Rightarrow  & & \left[\begin{array}{c}
    B_0^{\prime \H}  \\
    0 \
  \end{array}\right]+\left[\begin{array}{cc}
    B'_0 &
    0 \
  \end{array}\right]=2\EssD'_0\\
\Rightarrow  & & B'_0=\left[\begin{array}{c}
    B'_{11}  \\
    0 \
  \end{array}\right]
\end{eqnarray}
where $B'_{11}$ is a $r \times r$ matrix and full-rank and
$A_k^{\prime \H}A_k'=\EssD_k', ~~~ k=0,\cdots,2K-1$. Therefore the
matrices $A'_0,A'_1$ are of the form
\begin{equation}\label{appa4}
 A'_0=\left[\begin{array}{cc}
    B'_{11} & 0 \\
    0 &0
  \end{array}\right], ~~~~ A'_1=\left[\begin{array}{cc}
   I_r- B'_{11} & 0 \\
    0 &I_{N-r}
  \end{array}\right].
\end{equation}
 Let
\[
D_1=\left[\begin{array}{cc}
    D_{11} & D_{12} \\
    D_{21} & D_{22}
  \end{array}\right]
  \]  be a matrix such that
\begin{equation}\label{appa4a}
A_i^{\prime \H}D_1+D_1^\H A'_i=0, ~~~~ i=0,1
\end{equation}
 where $D_{11} \in \cp ^{r \times r},~~~ D_{22} \in \cp^{N-r \times N-r}$.
  Substituting the structure of $A'_0$ we have
\begin{eqnarray}\label{appa5}
&&A_0^{\prime \H}D_{1}+D_1^\H A'_0=0\\
 \Rightarrow &&\left[\begin{array}{cc}
    B_{11}^{\prime \H}D_{11}+D_{11}^\H B'_{11} & B_{11}^{\prime \H}D_{12} \\
    D_{12}^\H B'_{11} &0
  \end{array}\right]=0.\label{appa5l}
\end{eqnarray}
As $B'_{11}$ is full-rank it follows that $D_{12}=0$. Substituting
the structure of $A'_1$ we have
\begin{eqnarray}\label{appa6}
  \left[\begin{array}{cc}
  (I_r-  B_{11}^{\prime \H})D_{11}+D_{11}^\H(I_r-B'_{11}) & D_{21}^\H \\
    D_{21} & D_{22}+D_{22}^\H
\end{array}\right]=0 \label{appa6l} \\
  \Rightarrow  D_{21}=0.
\end{eqnarray}
 It follows that  $D_1$ is block diagonal and consequently all
the $A'_k, ~~~ 2 \le k \le 2K-1$ are block diagonal of the form
$D_1$ as they satisfy (\ref{appa4a}). Consequently,
$C'_k={A'}_{2k}+{A'}_{2k+1}, ~~~ k=1,\cdots,K-1$ are also block
diagonal of the form $C'_k=\left[\begin{array}{cc}
  C'_{k1} & 0 \\
  0 & C'_{k2}
\end{array}\right]\mbox{ where } C'_{k1} \in \cp^{r \times r}, ~~ C'_{k2} \in \cp^{N-r \times
N-r}$.  Also, from (\ref{appa5l}), (\ref{appa6l}) we have
\begin{equation}\label{appa6a}
 D_{11}=-D_{11}^\H, D_{22}=-D_{22}^\H.
\end{equation}
Now, in addition to this block diagonal structure the matrices
$A'_k, ~~ 2 <k \le K-1$ have to satisfy (\ref{cond1l}) among
themselves. It follows that the two sets of square matrices
$\{C'_{k1},k=0,\dots,K-1\}$ and  $\{ C'_{k2},k=0\dots,K-1\}$
satisfy
\begin{eqnarray}\label{6b}
  & & C_{ki}^{\prime 2}=-\EssD_{ki}, ~~~~ k=1,\cdots,K-1, ~~~~ i=1,2;  \\
   & & C'_{ki}C'_{li}=-C'_{li}C'_{ki}, ~~~~   1\le k\ne l \le
   K-1, ~~~~ i=1,2,
\end{eqnarray}
where $-\EssD_{ki}$ are diagonal and full-rank $\forall k, i$.
Define
\begin{equation}\label{appa1111}
  \hat{C}_{ki}=C_{ki}'\EssD_{ki}^{-1/2}, ~~~~ k=1,\cdots,K-1, ~~~~ i=1,2;
\end{equation}
then from Theorem \ref{c1a2prop1ca},
\begin{eqnarray}\label{6baaa}
  & & \hat{C}_{ki}^{ 2}=-I, ~~~~ k=1,\cdots,K-1, ~~~~ i=1,2;  \\
   & & \hat{C}_{ki}\hat{C}_{li}=-\hat{C}_{li}\hat{C}_{ki}, ~~~~   1\le k\ne l \le K-1, ~~~ i=1,2 .
\end{eqnarray}
 and the sets of square matrices
$\{\hat{C}_{k1},k=1,2,\dots,K-1\}$ and  $\{
\hat{C}_{k2},k=1,2,\dots,K-1\}$ constitute  Hurwitz families of
order $r, N-r$ corresponding to $i=1,2$ respectively.
 Let $\hn{N}-1$ be the maximum number of matrices in a Hurwitz family of order $N$,
 then from the Hurwitz Theorem \cite{Jos} , $N=2^ab, b$ odd and
\begin{equation}\label{6c}
  \hn{N}=2a+2.
\end{equation}
Observe that due to the block diagonal structure of $C'_{k}$,
$K=\min\{\hn{r_i},\hn{N-r_i}\}$. Following  the Hurwitz Theorem it
is sufficient to consider both $r,N-r$ to be of the form $2^{a}$,
say $2^{a_1},2^{a_2}$ respectively. It follows that $K$ is
maximized iff $r= N-r=2^{a'}\Rightarrow N=2^{a'+1}$. It follows
that the maximum rate of RFSDD of size $N=2^a$ ($a=a'+1$) is
\begin{equation}\label{appa6d}
  \R=\frac{2a}{2^a}.
\end{equation}
\end{proof}
An important observation regarding square RFSDDs is summarized in
the following Corollary:
\begin{cor}
A maximal rate square RFSDD,
$S=\sum_{k=0}^{K-1}x_{kI}A_{2k}+x_{kQ}A_{2k+1}$ exists iff both
$\EssD_{2k}$ and $\EssD_ {2k+1}$ are not full-rank for all $k$.
\end{cor}
\begin{proof}
Immediate from the proof of above theorem.
\end{proof}

\noindent An immediate consequence of this characterization of
maximal rate RFSDDs is:
\begin{thm}\label{thmex3}
 A square RFSDD of rate-one, exists iff $N=2,4$.
\end{thm}
\begin{proof}
From (\ref{appa6d}) $\R=1$ iff $N=2,4$
\end{proof}
It follows that

\begin{thm}\label{thmex23}
 The maximal rate, $\R$, achievable by a square FSDD with
$N=2^ab, b$ odd (where $a,b > 0$) transmit antennas is
\begin{equation}\label{eqnthm23}
\R=\frac{2a}{2^ab}.
\end{equation}
Furthermore square GLCODs are not maximal rate FSDD except for
$N=2$.
\end{thm}

 Next we give a construction of square RFSDD that
achieves the maximal rates obtained in Theorem \ref{thmex2}.

\begin{thm}\label{thmex4}
 A square RFSDD $S$, of size $N$, in variables $x_{i},
i=0,\cdots, K-1$ achieving the rate of Theorem \ref{thmex2} is
given by
\begin{equation}\label{lb1}
S=\left[\begin{array}{cc} \underbrace{\Theta(\tilde{x}_0,\cdots,\tilde{x}_{K/2})}_{\Theta_1} &0\\
0 &
\underbrace{\Theta(\tilde{x}_{K/2},\cdots,\tilde{x}_{K-1})}_{\Theta_2}
\end{array}\right]
\end{equation}
where $\Theta(x_0,\cdots,x_{K/2-1})$ is a maximal rate square
 GLCOD of size $N/2$ \cite{TiH1,SuX2}, $\tilde{x}_i=\RE\{x_i\}+\J
\IM\{x_{({i+K/2})_K}\}$
 and where $(a)_K$ denotes $a \pmod K$.
\end{thm}
\begin{proof}
The proof is by direct verification.  As the maximal rate of
square GLCOD of size $N/2$ is $\frac{a}{2^{a-1}b}$
\cite{TiH1,SuX2} the rate of $S$ in (\ref{lb1}) is
$2\frac{a}{2^{a}b}=\frac{2a}{2^{a}b}$ and hence $S$ is maximal
rate. Next we show that $S$ is a RFSDD. Consider
\[
S^{\cal H}S=\left[\begin{array}{cc}
  \Theta_1^{\cal H}\Theta_1 & 0 \\
  0 &  \Theta_2^{\cal H}\Theta_2
\end{array}\right],
\]
by construction, the sum of weight matrices of $x_{kI}^2,x_{kQ}^2$
for any symbol $x_k$ is $I_N$ and (\ref{cond1})-(\ref{cond1l}) are
satisfied as $\Theta$ is a GLCOD. Therefore $S$ is a RFSDD.
\end{proof}

 Other square RFSDDs can be
constructed from (\ref{lb1}) by applying some of the following
\begin{description}
  \item[-] permuting rows and/or columns of (\ref{lb1}),
  \item[-] permuting the real symbols
  $\{x_{kI},x_{kQ}\}_{k=0}^{K-1}$,
  \item[-] multiplying a symbol by -1 or $\pm \J$
  \item[-] conjugating a symbol in (\ref{lb1}).
\end{description}
Following \cite[Theorem 2]{TiH1} we have

\begin{thm}\label{thmex5}
All square RFSDDs can be constructed from RFSDD $S$ of (\ref{lb1})
by possibly deleting rows from a matrix of the form
\begin{equation}\label{eqn7}
  S'=USV
\end{equation}
where $U, V$ are unitary matrices, up to permutations and possibly
sign change in the set of real and imaginary parts of the symbols.
\end{thm}
\begin{proof}
This follows from the observation  after (\ref{6b})  that the pair
of sets $\{C'_{ki}\}_{k=0}^{K-1}, i=1,2$ constitute a Hurwitz
family and Theorem 2 of \cite{TiH1} which applies to Hurwitz
families.
\end{proof}
It follows that the
 CIODs presented in Example \ref{ssdds1}
  are unique up to multiplication by unitary matrices. Moreover, observe that
   the square RFSDDs of Theorem \ref{thmex4} can be
thought of as designs combining co-ordinate interleaving and
GLCODs. We therefore, include such RFSDDs in the class of
co-ordinate interleaved orthogonal designs (CIODs), studied in
detail in the next section.
%%%%%%%%% non-square
%%%%%%%%%%%%%%%%%%%%Chapter 2 ends %%%%%%%%%%%%%%%%%%%%%%%%%%%%%%%%
%%%%%%%%%%%%%%%%%%%% Section  3 GCIOD %%%% begins %%%%%%%%%%%%%%%%%%
\section{Co-ordinate Interleaved Orthogonal Designs}\label{chap3}
In the Section \ref{ssdd} we characterized SDDs in terms of the
weight matrices. Among these we characterized a class of full-rank
SDD called FSDD and classified it into UFSDD and RFSDD. In the
previous section we derived and constructed maximal rate FSDDs.
However, we have not been able to derive the coding gain of the
either the class SDD or FSDD in general; the coding gain of GLCODs
is well-known. This section is devoted to an interesting class of
RFSDD $\subset$ FSDD called Co-ordinate Interleaved Orthogonal
Designs (CIODs) for which we will not only be able to derive the
coding gain but also the Maximum Mutual Information.

 We first give
an intuitive construction of the CIOD for two transmit antennas
and then formally  define the class of Co-ordinate Interleaved
Orthogonal Designs (CIODs) comprising of only symmetric designs
and its generalization, Generalized CIOD (GCIOD) which includes
both symmetric and non-symmetric (as special cases) designs in
Sub-section \ref{sec31}. Also, we show that rate-one GCIODs exist
for 2, 3 and 4 transmit antennas and for all other antenna
configurations the rate is strictly less than 1. A construction of
GCIOD is then presented which results in  rate 6/7 designs for 5
and 6 transmit antennas, rate 4/5 designs for 7 and 8 transmit
antennas and  rate $\frac{2(m+1)}{3m+1}$ GCIOD for $N=2m-3,2m-2
\ge 8$ corresponding to whether $N$ is odd or even. In Subsection
\ref{sec32} the signal set expansion associated with the use of
STBC from any co-ordinate interleaving when the uninterleaved
complex variables take values from a signal set is highlighted and
the notion of co-ordinate product distance (CPD) is discussed. The
coding gain aspects of the STBC from CIODs constitute Subsection
\ref{sec33} and we show that, for lattice constellations, GCIODs
have higher coding gain as compared to GLCODs. Simulation results
are presented in Subsection \ref{sec34}. The Maximum Mutual
Information (MMI) of GCIODs is discussed in Subsection \ref{sec35}
and is compared with that of GLCODs to show that, except for
$N=2$, CIODs have higher MMI. \textbf{In a nutshell this section
shows that, except for $N=2$ (the Alamouti code), CIODs are better
than GLCODs in terms of rate, coding gain, MMI and BER.}

%A brief discussion on the results of
%this section  constitute Section \ref{sec36}.
%%%%%%%%%%% Section 3.1 begins %%%%%%%%%%%%%%%%%%%%%%%%%%%%
\subsection{Co-ordinate Interleaved Orthogonal Designs}\label{sec31}
We begin from an intuitive construction of
the CIOD for two transmit antennas before giving a formal definition (Definition \ref{c2def1}).  Consider the Alamouti code
\[
S=\begin{bmatrix}
  x_0 & x_1 \\
-x_1^*& x_0^*
\end{bmatrix}.
\]
When the number of receive antennas $M=1$, observe that the diversity gain in the Alamouti code
is due to the fact that each symbol sees two different channels $h_{0}$ and
$h_{1}$ and the low ML decoding complexity is due to the use of the orthogonality of columns of signal transmission matrix, by the receiver,
over two symbol periods to form an estimate of each symbol.

Alternately, diversity gain may still be achieved by transmitting
quadrature components of each symbol separately on different
antennas. More explicitly, consider that the in-phase component,
$x_{0I}$, of a symbol, $x_0=x_{0I}+\J x_{0Q}$, is transmitted on
antenna zero and in the next symbol interval the quadrature
component, $x_{0Q}$, is transmitted from antenna one as shown in
Table \ref{c2tab1}.

 It is apparent that this procedure
is similar to that of co-ordinate interleaving (see Remark \ref{c2rem2} for references)
and that the symbol
has diversity two if the difference of the in-phase and quadrature
components is not-zero, but the rate is half. This loss of rate
can be compensated by choosing two symbols and exchanging their
quadrature components so that one co-ordinate of each symbol is
transmitted on one of the antennas as shown in Table \ref{c2tab2}.

As only one antenna is used at a time for transmission, the only
operation required at the receiver to decouple the symbols is to
exchange the quadrature components of the received signals for two
symbol periods after phase compensation.

The CIOD for four antennas is linked to the CIOD for two antennas in
a simple manner. The CIOD for two antennas uses complex symbols and
uses antenna cycling between antennas 0 and 1. For four antennas
consider antennas 0 and 1 as one set and antennas 2 and 3 as another
set. Using two antennas and complex symbols, we can transmit a
quaternion symbol (four co-ordinates) rather than a complex symbol
(two co-ordinates). After interleaving the co-ordinates of the
quaternion symbol we cycle between the first and second set of
antennas.

That the decoding
 is single-symbol decoding with the in-phase and quadrature-phase components
 having got affected by  noise components of different variances for any
 GCIOD is shown in Subsection \ref{c2coddec}. In the same subsection the
 full-rankness of GCIOD is also proved. If we combine, the Alamouti scheme
with co-ordinate interleaving we have the scheme for  4 transmit
antennas of Example \ref{ssdds1},% and whose baseband
%representation is shown in Fig. \ref{c2fig1}
and whose receiver structure is explained in detail in Example
\ref{4antreceiver}. Now, a formal definition of GCIODs follows:

%\begin{figure}
%\centerline{\psfig{figure=cciphfee.eps}} \caption{ Baseband
%representation of the CIOD for four transmit and the $j$-th
%receive antennas.} \label{c2fig1}
%\end{figure}

\begin{defn}
[GCIOD]
\label{c2def1} A  Generalized Co-ordinate Interleaved
Orthogonal Design (GCIOD) of size $N_1\times N_2$ in variables
$x_{i}, i=0,\cdots, K-1$ (where $K$ is even) is a $L \times N$
matrix $S({x}_{0},\cdots,{x}_{K-1})$, such that
\begin{equation}\label{c2eq1}
S=\left[\begin{array}{cc} \Theta_1(\tilde{x}_0,\cdots,\tilde{x}_{K/2-1}) &0\\
0 & \Theta_2(\tilde{x}_{K/2},\cdots,\tilde{x}_{K-1})
\end{array}\right]
\end{equation}
where $\Theta_1(x_0,\cdots,x_{K/2-1})$ and $\Theta_2(x_{K/2},\cdots,x_{K-1})$ are
GLCODs of size $L_1\times N_1$ and $L_2\times N_2$ respectively, with rates $K/2L_1,K/2L_2$  respectively, where $N_1+N_2=N$, $L_1+L_2=L$, $\tilde{x}_i=\RE\{x_i\}+\J
\IM\{x_{({i+K/2})_K}\}$
 and $(a)_K$ denotes $a \pmod K$.
 If $\Theta_1=\Theta_2$ then we call this design a
 \textbf{Co-ordinate interleaved orthogonal  design(CIOD)}\footnote{These designs were named as Co-ordinate interleaved orthogonal
design (CIOD) in \cite{KhR2,KhR3} since two different columns are
indeed orthogonal. However, the standard dot product of different
columns may be different whereas in conventional GLCODs apart from
orthogonality for two different columns, all the columns will have
the same dot product.}.
\end{defn}
Naturally, the theory of CIODs is simpler as compared to that
of GCIOD. Note that when $\Theta_1=\Theta_2$ and $N=L$ we
have the construction of square RFSDDs given in Theorem
\ref{thmex4}. Examples of square CIOD for $N=2,4$ were presented
in Example \ref{ssdds1}. 
\begin{eg}\label{c2ex1}
An example of GCIOD, where $\Theta_1 \ne
\Theta_2$ is  $S(x_0,\cdots,x_3)$
\begin{equation}\label{c2eq2} S= \left[\begin{array}{ccc}
x_{0I}+\J x_{2Q} & x_{1I}+\J x_{3Q} & 0 \\
-x_{1I}+\J x_{3Q} & x_{0I}-\J x_{2Q} & 0 \\
 0 &0 & x_{2I}+\J x_{0Q}  \\
 0 &0 & -x_{3I}+\J x_{1Q}
\end{array}\right]
\end{equation}
where $\Theta_1$ is the rate-one Alamouti code and $\Theta_2$ is the
trivial, rate-one, GLCOD for $N=1$ given by
\[
\Theta_2=\begin{bmatrix}
 x_0 \\
  -x_1^*
\end{bmatrix}.
\]
Observe that $S$ is non-square and rate-one.
 This code can also be thought  of as being obtained by dropping the last
column of the CIOD in (\ref{exeq1}). Finally, observe that
(\ref{c2eq2}) is not unique and we have different designs as we
take
\[
\Theta_2=\begin{bmatrix}
 x_0 \\
  x_1
\end{bmatrix},
\begin{bmatrix}
 x_0 \\
  -x_1
\end{bmatrix}
\] etc. for the second GLCOD.
\end{eg}
%%%%%%%%%%%%%%%%%%%%%%%%%%
\subsubsection{Coding and Decoding for STBCs from GCIODs}\label{c2coddec}
First, we show that every GCIOD is a RFSDD and hence is SD and achieves full diversity if the indeterminates take
values from a signal set with  non-zero CPD.
\begin{thm}\label{c2thmex4}
 Every GCIOD is an RFSDD.
\end{thm}
\begin{proof}
Let $S$ be a GCIOD defined in (\ref{c2eq1}). We have
\begin{eqnarray}
S^\h S&=&\left[\begin{array}{cc}
  \Theta_1^\h\Theta_1 & 0 \\
  0 &  \Theta_2^\h\Theta_2
\end{array}\right]\\
&=&\begin{bmatrix}
 a_kI_{N_1} & 0 \\
  0 & b_kI_{N_2}
\end{bmatrix}
\end{eqnarray}
where $a_k=\left(\sum_{k=0}^{K/2-1} x_{kI}^2+x_{(k+K/2)_KQ}^2\right)$ and $b_k=\left(\sum_{k=K/2}^{K-1} x_{kI}^2+x_{(k+K/2)_KQ}^2\right)$.
Observe that there are no terms of the form
$x_{kI}x_{kQ},x_{kI}x_{lQ}$ etc. in $S^\h S$, and therefore $S$ is
a SDD (this is clear from (\ref{crossterms})). Moreover, by
construction, the sum of weight matrices of $x_{kI}^2$ and
$x_{kQ}^2$ for any symbol $x_k$ is $I_N$ and hence $S$ is a FSDD.
Furthermore, for any given $k, 0 \le k \le K-1$ the  weight
matrices of both $x_{kI}^2,x_{kQ}^2$ are not full-rank and
therefore, by Definition \ref{c1def5}, $S$ is a RFSDD.
\end{proof}

The transmission scheme for a GCIOD, $S({x}_{0},\cdots,{x}_{K-1})$
of size $N$, is as follows: let $Kb$
 bits arrive at the encoder in a given time
slot. The encoder selects $K$ complex symbols, $s_i,i=0,\cdots,K-1$ from a
complex constellation $\A$ of size $|\A|=2^b$. Then setting
$x_i=s_i,i=0,\cdots,K-1$, the encoder populates the transmission
matrix with the complex symbols for the corresponding number of transmit antennas. The
corresponding transmission matrix is given by
$S({s}_{0},\cdots,{s}_{K-1})$.
 The received signal  matrix  (\ref{c0eq6}) is given by,
\begin{equation}\label{chmod1a}
  \bV=S \bH+\bW.
\end{equation}
Now as every GCIOD is a RFSDD (Theorem \ref{c2thmex4}), it is
SD and the receiver uses (\ref{met2}) to form
an estimate of each $s_i$ resulting in  the ML rule for each
$s_i,i=0,\cdots,K-1$, given by
\begin{equation}\label{c2dec1}
\min_{s_i \in \A}M_i(s_i)=\min_{s_i \in \A}\left\|{\mathbf
V}-(A_{2i}s_{iI}+A_{2i+1}s_{iQ}){\mathbf H}\right\|^2.
\end{equation}
\begin{rem}\label{c2decod}
Note that forming the ML metric for each variable  in (\ref{c2dec1}),
implicitly involves co-ordinate de-interleaving, in the same way
as the coding involves co-ordinate interleaving. Also notice that the components $s_{iI}$ and $s_{iQ}$ (i.e., the weight matrices that are not full-rank) have been weighted differently - something that does not happen for GLCODs. We elaborate these aspects of decoding GCIODs by considering the decoding of rate-one, CIOD for $N=4$ in detail.
\end{rem}
\begin{eg}[Coding and Decoding for CIOD for $N=4$]\label{4antreceiver}
\renewcommand{\T}{{\mathcal{T}}} Consider the CIOD for $N=4$ given
in (\ref{exeq1}). If the signals $s_0, s_1, s_2, s_3 \in {\cal A}$
are to be communicated, their interleaved version as given in
Definition \ref{c2def1} are transmitted. The signal transmission
matrix, $S$,  {\small
\begin{equation}\label{4mnew}
  S=\left[\begin{array}{cccc}
  \underbrace{s_{0I}+\J s_{2Q}}_{\tilde{s}_{0}} & \underbrace{s_{1I}+\J s_{3Q}}_{\tilde{s}_{1}} & 0 &0\\
-s_{1I}+\J s_{3Q} & s_{0I}-\J s_{2Q} & 0 &0\\
 0 &0 & \underbrace{s_{2I}+\J s_{0Q}}_{\tilde{s}_{2}} & \underbrace{s_{3I}+\J s_{1Q}}_{\tilde{s}_{3}} \\
 0 &0 & -s_{3I}+\J s_{1Q} &  s_{2I}-\J s_{0Q}
  \end{array}\right]
\end{equation}
}
is obtained by replacing $x_i$ in the CIOD by $s_i$ where
each $s_i,~i=0,1,2,3$ takes values from a signal set $\A$ with
$2^b$ points.

The received signals at the different time slots, $v_{jt},$
$t=0,1,2,3$ and $j=0,1,\cdots,M-1$ for the M receive antennas are
given by
\begin{eqnarray}\label{cite13}
v_{j0}&=&h_{0j}\tilde{s}_{0}+h_{1j}\tilde{s}_{1}+n_{j0}; \nonumber \\v_{j1}&=&-h_{0j}\tilde{s}_{1}^*+h_{1j}\tilde{s}_{0}^*+n_{j1}; \nonumber \\
 v_{j2}&=&h_{2j}\tilde{s}_{2}+h_{3j}\tilde{s}_{3}+n_{j2}; \nonumber \\ v_{j3}&=&-h_{2j}\tilde{s}_{3}^*+h_{3j}\tilde{s}_{2}^*+n_{j3}
\end{eqnarray}
where $n_{ji}~,i=0,1,2,3$ and $j=0,\cdots,M-1$ are complex
independent Gaussian random variables.

Let ${\bV}_j=[v_{j0},~v_{j1}^* ,~v_{j2},~ v_{j3}^*]^\T$,
$\tilde{{S}}=[\tilde{s}_{0},~\tilde{s}_{1},~\tilde{s}_{2},~\tilde{s}_{3}]^\T$,
${\bW}_j=[n_{j0},~n_{j1}^* ,~n_{j2},~ n_{j3}^*]^\T$ and $$
{\bH}_j=\left[\begin{array}{cccc}
    h_{0j}&h_{1j} &0&0\\
     h_{1j}^* &-h_{0j}^*&0&0\\
     0&0& h_{2j} &h_{3j}\\
     0&0&h_{3j}^*&-h_{2j}^*\
  \end{array}\right]
  $$ where $j=0,1,\cdots,M-1$.
  Using this notation,  (\ref{cite13}) can be written as
\begin{equation}\label{eqn41}
{\bV}_j={\bH}_j\tilde{{S}}+{\bW}_j.
\end{equation}
\noindent
Let
$$
\tilde{{\bV}}_j=[\tilde{v}_{j0},~\tilde{v}_{j1},~\tilde{v}_{j2},~
\tilde{v}_{j3}]^\T ={\bH}_j^\h   {\bV}_j. $$
 Then, we have
\begin{eqnarray}\label{eqn42}
\tilde{{\bV}}_j&=&\left[\begin{array}{cccc}
    \left(|h_{0j}|^2+|h_{1j}|^2\right)I_{2}&0 \\
     0&\left(|h_{2j}|^2+ |h_{3j}|^2\right)I_{2}\\
  \end{array}\right]\tilde{{S}} \nonumber \\
&&+{\bH}_j^\h{\bW}_j.
\end{eqnarray}
\noindent
Rearranging the in-phase and quadrature-phase components of
$\tilde{v}_{ji}$'s, (which corresponds to deinterleaving) define, for $i=0,1$,
\begin{eqnarray}\label{rec4M}
\hat{v}_i =\sum_{j=0}^{M-1}\tilde{v}_{ji,I}+\J\tilde{v}_{ji+2,Q}=a s_{i,I}+\J bs_{i,Q}+u_{0i} \\
\hat{v}_{i+2}=\sum_{j=0}^{M-1}\tilde{v}_{ji+2,I}+\J\tilde{v}_{ji,Q}=bs_{i+2,I}+\J
as_{i+2,Q}+u_{1i}\label{rec4Ml}
\end{eqnarray}
where $a=\sum_{j=0}^{M-1}\{|h_{0j}|^2+|h_{1j}|^2\}$,
 $b=\sum_{j=0}^{M-1}\{|h_{2j}|^2+|h_{3j}|^2\}$ and  $u_{0i}, u_{1i}$
are complex Gaussian random variables. Let $\tilde{{\bW}_j}=[\tilde{n}_{j0}~\tilde{n}_{j1}
~\tilde{n}_{j2} ~\tilde{n}_{j3}]^T={{\bH}_j^\h  {\bW}_j}$. Then $
u_{0i}=\sum_{j=0}^{M-1}\tilde{n}_{ji,I}+\J \tilde{n}_{ji+2,Q}$
 and $
u_{1i}=\sum_{j=0}^{M-1}\tilde{n}_{ji+2,I}+\J \tilde{n}_{ji,Q}$
where $i=0,1$. Note that $u_{00}$ and $u_{01}$ have the same
variance and
 similarly $u_{10}$ and $u_{11}$. The variance of
the in-phase component of $u_{00}$ is $a$ and that of the
quadrature-phase component is $b$. The in-phase component of
$u_{10}$ has the same variance as that of the quadrature-phase
component of $u_{00}$ and vice versa.
 The ML decision rule   for such a situation, derived
 in a general setting is:
Consider the received signal $r$, given by
\begin{equation}\label{mlder1}
r=c_1s_I+\J c_2s_Q+n
\end{equation}
 where $c_1, c_2$ are {\it real} constants and  $s_I,~s_Q$ are
in-phase and quadrature-phase components of transmitted signal
$s$.  The ML decision rule when the in-phase, $n_I$, and
quadrature-phase component, $n_Q$, of the Gaussian  noise, $n$
have different variances $c_1\sigma^2$ and $c_2\sigma^2$ is
derived by considering the pdf of $n$, given by
\begin{equation}\label{mlder1a}
p_{n}(n)=\frac{1}{2\pi\sigma^2\sqrt{c_1c_2}}e^{-\frac{n_I^2}{2c_1\sigma^2}}e^{-\frac{n_Q^2}{2c_2\sigma^2}}.
\end{equation}
The ML rule is: decide in favor of $s_i$, if and only if
\begin{equation}\label{mlder2}
  p_n(r/s_i) \ge p_n(r/s_k),~\forall ~i \ne k.
\end{equation}
Substituting from (\ref{mlder1}) and (\ref{mlder1a}) into
(\ref{mlder2}) and simplifying we have
\begin{eqnarray}\label{mlder3}
  c_2|r_I-as_{i,I}|^2&+&c_1|r_Q-bs_{i,Q}|^2 \le
  c_2|r_I-as_{k,I}|^2 \nonumber \\&&+c_1|r_Q-bs_{k,Q}|^2,  ~~ \forall ~i \ne k.
\end{eqnarray}
We use this by substituting $c_1=a$ and $c_2=b$, to obtain
(\ref{met41n}) and $c_1=b$ and $c_2=a$, to obtain (\ref{met42n}).
For $\hat{v}_j,~j=0,1$, choose signal $s_i \in \A$ iff
%\vspace*{-0.7cm}
\begin{eqnarray}\label{met41n}
  b|\hat{v}_{j,I}- a s_{i,I}|^2&+&
  a|\hat{v}_{j,Q}- bs_{i,Q}|^2   \le b|\hat{v}_{j,I}-
a s_{k,I}|^2 \nonumber \\ &+& a|\hat{v}_{j,Q}-bs_{k,Q}|^2, ~~ \forall ~i  \ne k
\end{eqnarray}
%\vspace*{-0.7cm}
and for $\hat{v}_j,~j=2,3$, choose signal $s_i$ iff %\vspace*{-0.7cm}
\begin{eqnarray}\label{met42n}
 a|\hat{v}_{j,I}- b s_{i,I}|^2&+&
  b|\hat{v}_{j,Q}- as_{i,Q}|^2   \le
  a|\hat{v}_{j,I}-b s_{k,I}|^2 \nonumber \\&&+ b|\hat{v}_{j,Q}-as_{k,Q}|^2, ~~~~ \forall ~i  \ne k.
\end{eqnarray}
From the above two equations it is clear that decoupling of the
variables is achieved by involving the de-interleaving operation
at the receiver in (\ref{rec4M}) and (\ref{rec4Ml}). Remember that
the entire decoding operation given in this example is equivalent
to using (\ref{c2dec1}). We have given this example only to bring
out the de-interleaving operation involved in the decoding of
GCIODs.
\end{eg}

Next we show that rate-one, GCIODs (and hence CIODs) exist for
$N=2,3,4$ only.
 \begin{thm}\label{c2thm1}
 A rate-one, GCIOD exists iff  $N=2,3,4$.
\end{thm}
\begin{proof}
First observe from (\ref{c2eq1}) that the GCIOD is rate-one iff the
GLCODs $\Theta_1, \Theta_2$ are rate-one. Following, Theorem
\ref{lix}, we have that a rate-one non-trivial GLCOD exist iff
$N=2$. Including the trivial GLCOD for $N=1$, we have that rate-one
GCIOD exists iff $N=1+1,1+2,2+2$, i.e. $N=2,3,4$.
\end{proof}
Next we construct GCIODs of rate greater than 1/2 for $N>4$. Using
the rate 3/4 GLCOD  i.e. by substituting $\Theta_1=\Theta_2$ by
the rate 3/4 GLCOD in (\ref{c2eq1}), we have rate 3/4 CIOD for 8
transmit antennas which is given in (\ref{c2eq24}).
\begin{table*}[!t]
\begin{equation}\label{c2eq24}
S(x),\cdots,x_5)=\begin{bmatrix}
   \Theta_4(x_{0I}+\J x_{3Q} , x_{1I}+\J x_{4Q}, x_{2I}+\J x_{5Q}) & 0 \\
  0 & \Theta_4(x_{3I}+\J x_{0Q}  , x_{4I}+\J x_{1Q}, x_{5I}+\J x_{2Q})
\end{bmatrix}.
\end{equation}
\end{table*}
Deleting one, two and three columns from $S$ we have rate 3/4 GCIODs for
$N=7,6,5$ respectively. Observe that by dropping columns of a CIOD
we get GCIODs and not CIODs. But the GCIODs for $N=5,6,7$ are not
 maximal rate designs that can be constructed from the Definition
\ref{c2def1} using known GLCODs.

Towards constructing higher rate GCIODs for $N=5,6,7$, observe
that the number of indeterminates of GLCODs $\Theta_1, \Theta_2$
in Definition \ref{c2def1} are equal. This is necessary for full-diversity so that the in-phase or the quadrature component of each
indeterminate, each seeing a different channel, together see all
the channels. The construction of such GLCODs for $N_1 \ne N_2$,
in general, is not immediate. One way is to set some of the
indeterminates in the GLCOD with higher number of indeterminates
to zero, but this results in loss of rate. We next give the
construction of such GLCODs which does not result in loss of
rate.
\begin{cons}\label{c2cons1}
Let $\Theta_1$ be a GLCOD of size $L_1 \times N_1$, rate
$r_1=K_1/L_1$ in $K_1$ indeterminates $x_0,\cdots,x_{{K_1}-1}$ and
similarly let $\Theta_2$ be a GLCOD of size $L_2 \times N_2$,
rate $r_2=K_2/L_2$ in $K_2$ indeterminates
$y_0,\cdots,y_{{K_2}-1}$. Let $K=\mathrm{lcm}(K_1,K_2)$,
$n_1=K/K_1$ and $n_2=K/K_2$. Construct
\begin{equation}\label{c2eq3}
  \hat{\Theta}_1=\left[\begin{array}{c}
    \Theta_1(x_0,x_1,\cdots,x_{{K_1}-1}) \\
    \Theta_1(x_{K_1},x_{K_1+1},\cdots,x_{2K_1-1}) \\
     \Theta_1(x_{2{K_1}},x_{{2K_1}+1},\cdots,x_{3{K_1}-1}) \\
    \vdots\\
    \Theta_1(x_{({n_1}-1){K_1}},x_{({n_1}-1){K_1}+1},\cdots,x_{{n_1}{K_1}-1}) \
  \end{array}\right]
\end{equation}
and
\begin{equation}\label{c2eq31}
  \mbox{  }
   \hat{\Theta}_2=\left[\begin{array}{c}
    \Theta_2(y_0,y_1,\cdots,y_{{K_2}-1}) \\
    \Theta_2(y_{K_2},y_{{K_2}+1},\cdots,y_{2{K_2}-1}) \\
     \Theta_2(y_{2{K_2}},y_{{2K_2}+1},\cdots,y_{3{K_2}-1}) \\
    \vdots\\
    \Theta_2(y_{({n_2}-1){K_2}},y_{({n_2}-1){K_2}+1},\cdots,y_{{n_2}{K_2}-1}) \
  \end{array}\right].
\end{equation}
\noindent
Then $\hat{\Theta}_1$ of size ${n_1L_1 \times N_1}$ is a GLCOD in
indeterminates $x_0,x_1, \cdots,x_{K-1}$ and $\hat{\Theta}_2$ of size ${n_2L_2 \times N_2}$ is a GLCOD in indeterminates $y_0,y_1,
\cdots,y_{K-1}$. Substituting these GLCODs in (\ref{c2eq1}) we
have a GCIOD of rate
\begin{eqnarray}\label{c2eq32}
\R&=&\frac{2K}{n_1L_1+n_2L_2} \nonumber \\&=&\frac{2\mathrm{lcm}(K_1,K_2)}{n_1L_1+n_2L_2}\nonumber\\
&=&\frac{2\mathrm{lcm}(K_1,K_2)}{\mathrm{lcm}(K_1,K_2)(L_1/K_1+L_2/K_2)} \nonumber \\&=&H(r_1,r_2)
\end{eqnarray}
where $H(r_1,r_2)$ is the Harmonic mean of $r_1,r_2$ with
$N=N_1+N_2$ and delay, $L=n_1L_1+n_2L_2$.
\end{cons}
We illustrate the Construction \ref{c2cons1} by
constructing a rate $6/7$ GCIOD for six transmit antennas in the
following example.
\begin{eg}\label{c2ex2}
Let
 \[
\Theta_1=\left[\begin{array}{cc} x_0 & x_1\\ -x_1^* &
  x_0^*\end{array}\right]
  \]
   be the Alamouti code. Then  $L_1=
  N_1=K_1=2$.
Similarly let
\[
\Theta_2=\begin{bmatrix}
   x_0 & x_1 & x_2 & 0 \\
  -x_1^* & x_0^* & 0 & x_2 \\
  -x_2^* & 0 & x_0^* & -x_1 \\
  0 & -x_2^* & x_1^* & x_0
\end{bmatrix}.
 \]
Then $L_2= N_2=4$, $K_2=3$ and the rate is $3/4$.
$K=\mathrm{lcm}(K_1,K_2)=6$, $n_1=K/K_1=3$ and $n_2=K/K_2=2$.
\begin{equation}\label{c2eq4}
  \hat{\Theta}_1=\left[\begin{array}{c}
    \Theta_1(x_0,x_1) \\
    \Theta_1(x_2,x_3) \\
    \Theta_1(x_4,x_5)
  \end{array} \right]=\begin{bmatrix}
    x_0 & x_1\\
    -x_1^* & x_0^* \\
    x_2 & x_3\\
     -x_3^* & x_2^*\\
     x_4 & x_5\\
     -x_5^* & x_4^*
  \end{bmatrix}.
\end{equation}
% \|^{ly}~
Similarly,
\begin{equation}
   \hat{\Theta}_2=\left[\begin{array}{cccc}
     x_0 & x_1 & x_2 & 0 \\
  -x_1^* & x_0^* & 0 & x_2 \\
  -x_2^* & 0 & x_0^* & -x_1 \\
  0 & -x_2^* & x_1^* & x_0\\
     x_3 & x_4 & x_5 & 0 \\
  -x_4^* & x_3^* & 0 & x_5 \\
  -x_5^* & 0 & x_3^* & -x_4 \\
  0 & -x_5^* & x_4^* & x_3
   \end{array}\right].
\end{equation}
The GCIOD for $N=N_1+N_2=6$ is given in (\ref{c2eq5}).
\begin{table*}[!t]
\begin{equation}\label{c2eq5}
S=\left[\begin{array}{cccccc}
    x_{0I}+\J x_{6Q} & x_{1I}+\J x_{7Q} & 0 & 0 & 0 & 0 \\
    -x_{1I}+\J x_{7Q} & x_{0I}-\J x_{6Q} & 0 & 0 & 0 & 0 \\
    x_{2I}+\J x_{8Q} & x_{3I}+\J x_{9Q} & 0 & 0 & 0 & 0 \\
     -x_{3I}+\J x_{9Q} & x_{2I}-\J x_{8Q} & 0 & 0 & 0 & 0 \\
     x_{4I}+\J x_{10Q} & x_{5I}+\J x_{11Q} & 0 & 0 & 0 & 0 \\
     -x_{5I}+\J x_{11Q} & x_{4I}-\J x_{10Q} & 0 & 0 & 0 & 0 \\
 0 & 0 &  x_{6I}+\J x_{0Q} & x_{7I}+\J x_{1Q} & x_{8I}+\J x_{2Q} & 0 \\
 0 & 0 & -x_{7I}+\J x_{1Q} & x_{6I}-\J x_{0Q} & 0 & x_{8I}+\J x_{2Q} \\
 0 & 0 & -x_{8I}+\J x_{2Q} & 0 & x_{6I}-\J x_{0Q}& -x_{7I}-\J x_{1Q} \\
 0 & 0 & 0 & -x_{8I}+\J x_{2Q} & x_{7I}-\J x_{1Q} & x_{6I}+\J x_{0Q}\\
 0 & 0 &x_{9I}+\J x_{3Q}& x_{10I}+\J x_{4Q} & x_{11I}+\J x_{5Q} & 0 \\
 0 & 0 & -x_{10I}+\J x_{4Q} & x_{9I}-\J x_{3Q} & 0 & x_{11I}+\J x_{5Q} \\
 0 & 0 & -x_{11I}+\J x_{5Q} & 0 & x_{9I}-\J x_{3Q} & -x_{10I}-\J x_{4Q} \\
 0 & 0 & 0 &  -x_{11I}+\J x_{5Q} & x_{10I}-\J x_{4Q} & x_{9I}+\J x_{3Q}
\end{array}\right].
\end{equation}
\end{table*}
The rate of the GCIOD in (\ref{c2eq5}) is $\frac{12}{14}=\frac{6}{7}=0.8571
> 3/4 $. This increased rate comes at the cost of additional delay. While the rate
3/4 CIOD for $N=6$ has a delay of 8 symbol durations, the rate 6/7
GCIOD has a delay of 14 symbol durations. In other words, the rate
$3/4$ scheme is \textbf{delay-efficient}, while the rate 6/7
scheme is \textbf{rate-efficient}\footnote{Observe that we are not
in a position to  comment on the optimality of both the delay and
the rate.}. Deleting one of the columns we have a rate 6/7 design
for 5 transmit antennas.
 \end{eg}
Similarly, taking $\Theta_1$ to be the Alamouti code and
$\Theta_2$ to be the rate $2/3$ design of \cite{Lia} in
Construction \ref{c2cons1}, we have a CIOD for $N=7$ whose rate is
given by
\[
\R=\frac{2}{3/2+1}=\frac{4}{5}=0.8.
\]
%The delay for this scheme is 36 symbol durations.
We have the following theorem:
\begin{thm}\label{thmgciodr}
 The maximal rate of GCIOD for $N=n+2$ antennas, $\R$ is lower bounded
 as $\R \ge \frac{2(m+1)}{3m+1}$ where $m=n/2$ if $n$ is even or
$m=(n+1)/2$ if $n$ is odd.
\end{thm}
\begin{proof}
We need to prove that a GCIOD of  rate $\R \ge
\frac{2(m+1)}{3m+1}$ where $m=n/2$ if $n$ is even or $m=(n+1)/2$
if $n$ is odd exists.

Consider Construction \ref{c2cons1}. For a given $N$, Let $
\Theta_1$
   be the Alamouti code. Then  $L_1=
  N_1=K_1=2$ and $N_2=N-2$.
Let $ \Theta_2$ be the GLPCOD for $n=N-2$ transmit antennas with
rate $r_2=\frac{m+1}{2m}$ where $m=n/2$ if $n$ is even or
$m=(n+1)/2$ if $n$ is odd \cite{Lia}.  The corresponding rate of
the GCIOD is given by
\[
\R=\frac{2}{\frac{2m}{m+1}+1}=\frac{2(m+1)}{3m+1}.
\]
\end{proof}
%For, $N=8$ the maximum rate obtained using known GLCODs is $5/8$.
Significantly, \textbf{there exist CIOD and GCIOD  of rate greater
that 3/4 and less than 1, while no such GLCOD is known to exist}.
Moreover for different choice of $\Theta_1$ and $\Theta_2$ we have
GCIODs of different rates. For example:
\begin{eg}\label{c2ex3}
For a given $N$, Let $ \Theta_1$
   be the Alamouti code. Then  $L_1=
  N_1=K_1=2$ and $N_2=N-2$.
Let $ \Theta_2$ be the rate  1/2 GLPCOD for $N-2$ transmit
antennas (either using the construction of \cite{TJC} or
\cite{GaS3}). Then $r_2=1/2$. The corresponding rate of the GCIOD
is given by
\[
\R=\frac{2}{2+1}=\frac{2}{3}.
\]
\end{eg}
 In Table \ref{c2tab3}, we present the
rate comparison between GLCODs and CIODs-both rate-efficient and
delay efficient; and in Table \ref{c2tab4}, we present the delay
comparison.

Observe that both in terms of delay and rate GCIODs are superior
to GLCOD.
%%%%%%%  Section 3.2  begins %%%%%%%%%%%%%%%%%%%%%%%%%%%%%%%
\subsubsection{GCIODs  vs. GLCODs}\label{sec32}
In this subsection we summarize the differences between the GCIODs
and GLCODs with respect to different aspects including signal set
expansion, orthogonality and peak to average power ratio (PAPR).
Other aspects like coding gain,  performance comparison using
simulation results and maximum mutual information are presented in
subsequent sections.

As observed earlier,  a STBC is obtained from the GCIOD by
replacing $x_i$ by $s_i$ and allowing each $s_i$,
$i=0,1,\cdots,K-1$, to take values from a signal set ${\cal A}$.
 For notational simplicity we will use only $S$ for $S(x_0,\cdots,x_{K-1})$
dropping the arguments, whenever they are clear from the context.

The following list highlights and compares the salient features of GCIODs and GLCODs:

\begin{itemize}
\item Both GCIOD and GLCOD are FSDD and hence STBCs from these designs are SD.
\item GCIOD is a RFSDD and hence STBCs from GCIODs achieve full-diversity iff $CPD$ of $\A$ is not
equal to zero. In contrast STBCs from GLCODs achieve full-diversity for all
$\A$.
\item \textbf{Signal Set Expansion:}
For STBCs from GCIODs, it is important to note that when the variables
$x_i,~ i=0,1,\cdots, K-1$, take values from a complex signal set ${\cal
A}$ the transmission matrix have entries which are co-ordinate
interleaved versions of the variables and hence the actual signal
points transmitted are not from  ${\cal A}$ but from an {\em
expanded version of ${\cal A}$} which we denote by $ \tilde {\cal
A}$. Figure \ref{c2fig2}(a) shows $ \tilde {\cal A}$ when ${\cal
A}=\{1, -1, {\mathbf j}, -{\mathbf j} \}$ which is shown in Figure
\ref{c2fig2}(c). Notice that $ \tilde {\cal A}$ has 8 signal
points whereas  ${\cal A}$ has 4. Figure \ref{c2fig2}(b) shows $
\tilde {\cal A^\prime}$ where ${\cal A}^\prime$ is the four point
signal set obtained by rotating ${\cal A}$ by 13.2825 degrees
counter clockwise i.e., ${\cal A}^\prime = \{e^{{\mathbf
j}\theta}, -e^{{\mathbf j}\theta},{\mathbf j}e^{{\mathbf
j}\theta}, -{\mathbf j}e^{{\mathbf j}\theta}\}$ where
$\theta=13.2825$ degrees as shown in Figure \ref{c2fig2}(d).
Notice that now the expanded signal set has 16 signal points  (The
value $\theta=13.2825$ has been chosen so as to maximize the
parameter called Co-ordinate Product Distance of the signal set
which is related to diversity and coding gain of the STBCs from GCIODs,
discussed in detail in Section \ref{sec33}). It is easily seen
that $|\A'|\le|\A|^2$.

Now for GLCOD, there is  an expansion of signal set, but $|\A
'|\le 2|\A|$. For example consider the Alamouti scheme, for the
first time interval  the symbols are from the signal set $\A$ and
for the next time interval symbols are from $\A^*$, the conjugate
of symbols of $\A$. But for constellations derived from the square
lattice $|\A '| << 2|\A|$ and in particular for square QAM $|\A '|
= |\A|$. So the transmission is from a larger signal set for
GCIODs as compared to GLCODs.

\item Another important aspect to notice  is that for GCIODs,
during the first $L/2$ time intervals $N_1 < N $ of the $N$
antennas transmit and the remaining $N_2=N-N_1$ antennas  transmit
nothing and vice versa. So, on an average half of transmit
antennas are idle.

\item For GCIODs, $S$, is not an scaled orthonormal  matrix but is an orthogonal matrix
while for  square GLCODs, $S$, is scaled orthonormal. For example
when $S$ is the CIOD  given by (\ref{4mnew}) for $N=4$ transmit
antennas,
\begin{table*}[!t]
\begin{equation}\label{ciodnotorthogonal}
  { S}^{\cal H}{ S}=\left[\begin{array}{cccc}
    |\tilde{x}_0|^2+ |\tilde{x}_1|^2    & 0 &0             &0\\
     0 &|\tilde{x}_0|^2+ |\tilde{x}_1|^2   &0          &0\\
     0              &0            & |\tilde{x}_2|^2+ |\tilde{x}_3|^2  &0  \\
     0              &0            &0  & |\tilde{x}_2|^2+ |\tilde{x}_3|^2     \\
  \end{array}\right].
\end{equation}
\end{table*}
\item GCIODs out perform GLCODs for $N > 2$ both in terms of rate and delay
as shown in Tables \ref{c2tab3} and \ref{c2tab4}.

\item Due to the fact that
at least half of the entries of GCIOD are zero, the peak-to-average
power ratio for any one antenna is high compared to those STBCs
obtained from GLCODs. This can be taken care of by ``power
uniformization'' techniques as discussed in \cite{TiH1} for GLCODs
with some zero entries.
\end{itemize}

%%%%%%%  Section 3.3 begins  %%%%%%%%%%%%%%%%%%%
\subsection{Coding Gain and Co-ordinate Product Distance (CPD)}\label{sec33}
%\vspace*{-1.2cm}
In this section we  derive the conditions under which the coding
gain of the STBCs from GCIODs is maximized. Recollect from Section
\ref{ssdd} that since GCIOD and CIOD are RFSDDs, they achieve
full-diversity iff CPD of $\A$ is non-zero. Here, in Subsection
\ref{cod}  we show that the coding gain defined in (\ref{pr3}) is
equal to a quantity, which we call, the Generalized CPD (GCPD)
which is a generalization of CPD. In Subsection \ref{cpdmax} we
maximize the CPD for lattice constellations by rotating the
constellation\footnote{ The optimal rotation for 2-D QAM signal
sets is derived in \cite{BoV} using Number theory and Lattice
theory. Our proof is simple and does not require mathematical
tools from Number theory or Lattice theory. }. Similar results are
also obtained for the GCPD for some particular cases.
 We then  compare the coding gains of STBCs from both GCIODs and GLCODs in Subsection  \ref{codcompare}
 and
show that, except for $N=2$, GCIODs have higher coding gain as
compared to GLCODs for lattice constellations at the same spectral
efficiency in bits/sec/Hz.
%%%%%%%%%%%% Subsection 331} begins %%%%%%%%%%%%%%%%%%%%%%%%%%%%%
\subsubsection{Coding Gain of GCIODs}\label{cod}
Without loss of generality, we assume that the GLCODs  $\Theta_1,\Theta_2$  of Definition
\ref{c2def1} are such that their weight matrices are unitary. Towards obtaining an expression for the coding gain of CIODs, we first
introduce
\begin{defn}[Generalized Co-ordinate Product Distance]\label{c2def3}
For arbitrary positive integers $N_1$ and $N_2$, the Generalized Co-ordinate Product Distance (GCPD) between any two signal points
 $u=u_I+{\mathbf j}u_Q$ and $v=v_I+{\mathbf j}v_Q$, $u \ne v$ of the signal set
  ${\cal A}$ is defined in (\ref{gcpddef1})
\begin{table*}[t]
\begin{eqnarray}\label{gcpddef1}
GCPD_{N_1,N_2}(u,v)&=&\min\left\{|u_I - v_I|^{\frac{2N_1}{N_1+N_2} }|u_Q
- v_Q|^{\frac{2N_2}{N_1+N_2}  },%\right.\nonumber\\&&\left.
 |u_I - v_I|^{\frac{2N_2}{N_1+N_2} }|u_Q - v_Q|^{ \frac{2N_1} {N_1+N_2}}\right\}
\end{eqnarray}
\end{table*}
and the minimum of this value among all possible pairs of distinct signal points of the signal set ${\cal A}$ is defined
as the GCPD of the signal set and will be denoted by $GCPD_{N_1,N_2}({\cal A})$ or simply by $GCPD_{N_1,N_2}$ when the signal set under consideration is clear from the context.
\end{defn}
\begin{rem}\label{c2rem1}
Observe that
\begin{enumerate}
  \item  When $N_1=N_2$,  the GCPD reduces to the
CPD defined in Definition \ref{c1def3} and is independent of both $N_1$ and $N_2$.
  \item $GCPD_{N_1,N_2}(u,v)$ = $GCPD_{N_2,N_1}(u,v)$ for any two signal points $u$ and $v$ and hence $GCPD_{N_1,N_2}({\cal A})$ = $GCPD_{N_2,N_1}({\cal A})$.

\end{enumerate}
\end{rem}
We have,
\begin{thm}\label{c2thm2}
The coding gain of a full-rank GCIOD with the variables taking
values from a signal set, is equal to the $GCPD_{N_1,N_2}$ of that
signal set.
\end{thm}
\begin{proof}
For a GCIOD  in Definition \ref{c2def1} we have,
\begin{eqnarray}\label{c2eq6}
S^{\cal H}S &=&\left[\begin{array}{cc}a_KI_{N_1} &0\\
0 &b_KI_{N_2}
\end{array}\right],\\
 a_K&=&|\tilde{x}_0|^2+\cdots+|\tilde{x}_{K/2-1}|^2,\nonumber\\
b_k&=&|\tilde{x}_{K/2}|^2+\cdots+|\tilde{x}_{K-1}|^2\nonumber
\end{eqnarray}
where $\tilde{x}_i=\RE\{x_i\}+\J \IM\{x_{({i+K/2})_K}\}$
 and where $(a)_K$ denotes $a \pmod K$. Consider the codeword difference matrix $B({\mathbf
S},{\mathbf S}^\prime)={\mathbf S}- {{\mathbf S}^\prime}$ which is
of full-rank for two distinct codeword matrices ${\mathbf
S},{{\mathbf S}^\prime}$. We have 

\begin{eqnarray}\label{c2eq7}
B^{\cal H}({\mathbf S},{\mathbf S}^\prime)B({\mathbf S},{\mathbf
S}^\prime)&=&  
 \left[\begin{array}{cc}\triangledown a_KI_{N_1} &0\\
0 &
\triangledown b_K I_{N_2}
\end{array}\right],
\end{eqnarray}
\begin{eqnarray*}
\triangledown a_K &=& |\tilde{x}_0 -\tilde{x}_0'|^2+\cdots+|\tilde{x}_{K/2-1}-\tilde{x}_{K/2-1}'|^2),\nonumber\\
\triangledown b_K& =& (|\tilde{x}_{K/2}-\tilde{x}_{K/2}'|^2 +\cdots+|\tilde{x}_{K-1}-\tilde{x}_{K-1}'|^2)\nonumber
\end{eqnarray*}
 where at least one $x_k$ differs from  $x_k'$, $k=0,\cdots,K-1$.
Clearly, the terms $(|\tilde{x}_0
-\tilde{x}_0'|^2+\cdots+|\tilde{x}_{K/2-1}-\tilde{x}_{K/2-1}'|^2)$
and
$(|\tilde{x}_{K/2}-\tilde{x}_{K/2}'|^2+\cdots+|\tilde{x}_{K-1}-\tilde{x}_{K-1}'|^2)
$ are both minimum iff $x_k$ differs from  $x_k'$ for only one
$k$. Therefore assume, without loss of generality, that the
codeword matrices ${\mathbf S}$ and ${\mathbf S^\prime}$ are such
that they differ by only one variable, say $x_0$ taking different
values from the signal set ${\cal A}$. Then, for this case,
\begin{eqnarray*}
\Lambda_1&=&\det\left\{B^{\cal H}({\mathbf S},{\mathbf S}^\prime)B({\mathbf
S},{\mathbf S}^\prime)\right\}^{1/N}\\
&=&|x_{0I} -
{x^\prime}_{0I}|^\frac{2N_1}{N_1+N_2}|x_{0Q} -
{x^\prime}_{0Q}|^\frac{2N_2}{N_1+N_2}.
\end{eqnarray*}
Similarly, when ${\mathbf S}$ and ${\mathbf S^\prime}$ are such
that they differ by only in $x_{K/2}$  then
\begin{eqnarray*}
\Lambda_2&=&\det\left\{B^{\cal H}({\mathbf S},{\mathbf S}^\prime)B({\mathbf
S},{\mathbf S}^\prime)\right\}^{1/N}\\
&=&|x_{K/2I} -
{x^\prime}_{K/2I}|^\frac{2N_2}{N_1+N_2}|x_{K/2Q} -
{x^\prime}_{K/2Q}|^\frac{2N_1}{N_1+N_2}
\end{eqnarray*}
 and the
coding gain is given by $
\min_{x_0,x_{K/2} \in \A} \left\{\Lambda_1,\Lambda_2 \right\}=GCPD_{N_1,N_2}$.
\end{proof}
An important implication of the above result is,
\begin{cor}\label{c2cor1}
The coding gain of a full-rank STBC from a CIOD with the variables
taking values from a signal set, is equal to the CPD of that
signal set.
\end{cor}
\begin{rem}\label{c2rem1a}
Observe that  the CPD is independent of the  parameters $N_1,N_2$
and is dependent only on the elements of the signal set. Therefore
the coding gain of STBC from CIOD is independent of the CIOD. In
contrast, for GCIOD the coding gain is a function of $N_1, N_2$.
\end{rem}
The full-rank condition of RFSDD i.e. $CPD \ne 0$ can be
restated for GCIOD as
\begin{thm}\label{c2thm23}
The  STBC  from GCIOD with variables taking values from a signal
set achieves full-diversity  iff the $GCPD_{N_1,N_2}$ of that
signal set is non-zero.
\end{thm}
It is important to note that the $GCPD_{N_1,N_2}$ is non-zero iff
the CPD is non-zero and consequently, {\bf this is not at all a
restrictive condition, since given any signal set ${\cal A}$, one
can always get the above condition satisfied by rotating it. In
fact, there are infinitely many angles of rotations that will
satisfy the required condition and only finitely many which will
not. Moreover, appropriate rotation leads to more coding gain
also.} From this observation it follows that  signal constellations with
$CPD=0$ and hence $GCPD=0$ like regular $M-ary~QAM$, symmetric $M-ary~ PSK$  will not
achieve full-diversity.  But the situation gets  salvaged by
simply rotating the signal set to get this condition satisfied as
also indicated in  \cite{JeR1,JeR2,BoB}. This result is similar to the ones
on co-ordinate interleaved schemes like co-ordinate interleaved
trellis coded modulation \cite{JeR1,JeR2} and bit and co-ordinate
interleaved coded modulation \cite{Sli1}-\cite{ChR}, \cite{KhR9}
for single antenna transmit systems.

\subsubsection{Maximizing CPD and GCPD for Integer Lattice constellations}\label{cpdmax}
In this subsection  we derive the optimal angle of rotation for
QAM constellation so that the $CPD$ and hence the coding gain
 of CIOD is maximized.
 We then generalize the derivation
  so as to present a method to maximize the
 $GCPD_{N_1,N_2}$.

 \subsubsection{Maximizing CPD}
 In the previous section we showed that the coding gain of CIOD is
 equal to the CPD and that constellations with non-zero CPD can be
 obtained by rotating the constellations with zero CPD.
Here we obtain the optimal angle of rotation for lattice
constellations analytically. It is noteworthy that the  optimal
performance of
    co-ordinate interleaved TCM for the 2-D QAM constellations considered \cite{JeR1,JeR2}, using simulation
   results was observed at $32^\circ$; analytically, the optimal angle of rotation derived herein
   is  $\theta=\tan(2)/2=31.7175^\circ$ for 2-D QAM
   constellations. The error is probably due to the incremental
   angle being greater than or equal to 0.5.
   We first derive the result for square QAM.
\begin{thm}\label{c2thm4}
Consider a square QAM constellation $\A$, with signal points from
the square lattice $(2k-1-Q)d+\J(2l-1-Q)d$ where $k,l \in [1,Q]$
and $d$ is chosen so that the average energy of the QAM
constellation is 1. Let  $\theta$ be the angle of rotation. The
maximum $CPD$ of $\A$ is obtained at
$\theta_{opt}=\frac{\arctan(2)}{2}=31.7175^\circ$ and is given by
\begin{equation}\label{cpd1}
  CPD_{opt}=\frac{4 d^2}{\sqrt{5}}.
\end{equation}
\end{thm}
\begin{proof}
The proof is in three steps. First we derive the optimum value of
$\theta$ for 4-QAM, denoted as $\theta_{opt}$ (the corresponding
$CPD$ is denoted as $CPD_{opt}$).
 Second, we show that at $\theta_{opt}$, $CPD_{opt}$ is in-fact the $CPD$ for all other (square) QAM.
 Finally, we show that for any other value of $\theta \in [0,\pi/2]$, $CPD < CPD_{opt}$
 completing the proof.

\noindent
{\bf Step 1:} Any point P$(x,y) \in \Real^2$ rotated by an angle $\theta
\in [0,90^\circ]$ can be written as
\begin{equation}\label{cpd2}
  \left[\begin{array}{c}
    x_R \\
    y_R \
  \end{array}\right]=\underbrace{\left[\begin{array}{cc}
    \cos{\theta} & \sin{\theta} \\
    -\sin{\theta} & \cos{\theta} \
  \end{array}\right]}_R\left[\begin{array}{c}
    x \\
    y \
  \end{array}\right].
\end{equation}
Let $P_1(x_1,y_1), P_2(x_2,y_2)$ be two distinct points in $\A$
such that $\triangle x=x_1-x_2, \triangle y =y_1-y_2$. Observe
that $\triangle x, \triangle y = 0, \pm 2 d, \cdots, \pm 2(Q-1)d$.
We may write $\triangle x= \pm 2 m d,\triangle y=\pm 2 n d, m,n
\in [0,Q-1]$ but both $\triangle x, \triangle y$ cannot be zero
simultaneously, as $P_1,P_2$ are distinct points in $\A$.
 Since, rotation is a linear
operation,
\begin{equation}\label{cpd2a}
  \left[\begin{array}{c}
    \triangle x_r \\
    \triangle y_r \
  \end{array}\right]=R\left[\begin{array}{c}
    \triangle x \\
    \triangle y \
  \end{array}\right],
\end{equation}
where $\triangle x_r ={x_1}_R-{x_2}_R, \triangle y_r
={y_1}_R-{y_2}_R$. The CPD between points $P_1$ and $P_2$ after rotation, denoted by $CPD(P_{1r},P_{2r})$, is then given by
\begin{eqnarray}\label{cpd3}
  CPD(P_{1r},P_{2r})&=&|\triangle x_r||\triangle y_r|
= \large \left|\begin{array}{c}
    \\
\\
\end{array}\triangle x\triangle y \cos(2 \theta)\right.\nonumber\\
&& +\left.
  \frac{(\triangle x)^2-(\triangle y)^2}{2}\sin(2\theta)\right|.
\end{eqnarray}
For 4-QAM, possible values of $CPD(P_{1r},P_{2r})$ are
\begin{eqnarray}\label{cpd4}
   CPD_1(P_{1r},P_{2r})&=&2 d^2 |\sin(2 \theta)|,\nonumber\\
CPD_2(P_{1r},P_{2r})&=&4 d^2 |\cos(2 \theta)|.
\end{eqnarray}
Fig. \ref{c2fig3} shows the plots of both $CPD_1$ and $CPD_2$. As sine is an increasing function and cosine a decreasing function
of $\theta$ in the first quadrant,  equating $CPD_1, CPD_2$ gives
the optimal angle of rotation, $\theta_{opt}$. Let $CPD(\theta)$
be the $CPD$ at angle $\theta$ and
$CPD_{opt}=\max_{\theta}CPD(\theta)$. It follows that
$\theta_{opt}=\frac{\arctan(\pm 2)}{2}=31.7175^\circ,58.285^\circ$
and $CPD_{opt}=2d^2 \sin(2 \theta_{opt})=4d^2 \cos(2
\theta_{opt}) = \frac{4d^2}{\sqrt{5}}$.

\noindent
{\bf Step 2:} Substituting the optimal values of $\sin(2 \theta_{opt}),
\cos(2 \theta_{opt})$ in (\ref{cpd3}) we have for any two arbitrary points of a square QAM constellation,
\begin{equation}\label{cpd5}
  CPD(P_{1r},P_{2r})=\frac{4d^2}{\sqrt{5}}
  \left|\pm nm +
  n^2-m^2\right| \mbox{ where } n,m \in \Z
\end{equation}
and both $n,m$ are not simultaneously zero  and $\Z$ is the set of
integers. It suffice to show that $$|\pm nm +
  n^2-m^2| \ge 1  ~~~~ \forall n,m  $$ provided both $n,m$ are not simultaneously zero.
  We consider the $\pm$ case separately.
  We have
\begin{eqnarray*}
     |nm +n^2-m^2|&=&\left|\left(n+\frac{m}{2}\right)^2-\left(1+\frac{1}{4}\right)m^2\right|\\
      &=&\left|\left(n+\frac{m}{2}\{1+\sqrt{5}\}\right)\right. \\&&\left.\left(n+\frac{m}{2}\{1-\sqrt{5}\}\right)\right|,\\
\end{eqnarray*}
Similarly,
\begin{eqnarray*}
|-nm +n^2-m^2|&=&\left|\left(n-\frac{m}{2}\{1-\sqrt{5}\}\right)\right.\\&&\left.
   \left(n-\frac{m}{2}\{1-\sqrt{5}\}\right)\right|.
\end{eqnarray*}
The quadratic equation in $n,~~~~ |\pm nm +n^2-m^2|=0$ has roots
$$
n=\frac{m}{2}\{\pm  1\pm\sqrt{5}\}.
$$
Since $n,m \in \Z$,  $|\pm nm +n^2-m^2| \in \Z$ and is equal to
zero only if $ n=0,\frac{m}{2}\{\pm  1\pm\sqrt{5}\}.$ Necessarily,
$ |\pm nm +n^2-m^2|\ge 1$ for  $n,m \in \Z$    and both $n,m$ are
not simultaneously zero. Therefore  $\theta_{opt}$ and $CPD_{opt}$ continue to be the optimum values of angle and the CPD for any square QAM.

\noindent
{\bf Step 3:} Next we prove that for all other values of $\theta \in
[0,\frac{\pi}{2}]$, $CPD(\theta) < CPD_{opt}$. To this end,
observe that for any value of $\theta$ other than $\theta_{opt}$
either $CPD_1$ or $CPD_2$ is less than $CPD_{opt}$ (see the
attached plot of $CPD_1,CPD_2$ in Fig. \ref{c2fig3}). It follows
that
$$
CPD(\theta) \le CPD_{opt}
$$
with equality iff $\theta=\theta_{opt}$.
\end{proof}
 Observe that Theorem \ref{c2thm4} has application in
  all schemes where the performance
depends on the $CPD$ such as those in \cite{KhR4},
\cite{Goe}, \cite{ChR}, \cite{JeR1,JeR2}, etc. and the references
therein.

\noindent {\bf Remark:} The 4 QAM constellation in Fig. \ref{c2fig2}(c) is a
rotated version (45$^\circ$) of the QAM signal set considered in
Theorem \ref{c2thm4}.

Next we generalize Theorem \ref{c2thm4} to all integer lattice
constellations obtainable from a square lattice. We first find
constellations that have the same CPD as the square QAM of which it
is a subset. Towards that end we define,
\begin{defn}[NILC ] A  Non-reducible integer lattice constellation
(NILC) is a finite subset of the
 square lattice,
$(2k)d+\J(2l)d$ where $k,l \in \mathbb{Z}$, such that
there exists at least a pair of signal
points $p_1=(2k_1)d+\J(2l_1)d$ and
$p_2=(2k_2)d+\J(2l_2)d$ such that either $|k_1-k_2|=1, |l_1-l_2|=0$ or
$|l_1-l_2|=1,|k_1-k_2|=0 $.
\end{defn}
 We have,
\begin{cor}\label{corapp1}
 The
$CPD$ of a non-reducible integer lattice constellation, $\A$, rotated by
an angle $\theta$,
 is maximized at $\theta=\frac{\arctan(2)}{2}=31.7175^\circ$
and is given by
\begin{equation}\label{cpd6}
  CPD_{opt}=\frac{4 d^2}{\sqrt{5}}.
\end{equation}
\end{cor}
\begin{proof}
Since $\A$ is a subset of an appropriate square QAM constellation,
we immediately have from Theorem \ref{c2thm4}
\begin{equation}\label{cpd7}
  CPD_{opt}\ge \frac{4 d^2}{\sqrt{5}}.
\end{equation}
We only need to prove the equality condition. The CPD between any
two points NILC at $\theta_{opt}$ is given by (\ref{cpd5})
\begin{equation}\label{cpd8}
  CPD(P_1,P_2)=\frac{4d^2}{\sqrt{5}}
  \left|\pm nm +
  n^2-m^2\right| \mbox{ where } n,m \in \Z.
\end{equation}
Since for NILC there exists at least a pair of signal points
$p_1=(2k_1)d+\J(2l_1)d$ and
$p_2=(2k_2)d+\J(2l_2)d$ such that either $|k_1-k_2|=1, |l_1-l_2|=0$ or
$|l_1-l_2|=1, |k_1-k_2|=0$, we have $CPD(p_1,p_2)=\frac{4d^2}{\sqrt{5}}$.
\end{proof}
In addition to the NILCs, the lattice constellations that are a
proper subset of the scaled rectangular lattices,$(4k)
d+\J(2l) d$ and $(2k)d+\J(4l)d$ where $k,l \in
\mathbb{Z}$ have CPD equal to $\frac{4d^2}{\sqrt{5}}$. All other
integer lattice constellations have $CPD>\frac{4d^2}{\sqrt{5}}$.
%%%%%%%%%%%%%%%%%%
\subsubsection{Maximizing the GCPD of the QPSK signal set}
To derive the optimal angles of rotation for maximizing the GCPD we consider only QPSK, since the optimal angle is not the same for any square QAM, as is the case with CPD.
\begin{thm}\label{c2thm6}
Consider a QPSK constellation $\A$, with signal points
$(2k-3)d+\J(2l-3)d$ where $k,l \in [1,2]$ and $d=1/\sqrt{2}$,
rotated by an angle $\theta$ so as to maximize the $GCPD_{N_1,N_2}$.
The $GCPD_{N_1,N_2}({\cal A})$ is maximized at
$\theta_{opt}=\arctan(x_0)$ where $x_0$ is the positive root of
the equation
\begin{equation}\label{gcpd1}
  \left(1-\frac{1}{x} \right)^{2N_1}\left(1+{x} \right)^{2N_2}=1
\end{equation}
where $N_1 >N_2$ and the corresponding $GCPD_{N_1,N_2}({\cal A})$  is $4d^2
\left(\frac{x_0^{\frac{2N_1}{N_1+N_2}}}{1+x_0^2}\right)$.
\end{thm}
\begin{proof}
Following the same notations as in Step 1 of Theorem \ref{c2thm4}, we have
\begin{eqnarray}\label{cpd3a}
 |\triangle x_r|^{N_1}|\triangle y_r|^{N_2} &=& \left|2dm \cos( \theta) +2dn\sin(\theta)\right|^{N_1}\nonumber\\
  &&\left|-2dm \sin( \theta) +2dn\cos(\theta)\right|^{N_2}.
\end{eqnarray}
The possible values of $GCPD_{(N_1,N_2)}(P_1,P_2)$ are
\begin{eqnarray}\label{cpd4a}
   GCPD_1&=&4 d^2
\left|\sin(\theta)-\cos(\theta)\right|^\frac{2N_1}{N_1+N_2}\nonumber\\&&
\left|\sin(\theta)+\cos(\theta)\right|^\frac{2N_2}{N_1+N_2}\\
  GCPD_2& =&4 d^2
\left|\sin(\theta)+\cos(\theta)\right|^\frac{2N_1}{N_1+N_2}\nonumber\\&&
\left|\sin(\theta)-\cos(\theta)\right|^\frac{2N_2}{N_1+N_2}\\
   GCPD_3&=&4 d^2 \left|\sin(\theta)\right|^\frac{2N_1}{N_1+N_2}
\left|\cos(\theta)\right|^\frac{2N_2}{N_1+N_2}\\
   GCPD_4&=&4 d^2 \left|\cos(\theta)\right|^\frac{2N_1}{N_1+N_2}
\left|\sin(\theta)\right|^\frac{2N_2}{N_1+N_2}.
\end{eqnarray}
Now by symmetry it is sufficient to consider $\theta \in
[0,\pi/4)$. In this range $\sin(\theta)< \cos(\theta) \le 1$ and
accordingly, if $N_1> N_2$ then $GCPD_3 < GCPD_4$ and similarly
$GCPD_1 < GCPD_2$. Equating $GCPD_1, GCPD_3$ gives the optimal
angle of rotation, $\theta_{opt}$. We have
\begin{eqnarray*}\label{cpd4aa}
&&\hspace*{-5mm}   GCPD_1= GCPD_3 \\
&&\hspace*{-5mm}\Rightarrow\\
&&\hspace*{-5mm}\left(\sin(\theta_{opt})-\cos(\theta_{opt})\right)^\frac{2N_1}{N_1+N_2}
\left(\sin(\theta_{opt})+\cos(\theta_{opt})\right)^\frac{2N_2}{N_1+N_2}
\\
&&\hspace*{-5mm}= \left(\sin(\theta_{opt})\right)^\frac{2N_1}{N_1+N_2}
\left(\cos(\theta_{opt})\right)^\frac{2N_2}{N_1+N_2}\\
&&\hspace*{-5mm}\Rightarrow\\
&&\hspace*{-5mm}
\left(1-\cot(\theta_{opt})\right)^\frac{2N_1}{N_1+N_2}
\left(1+\tan(\theta_{opt})\right)^\frac{2N_2}{N_1+N_2} = 1.
\end{eqnarray*}
Substituting $\tan(\theta_{opt})=x$ we have that $x$ is the root
of (\ref{gcpd1}). The $GCPD_1$ and hence the GCPD  at
this value is
\begin{eqnarray}
   GCPD_1&=&4 d^2
\left|\sin(\theta_{opt})-\cos(\theta_{opt})\right|^\frac{2N_1}{N_1+N_2}\nonumber\\
&&
\left|\sin(\theta_{opt})+\cos(\theta_{opt})\right|^\frac{2N_2}{N_1+N_2}\nonumber\\
&=& 4d^2
\frac{(x_0-1)^\frac{2N_1}{N_1+N_2}(x_0+1)^\frac{2N_2}{N_1+N_2}}{1+x_0^2}\nonumber\\
&=& 4d^2 \left(\frac{x_0^\frac{2N_1}{N_1+N_2}}{1+x_0^2}\right).\label{gcpd2}
\end{eqnarray}

\end{proof}
Table \ref{c2tab5} gives the optimal angle of rotation for various
values of $N=N_1+N_2$ along with the normalized $GCPD_{N_1,N_2}$
($GCPD_{N_1,N_2}$)/4$d^2$). Observe that for any given $N$ the
coding gain is large if $N_1,N_2$ are of the same size i.e., nearly
equal. Also observe that the optimal angle of rotation lies in the
range (26.656, 31.7175] and the corresponding normalized coding
gain varies from (0.2,0.4472].

Note that the infimum corresponds to the limit where $N_1=N$, $N_2=0$ and the maximum corresponds to $N_1=N_2=N/2$. Unfortunately, the optimal angle varies with the constellation
size, unlike CPD. In the  next proposition we find upper and lower
bounds on $GCPD_{N_1,N_2}$ for rotated lattice constellations.
\begin{prop}\label{c2prop1}
The $GCPD_{N_1,N_2}$ for rotated NILC is bounded as
\[
CPD^{\frac{2N_2}{N_1+N_2}} \le GCPD_{N_1,N_2} \le
  CPD, ~~~~  N_2 > N_1\] with equality iff $N_1=N_2$.
\end{prop}
\begin{proof}
%From Definition \ref{c2def3} we have for a given signal set $\A$
%\begin{equation}
%GCPD_{N_1,N_2}=\min_{u \ne v \in \A} \left\{ |u_I
%-v_I|^\frac{2N_1}{N_1+N_2}|u_Q - v_Q|^\frac{2N_2}{N_1+N_2},
%|u_I
%-v_I|^\frac{2N_2}{N_1+N_2}|u_Q - v_Q|^\frac{2N_1}{N_1+N_2}
%\right\}.
%\end{equation}
Let $p,q$ be two signal points such that
\begin{equation}
GCPD_{N_1,N_2}=GCPD_{N_1,N_2}(p,q).
\end{equation}
When $N_1=N_2$ or $\triangle x = \triangle y$ there is nothing to
prove as the inequality is satisfied.

 Therefore let $N_1 \ne N_2$ and $\triangle x \ne \triangle y$.
When  the signal points are from the square lattice
$(2k)d+\J(2l)d$ where $k,l \in \Z$ and $d$ is chosen so that the
average energy of the QAM constellation is 1, rotated by an angle
$\theta$ then
\begin{eqnarray}\label{gcpd3}
  GCPD_{(N_1,N_2)}(p,q)&=&\min \left\{|\triangle x_r|^\frac{2N_1}{N_1
  +N_2}|\triangle y_r|^\frac{2N_2}{N_1
  +N_2},\right.\nonumber\\&&\left.
|\triangle x_r|^\frac{2N_2}{N_1
  +N_2}|\triangle y_r|^\frac{2N_1}{N_1
  +N_2}
\right\}
\nonumber\\
&=&
  4d^2\left|m \cos( \theta) +n\sin(\theta)\right|^\frac{2N_1}{N_1
  +N_2}\nonumber\\
  &&\left|-m \sin( \theta) +n\cos(\theta)\right|^\frac{2N_2}{N_1
  +N_2},  
\end{eqnarray}
where $m,n \in \mathbb{Z}$. For a NILC the $GCPD_{N_1,N_2}$ is upper bounded by the $GCPD_{N_1,N_2}$
for QPSK and is given by (\ref{gcpd2}). Now the root of
(\ref{gcpd1}), $x_0$, is such that $x_0 \in (0.5,1)$ and $N_2 >
N/2$ and we immediately have
\begin{equation}\label{gcpd4}
  4d^2 \frac{x_0^\frac{2N_2}{N_1+N_2}}{(1+x_0^2)} < 4d^2 \frac{x_0}{(1+x_0^2)}
\end{equation}
completing $GCPD_{N_1,N_2} \le CPD$. For the second part observe
that, for $N_2 >N_1$, $|m \cos( \theta) +n\sin(\theta)|^{N_2} < |m
\cos( \theta) +n\sin(\theta)|^{N_1}$ as $|m \cos( \theta)
+n\sin(\theta)| <1 $. Substituting this in (\ref{gcpd3}) we have the
lower bound.
\end{proof}
In Proposition \ref{c2prop1}, if we use $\theta=\arctan(2)$
for rotating the NILC then the GCPD is bounded as
\begin{eqnarray}
&&\hspace*{-9mm}CPD_{opt}^{\frac{2N_2}{N_1+N_2}} \le GCPD_{N_1,N_2} \le
CPD_{opt},  ~ N_2
> N_1,\\
&&\hspace*{-9mm}\Rightarrow\nonumber\\
&&\hspace*{-9mm}\left(\frac{4d^2}{\sqrt{5}}\right)^{\frac{2N_2}{N_1+N_2}}
\le GCPD_{N_1,N_2} \le \left(\frac{4d^2}{\sqrt{5}}\right), ~  N_2
> N_1.
\end{eqnarray}
\begin{rem}\label{remc2add1}
It is clear from Table \ref{c2tab5} and the above inequalities on
GCPD that the value of GCPD decreases as the QAM constellation
size increases and also as the difference between $N_1,N_2$
increases. Therefore, while Construction \ref{c2cons1} gives
high-rate designs, the coding gain decreases for QAM
constellations. %In Appendix \ref{appc2a1}, we present another
%class of non-square RFSDDs whose coding gain is greater than CPD
%and also a construction derived from  Construction \ref{c2cons1}
%where the coding gain is still given by GCPD but the difference
%between $N_1,N_2$ is smaller. These codes do not belong to the
%class of GCIODs considered in this section.
\end{rem}
%%%%%%%% Section 4 begins here %%%%%% Simulation %%%%%%%%%%
\subsubsection{Coding gain of GCIOD vs that of GLCOD}\label{codcompare}
In this subsection we compare the coding gains of GCIOD and GLCOD
for the same number of transmit antennas and the same spectral
efficiency in bits/sec/Hz-for same total transmit power. For the sake
of simplicity we assume that both GCIOD and GLCOD
use square QAM constellations.
\subsubsection{The number of transmit antennas N=2}
The total transmit power constraint is given by $\tr{S^{\cal H}S}=L=2$.
If the signal set has unit average energy then the Alamouti code
transmitted is
\[
S=\frac{1}{\sqrt{2}}\left[\begin{array}{cc} x_0 & x_1\\ -x_1^* &
  x_0^*\end{array}\right]
  \]
where the multiplication factor is for power normalization. For
the same average transmit power the rate-one CIOD is
\[
 S=\left[\begin{array}{cc} x_{0I}+\J x_{1Q} & 0\\ 0 &
 x_{1I}+\J x_{0Q} \end{array}\right].
\]
Therefore the coding gain of the Alamouti code for NILC is given by
$\frac{4d^2}{2}$ and that of CIOD is given by Theorem \ref{c2thm4}
as $\frac{4d^2}{\sqrt{5}}$. Therefore the coding gain of the CIOD
for N=2 is inferior to the Alamouti code by a factor of
$\frac{2}{\sqrt{5}}=\frac{2}{2.23}=0.894$, which corresponds to a
coding gain of 0.4 dB for the Alamouti code\footnote{ In Section
\ref{fastsdd}, we revisit these codes for their use in rapid-fading
channels where we show that this loss of coding gain vanishes and
the CIOD for $N=2$ is SD while the Alamouti code is not. }.
 \subsubsection{The number of transmit antennas N=4}
 The average transmit power constraint is given by $\tr{S^{\cal H}S}=L=4$.
If the signal set has unit average energy then the rate 3/4 COD
code transmitted is {\small \[ S=\frac{1}{\sqrt{3}}\begin{bmatrix}
   x_0 & x_1 & x_2 & 0 \\
  -x_1^* & x_0^* & 0 & x_2 \\
  -x_2^* & 0 & x_0^* & -x_1 \\
  0 & -x_2^* & x_1^* & x_0
\end{bmatrix}
 \]}
where the multiplication factor is for power normalization. For
the same average transmit power, the rate 1 CIOD is given in (\ref{ciodnorm}).
\begin{table*}
\begin{equation}\label{ciodnorm}
 S= \frac{1}{\sqrt{2}}\left[\begin{array}{cccc}
x_{0I}+\J x_{2Q} & x_{1I}+\J x_{3Q} & 0 &0\\
-x_{1I}+\J x_{3Q} & x_{0I}-\J x_{2Q} & 0 &0\\
 0 &0 & x_{2I}+\J x_{0Q} & x_{3I}+\J x_{1Q} \\
 0 &0 & -x_{3I}+\J x_{1Q} &  x_{2I}-\J x_{0Q}
\end{array}\right].
\end{equation}
\end{table*}
If the rate 3/4 code uses a $2^n$ square QAM and the rate 1 CIOD
uses a $2^\frac{3n}{4}$ square QAM, then they have same spectral
efficiency in bits/sec/Hz, and the possible values of $n$ for
realizable square constellations is $n=8i, i \in \mathbb{Z}^+$.
Let $d_1,d_2$ be the values of $d$ so that the average energy of
$2^n$ square QAM and $2^\frac{3n}{4}$ square QAM is 1. Therefore the coding gain of rate 3/4 COD for NILC is given by
$\Lambda_{COD}=\frac{4d_1^2}{3}$ and that of CIOD is given by
Theorem \ref{c2thm4} as $\Lambda_{CIOD}=\frac{4d_2^2}{2\sqrt{5}}$.
Using the fact that for unit average energy M-QAM square constellations
$d=\sqrt{\frac{6}{M-1}}$, we have
\[
\Lambda_{COD}=\frac{8}{ (2^{8i}-1)}\mbox{ and }
\Lambda_{CIOD}=\frac{12}{ \sqrt{5} (2^{6i}-1)} \mbox{ where } i
\in \mathbb{Z}^+
\]
for a spectral efficiency of $6i$ bits/sec/Hz. For $i=1,2,3$ we
have  $\Lambda_{COD}$ = 0.0314, 1.2207e-004, 4.7684e-007 and
$\Lambda_{CIOD}$ = 0.0422, 6.5517e-004, 1.0236e-005 respectively,
corresponding to a coding gain of $1.29,7.29,13.318$ dB
 for the CIOD code.
 Observe that in contrast to the coding gain for $N=2$ which is independent
 of the spectral efficiency, the  coding gain for $N=4$
 appreciates with spectral efficiency.
 \subsubsection{The number of transmit antennas N=8}
 The total transmit power constraint is given by $\tr{S^{\cal H}S}=L=8$.
If the signal set has unit average energy then the rate 1/2 COD
code has a multiplication factor of $1/2$ and for the same
transmit power, the rate 3/4 CIOD has a multiplication factor of
$1/\sqrt{3}$.  If the rate 1/2 COD code uses a $2^n$ square QAM and
the rate 3/4 CIOD uses a $2^\frac{3n}{2}$ square QAM, then they have
same spectral efficiency in bits/sec/Hz, and the possible values
of $n$ for realizable square constellations is $n=4i, i \in
\mathbb{Z}^+$. Let $d_1,d_2$ be the values of $d$ so that the
average energy of $2^n$ square QAM and $2^\frac{3n}{2}$ square QAM
is 1.  Therefore the coding gain of rate 1/2 COD for
NILC is given by $\Lambda_{COD}=\frac{4d_1^2}{4}$ and that of CIOD
is given by Theorem \ref{c2thm4} as
$\Lambda_{CIOD}=\frac{4d_2^2}{3\sqrt{5}}$. Using the fact that for
unit average energy M-QAM square constellations $d=\sqrt{\frac{6}{M-1}}$,
we have
\[
\Lambda_{COD}=\frac{ 6}{ (2^{4i}-1)}\mbox{ and }
\Lambda_{CIOD}=\frac{8}{ \sqrt{5} (2^{3i}-1)} \mbox{ where } i \in
\mathbb{Z}^+
\]
for a spectral efficiency of $3i$ bits/sec/Hz. For $i=1,2,3$ we
have  $\Lambda_{COD}$= 0.4, 0.0235, 0.0015 and $\Lambda_{CIOD}$ =
0.4737, 0.0563, 0.007 respectively, corresponding to a coding gain
of $0.734, 3.789, 6.788$ dB
 for the CIOD code.
 Observe that as in the case of $N=4$ the coding gain
 appreciates with spectral efficiency.

Next we compare the coding gains of some GCIODs.
 \subsubsection{The number of transmit antennas N=3}
 Both the GCIOD and GCOD for $N=3$ is obtained from
 the $N=4$ codes by dropping one of the columns, consequently the rates
  and the total transmit power constraint are same as for $N=4$.
Accordingly, the rate 3/4 GCOD code uses a $2^n$ square QAM and
the rate-one GCIOD uses a $2^\frac{3n}{4}$ square QAM where $n=8i, i
\in \mathbb{Z}^+$. The coding gain for the rate 3/4 GCOD for NILC is
given by $\Lambda_{GCOD}=\frac{4d_1^2}{3}$ and that of GCIOD is
lower bounded by Proposition \ref{c2prop1} as $\Lambda_{GCIOD} >
\left(\frac{4d_2^2}{2\sqrt{5}}\right)^\frac{4}{3}$. Using the fact
that for unit average energy M-QAM square constellations
$d=\sqrt{\frac{6}{M-1}}$, we have
\[
\Lambda_{GCOD}=\frac{8}{ (2^{8i}-1)}\mbox{ and }
\Lambda_{GCIOD}>\left(\frac{12}{ \sqrt{5}
(2^{6i}-1)}\right)^\frac{4}{3} \mbox{ where } i \in \mathbb{Z}^+
\]
for a spectral efficiency of $6i$ bits/sec/Hz. For $i=1,2,3$ we
have  $\Lambda_{GCOD}$ = 0.0314, 1.2207$e^{-4}$, 4.7684$e^{-7}$ and
$\Lambda_{GCIOD} > $ 0.0147, 5.69$e^{-5}$, 2.22$e^{-7}$ respectively.

 Observe that at high spectral rates, even the lower bound
is larger than the coding gain of GCOD. In practice, however, the
GCIOD performs better than GCOD at all spectral rates.
%%%%%%%%%%% Section 3.4 begins %%%%%%%%%%%%%%%%%%%%%%%%%%%%
\subsection{Simulation Results}\label{sec34}
In this section we present simulation results for
 4-QAM and 16-QAM  modulation over a quasi-static fading channel.
The fading is assumed to be constant over a fade length of 120 symbol durations.
%\newpage

First, we compare the CIOD for $N=4$, with (i) the STBC (denoted
by STBC-CR in Fig. \ref{c2fig4} and \ref{c2fig5}) of \cite{BoV},
(ii) rate 1/2, COD and (iii) rate 3/4 COD for four transmit
antennas for the {\bf identical  throughput} of 2 bits/sec/Hz.
For CIOD the transmitter chooses symbols from a QPSK signal set
rotated by an angle of $13.2825^\circ$ so as to maximize the
$CPD$. For STBC-CR the symbols are from a QPSK signal set and rate
1/2 COD from 16-QAM signal set. For rate 3/4 COD, the symbols are
chosen from 6-PSK for a throughput of 1.94 bits/sec/Hz which is
close to 2 bits/sec/Hz. The average transmitted power  is equal in
all the cases i.e. ${\mathrm E \{ tr }(S^{\cal H}S)\}=4$, so that
average energy per bit using the channel model   of (\ref{c0eq6})
is equal. The Fig. \ref{c2fig4}. shows the BER performance  for
these schemes. Observe that the scheme of this paper  outperforms
rate 1/2 COD by 3.0 dB, rate 3/4 COD by 1.3 dB and STBC-CR by 1.2
dB at $P_b=10^{-5}$. A comparison of the coding gain, $\Lambda$,
of these schemes  is given in tabular form in Table \ref{tab2}.

For CIOD, $\Lambda_{CIOD}=0.4478$ while for STBC-CR
$\Lambda_{STBC-CR}=0.5$ at $R=2$ bits/sec/Hz, but still CIOD
out-performs STBC-CR because the coding gain is derived on the
basis of an upper bound. If we take into consideration the
\textit{kissing number} i.e. the number of codewords at the given
minimum coding gain, then we clearly see that though STBC-CR has
higher coding gain, it has more than double the kissing number of
CIOD. The results for rest of the schemes are in accordance with
their coding gains;
$$10\log_{10}\left(\frac{\Lambda_{CIOD}}{\Lambda_{rate~1/2~COD}}\right)
=3.5$$  and
$$10\log_{10}\left(\frac{\Lambda_{CIOD}}{\Lambda_{rate~3/4~COD}}\right)=1.3.$$
Observe that rate 3/4 COD and STBC-CR have almost similar
performance at 2 bits/sec/Hz, and around 1.6 dB coding gain over
rate 1/2 COD. A possible
 apparent inconsistency of these with the  results in \cite{XWG1,XWG2}, which report coding gain of over 2 dB, is due to
the fact that  symbol error rate (SER) vs. $\rho$ is plotted in \cite{XWG1,XWG2}. As rate 1/2 COD chooses symbols from 16
QAM and STBC-CR from 4 QAM, SER vs. $\rho$ plot gives an overestimate of the errors for STBC-OD as compared to STBC-CR and
 bit error rate (BER) vs. $E_b/N_0$ is a more appropriate plot for comparison at the same through put (2 bits/sec/Hz).

From the Table \ref{tab2}, which gives the coding gains of various
schemes at spectral efficiencies of 2,3,4  bits/sec/Hz, we see
that the coding gain of STBC-CR and CIOD are nearly equal (differ
by a factor of 1.11) and significantly greater than other schemes.
But, the main factor in favor of CIOD as compared to STBC-CR (as
also any  STBC other than STBC-OD) is that CIOD allows linear
complexity ML decoding while STBC-CR has exponential ML decoding
complexity.  At a modest rate of 4 bits/sec/Hz, CIOD requires 64
metric computations while STBC-CR  requires $16^4=65,536$ metric
computations. Even the sphere-decoding algorithm is  quite complex
requiring exponential complexity when $M < N$ and polynomial
otherwise \cite{DMB2}.

For 4-QAM and 16-QAM constellations, Fig. \ref{c2fig5} shows the
performance for CIOD, STBC-CR and Diagonal Algebraic Space Time
(DAST) codes of \cite{DMB1}. As expected CIOD shows better
performance. Finally note that the performance of full-diversity
QODs \cite{SuX3,SuX4} is same as the performance of CIODs, however
QODs are not single-symbol decodable.
%%%%%%%%%%% Section 3.5 begins %%%%%%%%%%%%%%%%%%%%%%%%%%%%
\subsection{Maximum Mutual Information (MMI) of CIODs}\label{sec35}
In this Subsection we analyze the maximum mutual information (MMI)
that can be attained by  GCIOD schemes presented in this section.
We show that except for the Alamouti scheme all other GLCOD have
lower MMI than the corresponding GCIOD. We also compare the MMI of
rate-one STBC-CR with that of GCIOD to show that GCIOD have higher
MMI.

 It is very clear
from the number of zeros in the transmission matrices of
GCIODs, presented
in the previous sections, that these schemes do not achieve
capacity. This is because the emphasis is on low decoding
complexity rather than attaining capacity. Nevertheless we intend
to quantify the loss in capacity due to the presence of zeros in GCIODs.

We first consider the
$N=2, M=1$ CIOD. Equation (\ref{c0eq6}), for the CIOD code given in (\ref{nw21}) with power
normalization, can be written as
\begin{equation}\label{chmodch}
  {\mathbf V}=\sqrt{\rho}{ H} s+{\mathbf N}
\end{equation}
where
$${ H}=\left[\begin{array}{cc}h_{00} & 0\\ 0 &
h_{10}\end{array}\right]
$$
and $s=[\tilde{s}_0~~ \tilde{s}_1]^{\cal T}$, and where $\tilde{s}_0=s_{0I}+\J s_{1Q},\tilde{s}_1=s_{1I}+\J s_{0Q}, s_0,s_1 \in \A$.
If we define $C_D(N,M,\rho)$ as the maximum mutual information of
the GCIOD for $N$ transmit
and $M$ receive antennas at SNR, $\rho$, then
\begin{eqnarray}\label{cap1}
  C_D(2,1,\rho)&=&\frac{1}{2}E(\log\det(I_2+\rho{ H}^{\cal H}{ H}))\nonumber\\
&=&\frac{1}{2}E\log\{(1+\rho
  |h_{00}|^2)(1+\rho|h_{10}|^2)\}\nonumber\\
&=&\frac{1}{2}E\log\{1+\rho
  |h_{00}|^2\}+\frac{1}{2}E\log\{1+\rho|h_{10}|^2\}\nonumber\\
&=&C(1,1,\rho) < C(2,1,\rho).
\end{eqnarray}
 It is similarly seen for CIOD code for $N=4$ given in (\ref{exeq1}) that
for
$${H}=\left[
\begin{array}{cccc}
h_{00}    & h_{10}   & 0 & 0 \\
-h_{10}^* & h_{00}^* & 0 & 0 \\
0   & 0 &  h_{20} & h_{30} \\
0   & 0 & -h_{30}^* & h_{20}
\end{array}
\right],
$$

\begin{eqnarray}\label{cap14}
  C_D(4,1,\rho)&=&\frac{1}{4}E\left(\log\det\left[I_4+\frac{\rho}{2}{H}^{\cal H}  { H}\right]\right)\nonumber\\
&=&\frac{1}{2}E\log\left\{\left[1+\frac{\rho}{2}
  \left(|h_{00}|^2+|h_{10}|^2\right)\right]\right.\nonumber\\
&&\left.\left[1+\frac{\rho}{2}\left(|h_{20}|^2+|h_{30}|^2\right)\right]\right\}\nonumber\\
&=&\frac{1}{2}E\log\{1+\frac{\rho}{2}
  (|h_{00}|^2+|h_{10}|^2)\}\nonumber\\&&
+\frac{1}{2}E\log\{1+\frac{\rho}{2}(|h_{20}|^2+|h_{30}|^2)\}\nonumber\\
&=&C(2,1,\rho) < C(4,1,\rho)
\end{eqnarray}
and
\begin{equation}\label{cap2}
C_D(3,1,\rho)=\frac{1}{2}\{C(2,1,\rho)+ C(1,1,\rho)\}<
C(3,1,\rho).
\end{equation}

 Therefore CIODs do not achieve full channel capacity even for one receive
antenna.  The capacity loss  is negligible for one receiver as is
seen from Figures \ref{fcap1}, \ref{fcap3} and  \ref{fcap34}; this is because the
increase in capacity is small from two to four transmitters in
this case. The capacity loss is substantial when the number of
receivers is more than one, as these schemes achieve capacity
that could be attained with half the number of transmit antennas. This
is because half of the antennas are not used during any given
frame length.

Another important aspect is the comparison of MMI of CODs for three
and four transmit antennas with the capacity of CIOD and GCIOD for
similar antenna configuration-we already know that for two transmit
antennas and one receive antenna, complex orthogonal designs,
(Alamouti code) achieve capacity; no code can beat the performance
of Alamouti code.

 It is shown in \cite{San} that
\begin{equation}\label{mu1}
  C_O(3,M,\rho)=\frac{3}{4} C(3M,1, \frac{4}{3}M\rho)
\end{equation}
where  $C_O(N,M,\rho)$ is the MMI of GLCOD for $N$ transmit and
$M$ receive  antennas at a SNR of $\rho$. Similarly,
\begin{equation}\label{mu2}
  C_O(4,M,\rho)=\frac{3}{4} C(4M,1,\frac{4}{3}M\rho).
\end{equation}
Equation (\ref{mu2}) is plotted for $M=1,2$ in Fig. \ref{fcap1}
and (\ref{cap2}) is plotted in Fig. \ref{fcap3} along with the
corresponding plots for CIOD
 derived from (\ref{cap14}) and
 (\ref{cap2}).  We see from these plots that the
capacity of CIOD is just less than the actual capacity when there
is only one receiver and is considerably greater than the capacity
of code rate 3/4 complex orthogonal designs for  four
transmitters. When there are two receivers the capacity of CIOD is
less than the actual capacity but is considerably greater than the
capacity of code rate 3/4 complex orthogonal designs four
transmitters.

Next we present the comparison of GCOD and GCIOD for $N>4$.
Consider the MMI of GLCOD of rate $K/L$. The effective channel
induced by the GLCOD is given by \cite{San}
\begin{equation}\label{c2eqt1}
  \mathbf{v}={\frac{L\rho}{KN}}\|\mathbf{H}\|^2\mathbf{x}+\mathbf{n}
\end{equation}
where $\mathbf{v}$ is a $2K \times 1$ vector after linear
processing of the received matrix $\mathbf{v}$, $\mathbf{x}$ is a
$2K \times 1$ vector consisting of the in-phase and quadrature
components of the $K$ indeterminates $x_0,\cdots,x_{K-1}$ and
$\mathbf{n}$ is the noise vector with Gaussian iid entries with
zero mean and variance $\|\mathbf{H}\|^2/2$. Since (\ref{c2eqt1})
is a scaled AWGN channel with
$SNR=\frac{L\rho}{KN}\|\mathbf{H}\|^2$ and rate $K/L$, the average
MMI in bits per channel use of GLCOD can be written as
\cite{San}
\begin{equation}\label{c2eqt2a}
C_O(N,M,\rho)=\frac{K}{L}\E{\log_2\left(1+\frac{L\rho}{KN}\|\mathbf{H}\|^2\right)}
\end{equation}
observe that $\mathbf{H}$ is a $N \times M$ matrix. Since
$\|\mathbf{H}\|^2=\vec{\mathbf{H}}^{\cal H}\vec{\mathbf{H}}$ where
$\vec{\mathbf{H}}$ is the $ NM \times 1$  vector formed by
stacking the columns of $\mathbf{H}$, we have
\begin{eqnarray}\label{c2eqt2}
C_O(N,M,\rho)&=&\frac{K}{L}C(MN,1,\frac{ML}{K}\rho)\\
&=& \frac{K}{L} \frac{1}{\Gamma(MN)} \int_0^\infty
\log\left(1+\frac{L\rho \lambda}{K
N}\right)\nonumber\\&&
\lambda^{MN-1}\mathrm{e}^{-\lambda}\mathrm{d}\lambda\label{c2eqt2l}
\end{eqnarray}
where (\ref{c2eqt2l}) follows from \cite[eqn. (10)]{Tel}. For
GCIOD, recollect that it consists of two GLCODs,
$\Theta_1,\Theta_2$ of rate $K/2L_1,K/2L_2$ as defined in
(\ref{c2eq1}). Let $C_{1,O},C_{2,O}$ be the MMI of
$\Theta_1,\Theta_2$ respectively. Then the MMI of GCIOD is given
by
\begin{eqnarray}\label{c2eqt3}
  C_{D}(N,M,\rho)&=&\frac{1}{L}\left\{L_1 C_{1,O}+ L_2
C_{2,O}\right\}\\
&=& \frac{1}{L}\left\{L_1 C_{O}(N_1,M,{\rho})\right.\nonumber\\&&\left.+ L_2
C_{O}(N_2,M,{\rho})\right\}\\
&=&\frac{K}{2L}\left\{C\left(MN_1,1,\frac{2L_1M\rho}{K}\right)\right.\nonumber\\&& \left.+
C\left(MN_2,1,\frac{2L_2M\rho}{K}\right)\right\}.
\end{eqnarray}
The
above result follows from the fact that the GCIOD is block
diagonal with each block being a GLCOD. When $L_1=L_2$ i.e.
$\Theta_1=\Theta_2$ we have
\begin{equation}\label{c2eqt3a}
  C_{D}(N,M,\rho)= \frac{K}{L}C\left(\frac{MN}{2},1,\frac{LM\rho}{K}\right)
\end{equation}
as we have already seen for $N=2,4$.

Let $\triangle C =C_{D}- C_O$. For square designs ($N=L=2^ab,b$
odd) we have
\begin{eqnarray}\label{c2eqt4}
  \triangle C
&=&\frac{2a}{N}C(M2^{a-1}b,1,2^{a+1}M\rho)\nonumber\\&&-\frac{a+1}{N}C(M2^{a}b,1,2^{a}M\rho).
%&=&\frac{2a}{N}\left( \frac{1}{\Gamma(M2^{a-1}b)} \int_0^\infty
%\log\left(1+\frac{\rho
%\lambda}{M2^{a-1}b}\right)\lambda^{M2^{a-1}b-1}\mathrm{e}^{-\lambda}\mathrm{d}\lambda\right)
%-\frac{a+1}{N}\left( \frac{1}{\Gamma(M2^{a}b)} \int_0^\infty
%\log\left(1+\frac{\rho
%\lambda}{M2^{a}b}\right)\lambda^{M2^{a}b-1}\mathrm{e}^{-\lambda}\mathrm{d}\lambda\right)
\end{eqnarray}
It is sufficient to consider $b=1$. When $N=2$,
$\frac{2a}{N}=\frac{a+1}{N}=1$ and $ \triangle C
=C(M,1,M\rho)-C(2M,1,M\rho) < 0$, as seen from \cite[Figure 3: and
Table 2]{Tel}. When $N>2$, $2a > a+1$ and  $\lim_{a \rightarrow
\infty } \frac{2a}{a+1}=2$. Also $C(M2^{a-1},1,M\rho)$ is
marginally smaller than $C(M2^{a},1,M\rho)$ for $M>1,a>1$ as can
be seen from \cite[Figure 3: and Table 2]{Tel}. It therefore
follows that
\begin{thm}
The MMI of square CIOD is greater than MMI of square GLCOD except
when $N=2$.
\end{thm}

It can be shown that a similar result holds for GCIOD also, by carrying out the analysis for each $N$. We are omitting $N=5,6,7$. For $N \ge 8$ we compare rate 2/3 GCIOD
with the rate 1/2 GLCODs. The MMI of rate 1/2 GLCOD is given by
\begin{eqnarray}\label{c2eqt5}
C_O(N,M,\rho)&=&\frac{1}{2}C(MN,1,2M\rho).
%\\&=& \frac{K}{L}\left( \frac{1}{\Gamma(N)} \int_0^\infty
%\log\left(1+\frac{\rho
%\lambda}{N}\right)\lambda^{N-1}\mathrm{e}^{-\lambda}\mathrm{d}\lambda\right)\label{c2eqt2l}
\end{eqnarray}
The MMI of rate 2/3 GCIOD is given by,
\begin{equation}
C_{D}(N,M,\rho) =\frac{1}{3}\left\{C(2M,1,M\rho)+
C(M(N-2),1,2M\rho)\right\}.
\end{equation}
For reasonable values of $N$ that is $N\ge 8$, $C(MN,1,2M\rho)\approx
C(M(N-2),1,2M\rho)$ and $C(2M,1,M\rho) \approx C(MN,1,M\rho) \approx C(MN,1,2M\rho)$ and
it follows that
\begin{equation}
C_{D}(N,M,\rho) \approx \frac{2}{3} C(MN,1,2M\rho).
\end{equation}
Note that in arriving this approximation we have used the property of $C(M,N,L)$ that for $N=1$, as $M$ increases the increment in $C$ is small and also that for a given $M,N$,   $C$ saturates w.r.t. $\rho$.

Figure \ref{c2fig6} shows the capacity plots for $N=8$, observe
that the capacity of rate 2/3 GCIOD is considerably greater than
that of rate 1/2 GLCOD. At a capacity of 7 bits the gain is around
10 dB for $M=8$. Similar plots are obtained for all $N>8$ with
increasing coding gains and have been omitted. Finally, it is
interesting to note that the MMI of QODs is same as that of CIODs;
however QODs are not SD.

%%%%%%% Section 3.6 (discussion) begins %%%%%%%%%%%%
%%%%%%%%%%%%%%%%%%% fast-fading
\section{Single-symbol decodable designs for rapid-fading
channels}\label{fastsdd}
% Space-Time block codes (STBC) obtained
%from OD (Orthogonal Designs), QOD (Quasi-Orthogonal Designs) and
%their variations \cite{Ala}-\cite{SuX4} are  attractive due to
%their fast ML decoding (single/ double-symbol decoding) when used
%over quasi-static fading channels. However, these STBCs from
%Designs have not been studied well for use in rapid-fading
%channels.
  In this section, we study  STBCs for use in
rapid-fading channels by giving a matrix representation of the
multi-antenna rapid-fading channels. The emphasis is on finding
STBCs that are single-symbol decodable in both quasi-static and
rapid-fading channels, as performance of such STBCs will be
invariant to channel variations. Unfortunately, we show that such a
rate 1 design exists for only two transmit antennas.

We first characterize all linear STBCs that allow single-symbol ML
decoding when used in rapid-fading channels. Then, among these we
identify those with full diversity, i.e., those with diversity $L$
when the STBC is of size $ L \times N, (L \ge N)$, where $N$ is the
number of transmit antennas and $L$ is the time interval. The
maximum rate for such a full-diversity, single-symbol decodable code
is shown to be $2/L$ from which  it follows that rate 1 is possible
only  for 2 Tx. antennas. The co-ordinate interleaved orthogonal
design (CIOD) for 2 Tx (introduced in section \ref{ssdd}) is shown
to be one such full-rate, full-diversity and single-symbol decodable
code. (It turns out that Alamouti code is not single-symbol
decodable for rapid-fading channels.)

\subsection{Extended  Codeword Matrix and the Equivalent Matrix
Channel}
 The inability to write (\ref{c5chmod1}) in the matrix form
as in (\ref{c0eq6}) for rapid-fading channels seems to be the reason
for scarce study of STBCs for use in rapid-fading channels. In this
section we solve this problem by introducing proper matrix
representations for the codeword matrix and the channel. In what
follows we assume that $M=1$,
  for simplicity. For a rapid-fading channel (\ref{c5chmod1}) can be written as
\begin{equation}\label{c5chmod5}
   \bV=S H+\bW
\end{equation}
\noindent where ${ \bV} \in \cp^{L \times 1}$ ($\cp$ denotes the
complex field) is the received signal vector, ${ S} \in \cp^{L
\times NL}$ is the {\bf{Extended codeword  matrix (ExCM)}} (as
opposed to codeword matrix ${\mathbf S}$) given by
\begin{equation}\label{c5chmod3a}
S= \left[\begin{array}{ccccc}
  S_0 & 0 & 0 & 0  \\
  0 & S_1 & 0 & 0  \\
  \vdots& \ddots & \ddots & \ddots  \\
  0 & 0 & 0 & S_{L-1}
\end{array}\right]
\end{equation}
$\mbox{ where } S_t=\left[\begin{array}{ccccc} s_{0t} & s_{1t} &
\cdots & s_{(N-1)t}\end{array}\right],$  ${ H} \in \cp^{NL \times
1}$ denotes the {\bf equivalent channel matrix (EChM)}  formed by
stacking the channel vectors for different $t$ i.e.
$$
H=\left[\begin{array}{c} H_0\\ H_1\\\vdots\\H_{L-1}
\end{array}\right] \mbox{ where } H_t=\left[\begin{array}{c}
h_{0t} \\ h_{1t} \\ \cdots\\ h_{(N-1)t}\end{array}\right],
$$
 and ${\mathbf W}\in \cp^{L\times 1} $ has entries that are Gaussian distributed with zero mean and unit variance
and also are temporally and spatially white.
%Equation (\ref{chmod5}) will be referred to as equivalent channel model (ECM) for rapid-fading channels.
We  denote the  codeword matrices by boldface letters   and the
ExCMs  by normal letters. For example, the ExCM $S$ for the
Alamouti code, ${\small \mathbf{S}=\left[\begin{array}{rr}
    x_0&x_1\\
     -x_1^*& x_0^*
  \end{array}\right]
}$, is given by
\begin{equation}\label{c5chmod6}
 S= \left[\begin{array}{cccc}
    x_0&x_1&0&0\\
     0&0&-x_1^*& x_0^*
  \end{array}\right].
\end{equation}

Observe that for a linear space-time code, its ExCM $S$ is also
linear in the indeterminates $x_k, k=0,\cdots,K-1$ and can be
written as $S=\sum_{k=0}^{K-1}x_{kI}A_{2k}+x_{kQ}A_{2k+1}$, where
$A_k$ are  referred to as {\bf extended weight matrices} to
differentiate from weight matrices corresponding to the codeword
matrix $\mathbf{S}$.
\subsubsection{Diversity and Coding gain criteria for rapid-fading
channels}
 With  the notions of ExCM and EChM developed above and the similarity
between (\ref{c0eq6}) and (\ref{c5chmod5}) we observe that,
\begin{enumerate}
\item The {\bf distance criterion} on the difference of two distinct
codeword matrices is equivalent to the {\bf rank criterion} for
the difference of two distinct ExCM.
\item The {\bf product criterion} on the difference of two distinct
codeword matrices is equivalent to the {\bf determinant criterion}
for the difference of two distinct ExCM.
\item The trace criterion  on the difference of two distinct
codeword matrices derived for quasi-static fading in \cite{CYV}
applies to rapid-fading channels also-following the observation that
$\tr{{\mathbf S}^H{\mathbf S}}=\tr{S^HS}$.
\item The ML metric
(\ref{c5chmod2}) can again be represented as (\ref{c1eq3}) with
the code word $\mathbf{S}$ replaced by the ExCM, $S$ i.e.
\begin{equation}\label{c5chmod4a}
M( S)= \tr{( \bV-S  H)^H( \bV-S  H)}.
\end{equation}
This amenability to write the ML decoding metric in matrix form for
rapid-fading channels (\ref{c5chmod4a}) allows the results on
single-symbol decodable designs  of section \ref{ssdd} to be applied
to rapid-fading channels.
\end{enumerate}
\subsection{Single-symbol decodable codes}\label{fsdd}
 Substitution of the codeword
matrix $\mathbf{S}$ by the ExCM, $S$ in Theorem \ref{prop1a} leads
to characterization of single-symbol decodable STBCs for
rapid-fading channels. We have,
\begin{thm}\label{c5thm2} For a linear STBC in $K$ complex variables, whose ExCM is
given by,
 $S=\sum_{k=0}^{K-1}x_{kI}A_{2k}+x_{kQ}A_{2k+1}$, the ML metric,
 $M(S)$ defined in (\ref{c5chmod4a})  decomposes as
 $M( S)=\sum_k M_k(x_k) +M_C$ where $M_C=-(K-1)\tr{ V^H V}$,
 iff
 \begin{equation}\label{c5chr1a}
A_k^HA_l+A_l^HA_k=0, 0 \le k \ne l \le 2K-1.
\end{equation}
\end{thm}

Theorem \ref{c5thm2} characterizes all linear designs which admit
single-symbol decoding  over rapid-fading channels in terms of the
extended weight matrices.

\begin{eg}\label{c5ex1} The Alamouti code is not single-symbol decodable
for rapid-fading channels. The extended weight matrices are {\small
\begin{eqnarray*}
A_0=\left[\begin{array}{cccc}
1 & 0 & 0 &0\\
 0 &0 & 0 & 1
\end{array}\right],
 A_1=\left[\begin{array}{cccc}
\J & 0 & 0 &0\\
 0 &0 & 0 & -\J
\end{array}\right],\\
A_2=\left[\begin{array}{cccc}
 0 & 1 & 0 &0\\
 0 &0 & -1 & 0
\end{array}\right],
 A_3=\left[\begin{array}{cccc}
0 & \J & 0 &0\\
 0 &0 & \J & 0
\end{array}\right].
\end{eqnarray*}

} It is easily checked that the pair $ A_0,  A_2 $ does not
satisfy equation (\ref{c5chr1a}).
\end{eg}
%%%%%%%%%%%%%%%%%%%%%%%%%%%%%%%%%%%%%%%%%%%%%%%%%%%%%%%%%%
\subsection{Full-diversity, Single-Symbol decodable codes} In this
section we proceed to identify all full-diversity codes among
single-symbol decodable codes. Recall that for single-symbol
decodability in quasi-static fading the weight matrices have to
satisfy (\ref{chr}) while for rapid-fading the {extended weight
matrices}, have to satisfy (\ref{c5chr1a}).

In contrast to quasi-static fading (\ref{c5chr1a}) is not easily
satisfied for rapid-fading due to the structure of the equivalent
weight matrices imposed by the
 structure of $S$ given in (\ref{c5chmod3a}).
 The weight matrices $A_k$ are  block diagonal of the form
(\ref{c5chmod3a})
\begin{equation}\label{mat1}
A_k= \left[\begin{array}{ccccc}
  A_k^{(0)} & 0 & 0 & 0  \\
  0 & A_k^{(1)} & 0 & 0  \\
  \vdots& \ddots & \ddots & \ddots  \\
  0 & 0 & 0 & A_k^{(L-1)}
\end{array}\right]. \end{equation}
 $\mbox{ where } A_k^{(t)} \in \cp^{1 \times N}
$.
 In other words even for
square codeword matrix the equivalent transmission matrix is
rectangular. For example consider the Alamouti code, ${\small A_0
=\left[\begin{array}{rrrr}
    1&0 & 0    & 0\\
     0 & 0 &0 &1
  \end{array}\right]}$, ${\small A_1
=\left[\begin{array}{rrrr}
    1&0 & 0    & 0\\
     0 & 0 &0 &-1
  \end{array}\right]}$ etc., (\ref{c5chr1a}) is not satisfied as a result we have
\begin{equation}\label{c5chmod6a}
 S^HS= \left[\begin{array}{cccc}
    |x_0|^2&x_0^*x_1&0&0\\
    x_1^*x_0&|x_1|^2&0&0\\
     0&0&|x_1|^2& -x_1x_0^*\\
     0&0&-x_1^*x_0& |x_0|^2\
  \end{array}\right],
\end{equation}
and hence single-symbol decoding is not possible for the Alamouti
code over rapid-fading channels.

The structure of equivalent weight matrices that satisfy
(\ref{c5chr1a}) is given in Proposition \ref{c5prop1}.

\begin{prop}\label{c5prop1} All the  matrices $A_l$ that satisfy (\ref{c5chr1a}), with a
specified non-zero matrix $A_k$  in (\ref{mat1}) are of the form
\begin{equation}\label{mat2}
\left[\begin{array}{ccccc}
  a_0A_k^{(0)} & 0 & 0 & 0  \\
  0 & a_1A_k^{(1)} & 0 & 0  \\
  \vdots& \ddots & \ddots & \ddots  \\
  0 & 0 & 0 & a_{L-1}A_k^{(L-1)}
\end{array}\right]. \end{equation}
where $a_i=0, \J  ~\forall i$.
\end{prop}
\begin{proof}
 The the matrix $A_k$ can satisfy the condition of
Theorem \ref{c5thm2} iff
$A_k^{(t)H}A_l^{(t)}=-A_l^{(t)H}A_k^{(t)}$, $\forall t$. For a
given $t$, $A_k^{(t)H}A_l^{(t)}$ is skew-Hermitian and rank one,
it follows that $A_k^{(t)H}A_l^{(t)}=UDU^H$ where $U$ is unitary
and $D$ is diagonal with one imaginary entry only. Therefore
$A_k^{(t)}=\pm
 \J cA_l^{(t)}$ where $c$ is a real constant-in fact only the values $c=0,1$ are
of interest as other values can be normalized to 1, completing the
proof.
\end{proof}

We give a necessary condition, derived from the rank criterion for
ExCM, in terms of the extended weight matrices $A_k$ for the code
to achieve diversity $r \le L$. This necessary condition results
in ease of characterization.
\begin{lem}\label{c5lem1} If a linear STBC in $K$ variables, whose ExCM is
given by, $S=\sum_{k=0}^{K-1}x_{kI}$ $A_{2k}+x_{kQ}A_{2k+1}$,
achieves diversity $r$  then the  matrices $A_{2k},A_{2k+1}$
together have at least $r$ different non-zero rows for every $k$,
$0 \le k \le K-1$.
\end{lem}
\begin{proof}
This follows from the rank criterion of ExCM interpretation of the
distance criterion. If, for a given  $k$, $A_{2k},A_{2k+1}$
together have at less than $r$ different non-zero rows then the
difference of ExCMs, $S-\hat{S}$ which differ in $x_{k}$ only, has
rank less than $r$.
\end{proof}
 The conditions of  Lemma \ref{c5lem1} is only
a necessary condition since either $(x_{kI}-\hat{x}_{kI})$ or
$(x_{kQ}-\hat{x}_{kI})$ may be zero for $x_{k} \ne\hat{x}_{k}$.
 The
sufficient condition is obtained by a slight modification of
Theorem \ref{prop3} and is given by
\begin{cor}\label{c5prop3}
A linear STBC, $S=\sum_{k=0}^{K-1}x_{kI}A_{2k}$+$x_{kQ}A_{2k+1}$
where $x_k$ take values from a signal set $\A, \forall k$,
satisfying the necessary condition of Lemma \ref{c5lem1} achieves
diversity $r\ge N$ iff
\begin{enumerate}
  \item \label{c5sufcond1} either $A_k^HA_k$ is of rank $r$ ($r$ different non-zero rows)
  for all $k$
  \item \label{c5sufcond2} or the $CPD \mbox{ of } \A \ne 0$.
\end{enumerate}
\end{cor}

Using Lemma \ref{c5lem1} with $r=L$, we have

\begin{thm}\label{c5thm3}
For rapid-fading channel, the maximum rates possible for a
full-diversity single-symbol decodable STBC using $N$ transmit
antennas is $2/L$.
\end{thm}
\begin{proof}
We have two cases corresponding to the two cases of Corollary
\ref{c5prop3} and we consider them separately.

 \noindent {\it Case 1:   $A_k$ has $L$ non-zero
rows $\forall k$.} The number of matrices that satisfy
     Proposition \ref{c5prop1} are 2, and the maximal rate is $R=1/L$. The corresponding
     STBC is given by its equivalent transmission matrix $S=x_0A_0$, where
     $A_0$ is of the form given in (\ref{mat1}).

\noindent {\it Case 2: $A_k$ has less than $L$ non-zero rows for
some  $ k$.}  As Lemma 1 requires $A_{2k}, A_{2k+1}$ to have $L$
non-zero rows, we can assume that $A_{2k}$ has $r_1$ non-zero rows
and $A_{2k+1}$ has non-overlapping  $L-r_1$ non-zero rows.  The
number of such matrices that satisfy
     Proposition \ref{c5prop1} are 4, and hence the maximal rate is $R=2/L$.
\end{proof}

From Theorem \ref{c5thm3} it follows that the maximal rate
full-diversity single-symbol decodable code is given by its ExCM
     \begin{equation}\label{fdfssdd}
     S=x_{0I}A_0+x_{0Q}A_1+x_{1I}A_2+x_{1Q}A_3,
\end{equation}
 where
     $A_{2k},A_{2k+1},k=0,1$ are of the form
\begin{eqnarray}\label{dis1}
 \left[\begin{array}{cc}
  A & 0 \\
   0 & 0
\end{array}\right],\left[\begin{array}{cc}
  \J A & 0 \\
   0 & 0
\end{array}\right],
\left[\begin{array}{cc}
  0 & 0 \\
   0 & B
\end{array}\right]\mbox{ and }\left[\begin{array}{cc}
  0 & 0 \\
   0 & \J B
\end{array}\right]
\end{eqnarray}
where   $A, B$ are of the form given in (\ref{mat1}) with $L=r_1$
and $L=L-r_1$ respectively. Observe that
     other STBC's can be obtained from the above, by change of variables,
     multiplication by unitary matrices etc.
Of interest is the code for $L=2$ due to its  full rate. Setting
$A=[1 ~0], B=[0 ~1]$
  we have the ExCM,
\begin{equation}\label{nwetm}
  S=\left[\begin{array}{cccc} x_{0I}+\J x_{1Q} & 0 & 0 & 0\\ 0 & 0 &0 &
  x_{1I}+\J x_{0Q}\end{array}\right]
\end{equation}
and the corresponding codeword matrix is
\begin{equation}\label{nw21c}
  \mathbf{S}=\left[\begin{array}{cc} x_{0I}+\J x_{1Q} & 0\\ 0 &
  x_{1I}+\J x_{0Q}\end{array}\right].
\end{equation}
Observe that $\bS$ is the CIOD of size 2 presented in Section
\ref{ssdd}. Also observe that other full rate STBC's that achieve
full diversity can be achieved from $S$ by performing linear
operations (not necessarily unitary) on $S$ and/or permutation of
the real symbols(for each complex symbol there are two real
symbols). Consequently the most general full-diversity
single-symbol decodable code  for $N=2$ is given by the codeword
matrix
\begin{equation}\label{nw21cc}
  \mathbf{S}=\left[\begin{array}{cc} x_{0I}+\J x_{1Q} & b(x_{0I}+\J x_{1Q})\\
  c(x_{1I}+\J x_{0Q}) &
  x_{1I}+\J x_{0Q}\end{array}\right], b,c \in \cp.
\end{equation}

An immediate consequence is

\begin{thm}\label{c5thm4}
 A rate 1 full-diversity single-symbol decodable design for
rapid-fading channel  exists iff $L=N=2$.
\end{thm}
Following  the results of Section \ref{ssdd},

\begin{thm}\label{c5thm5}
The CIOD of size 2 is the only STBC that achieves full diversity
over both quasi-static and rapid-fading channels and provides
single-symbol decoding.
\end{thm}
Other  STBC that achieves full diversity over both quasi-static
fading channels and provides single-symbol decoding are unitarily
equivalent to the CIOD for two antennas. Note that the CIOD for two
antennas dose not have any advantage in rapid-fading channels over
other SD codes in rapid-fading channels.

\begin{rem}\label{c5rem13}
 Contrast the rates of single-symbol decodable codes for quasi-static
  and rapid-fading channels. From Theorem \ref{c5thm3} we have the maximal rate
  is $2/L$ for  rapid-fading channels, while that of  square matrix OD
\cite{TiH1} is given by
$\frac{\lceil\log_2N\rceil+1}{2^{\lceil\log_2 N\rceil}}$ and that of
square FRSDD is given by $\frac{\lceil\log_2
{N/2}\rceil+1}{2^{\lceil\log_2 N\rceil-1}}$ respectively. The
maximal rate is independent of the number of transmit antennas for
rapid-fading channels.
\end{rem}

\section{Discussions}\label{conc}
In this paper we have conducted extensive research on STBCs that
allow single-symbol decoding in both quasi-static and rapid-fading
channels. We have characterized all single-symbol decodable STBCs,
both for quasi-static and rapid-fading channels. Further, among the
class of single-symbol decodable designs, we have characterized
a class  that can achieve full-diversity.

 As a result of this characterization of SD codes
 for quasi-static fading channels, we observe that when there is no
 restriction on the signal
set then STBCs from orthogonal design (OD) are the only STBCs that
are SD and achieve full-diversity. But when there is a restriction
on the signal set, that the co-ordinate product distance is
non-zero (CPD $\ne 0$), then there exists a separate class of
codes, which we call Full-rank Single-symbol Decodable designs
(RFSDD), that allows single-symbol decoding and can achieve
full-diversity. This restriction on the signal set allows for
increase in rate (symbols/channel use), coding gain and maximum
mutual information over STBCs from ODs except for two transmit
antennas. Significantly, rate-one, STBCs from RFSDDs are shown to
exist for 2, 3, 4 transmit antennas while rate-one STBCs  ODs
exist only for 2 transmit antennas. The maximal rates of square
RFSDDs were derived and a sub-class of RFSDDs called generalized
co-ordinate interleaved orthogonal designs (GCIOD) were presented
and their performance analyzed.
  Construction of fractional rate
GCIODs has been dealt with thoroughly resulting in construction of
various high rate GCIODs. In particular a rate 6/7 GCIOD for
$N=5,6$, rate 4/5 GCIOD for $N=7,8$ and rate $>$2/3 GCIOD for $N
\ge 8$ have been presented. The expansion of signal constellation
due to co-ordinate interleaving has been brought out. The coding
gain of GCIOD is linked  to a new distance called generalized
co-ordinate product distance (GCPD) as a consequence the coding
gain of CIOD is linked to CPD. Both the GCPD and the CPD for
signal constellations derived from the square lattice have been
investigated. Simulation results are then presented for $N=4$ to
substantiate the theoretical analysis and finally the maximum
mutual information for GCIOD has been derived and compared with
GLCOD. It is interesting to note that except for $N=2$, the GCIOD
turns out to be superior to GLCOD in terms of rate, coding gain
and  MMI.   A significant drawback of GCIOD schemes is that half
of the antennas are idle, as a result  these schemes have higher
peak-to-average ratio (PAR) compared to the ones using Orthogonal
Designs.
 This problem can be solved by pre-multiplying with a
Hadamard matrix as is done for DAST codes in \cite{DMB1}. This
pre-multiplication by a Hadamard matrix will not change the
decoding complexity while more evenly distributing the transmitted
power across space and time.

An important contribution of this paper is the novel application of
designs to rapid-fading channels, as a result of which we find that
the CIOD for two transmit antennas is the only design that allows
single-symbol decoding over both rapid-fading and quasi-static
channel. It turns out that the single-symbol decodability criterion
is very restrictive in rapid-fading channels and results in constant
rate.

Though we have rigorously pursued single-symbol decodable STBCs
and,  in particular, square single-symbol decodable
 STBCs, much is left to be desired in non-square STBCs.
Although non-square STBCs are shown to be useless for rapid-fading
channels, Su, Xia  and Xue-bin-Liang \cite{SuX2,Lia} have shown for
STBCs from ODs in quasi-static channels, that higher rates can be
obtained from non-square designs. Here we list some open problems
that were not addressed, or partly addressed in this paper.
\begin{itemize}
  \item Construction of maximal-rate non-square UFSDDs, RFSDDs. However, the
  construction of maximal rate non-square GLCODs (not GCODs) is itself and
  open problem and any contribution in this direction will greatly
  enhance our understanding of non-square FSDDs.

  \item Proof (or contradiction) of existence of non-square FSDDs,
  $S$, such that $S^\h S$ is not unitarily-diagonalizable by a
  constant matrix. In Subsection \ref{sqssdd}, we have shown that
  such square designs do not exist. It would be interesting to see
  if we can obtain even an example of such a design. If such a
  design does not exist then class of UFSDDs reduces to GLCODs. In this case
  the classification of UFSDD is complete. Consequently,
    \item classification of non-square RFSDD, UFSDD is an open problem. In-fact complete classification of RFSDDs appears to be even more
    difficult. Interestingly, \cite{KRL} shows that there exist
     RFSDDs, that do not belong to the class of GCIODs.
    \item Even the smaller problem of maximal rates (and design) for non-square GCIOD is  an open
    problem.
    \item  the CPD of non-square lattice constellations and the GCPD for both square and non-square
    lattice constellations needs to be quantified. It is worth mentioning that the authors
    presented a class of non-square RFSDDs called ACIODs in \cite{KRL} whose
    coding gain depends on CPD and not on GCPD as is the case for GCIODs.
    \item Finally, characterization of non-linear
    STBCs with SD property is another open problem.
    One results in this direction is \cite{Mohd}.
\end{itemize}
Similar characterization of double-symbol decodable designs will
be reported in a future paper.

While the final version of the manuscript was
under preparation the authors became aware of the
work \cite{CYT} that claim to unify the results of
\cite{KhR3,KRL} which is incorrect. The class of codes
of \cite{CYT} do not intersect with the class of codes
of \cite{KhR3} and \cite{KRL} for the weight matrices of the
codes of \cite{CYT} are unitary matrices whereas that
of  the codes of \cite{KhR3} and \cite{KRL} are not.
Furthermore, the STBCs presented in \cite{CYT,CYT1} and
\cite{HWX} are SD STBCs that do not satisfy (\ref{ch}) and
such full-rank SD STBCs are not considered in
this paper. 

%%%%%%%%%%%%%%%%%
\section*{Acknowledgment}
% optional entry into table of contents (if used)
%\addcontentsline{toc}{section}{Acknowledgment}
The authors would like to thank Prof. ${\O}$yvind Ytrehus and the
anonymous  reviewers for valuable comments that helped  improve
the clarity and presentation of the paper.

\onecolumn
%%%%%%%% Begin Tables Captions%%%%%%%%%%%%%%%%%%%%%%%%%%%%%%%%%%%%%%%%%%%%%%%%%
\section*{List of Table Captions}

\begin{enumerate}
\item Table \ref{c2tab1}: The Encoding And Transmission Sequence For $N=$2, Rate
1/2 CIOD.
\\
\item Table \ref{c2tab2}: The Encoding And Transmission Sequence For $N=$2,
Rate 1 CIOD.
\\
\item Table \ref{c2tab3}: Comparison of rates of known   GLCODs and GCIODs for all
$N$
\\
\item Table \ref{c2tab4}: Comparison of delays of known  GLCODs and GCIODs  $N \le
8 $.
\\
\item Table \ref{c2tab5}: The optimal angle of rotation for QPSK and normalized
$GCPD_{N_1,N_2}$ for various values of $N=N_1+N_2$.

\end{enumerate}
%%%%%%%% End Tables Captions%%%%%%%%%%%%%%%%%%%%%%%%%%%%%%%%%%%%%%%%%%%%%%%%%
%%%%%%%% Begin Figure Captions%%%%%%%%%%%%%%%%%%%%%%%%%%%%%%%%%%%%%%%%%%%%%%%%%
\newpage
\section*{List of Figure Captions}
\begin{figure}[p]
\centerline{\psfig{figure=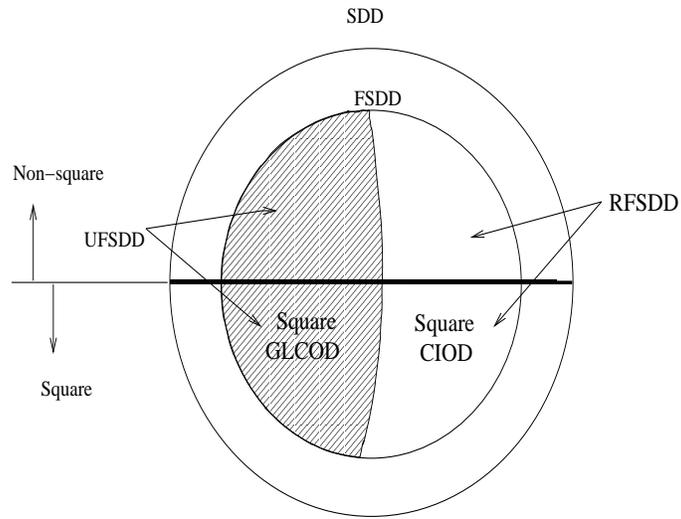,height=2.9in,width=3.7in}} \caption{The classes of
Full-rank Single-symbol Decodable Designs (FSDD).} \label{c1fig1}
\end{figure}

\begin{figure}[p]
\centerline{\psfig{figure=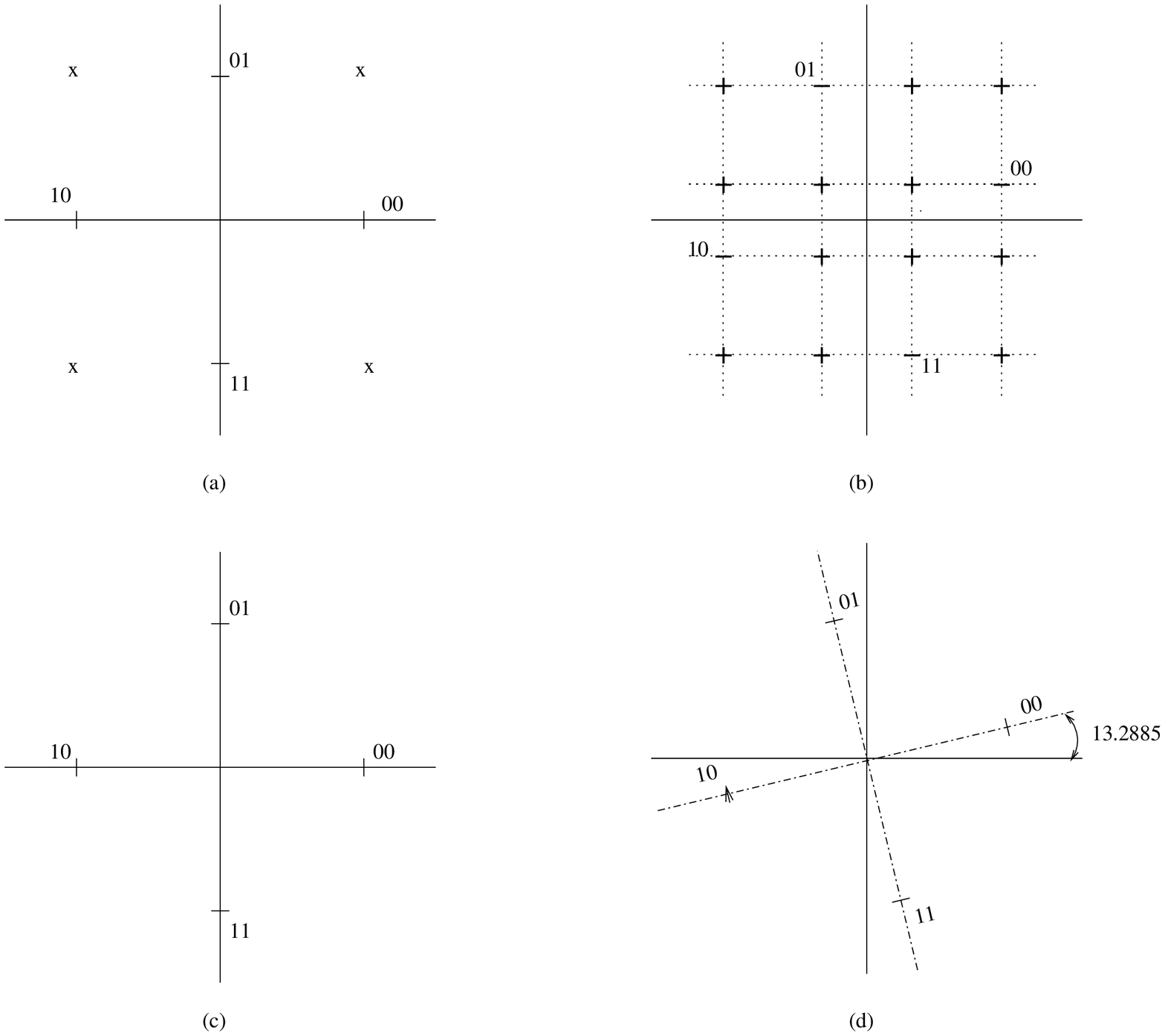,width=6.2in}} \caption{
Expanded signal sets $\tilde {\cal A}$ for ${\cal A}=\{1, -1,
{\mathbf j}, -{\mathbf j} \}$ and a rotated version of it.}
\label{c2fig2}
\end{figure}

\begin{figure}[p]
\centerline{\psfig{figure=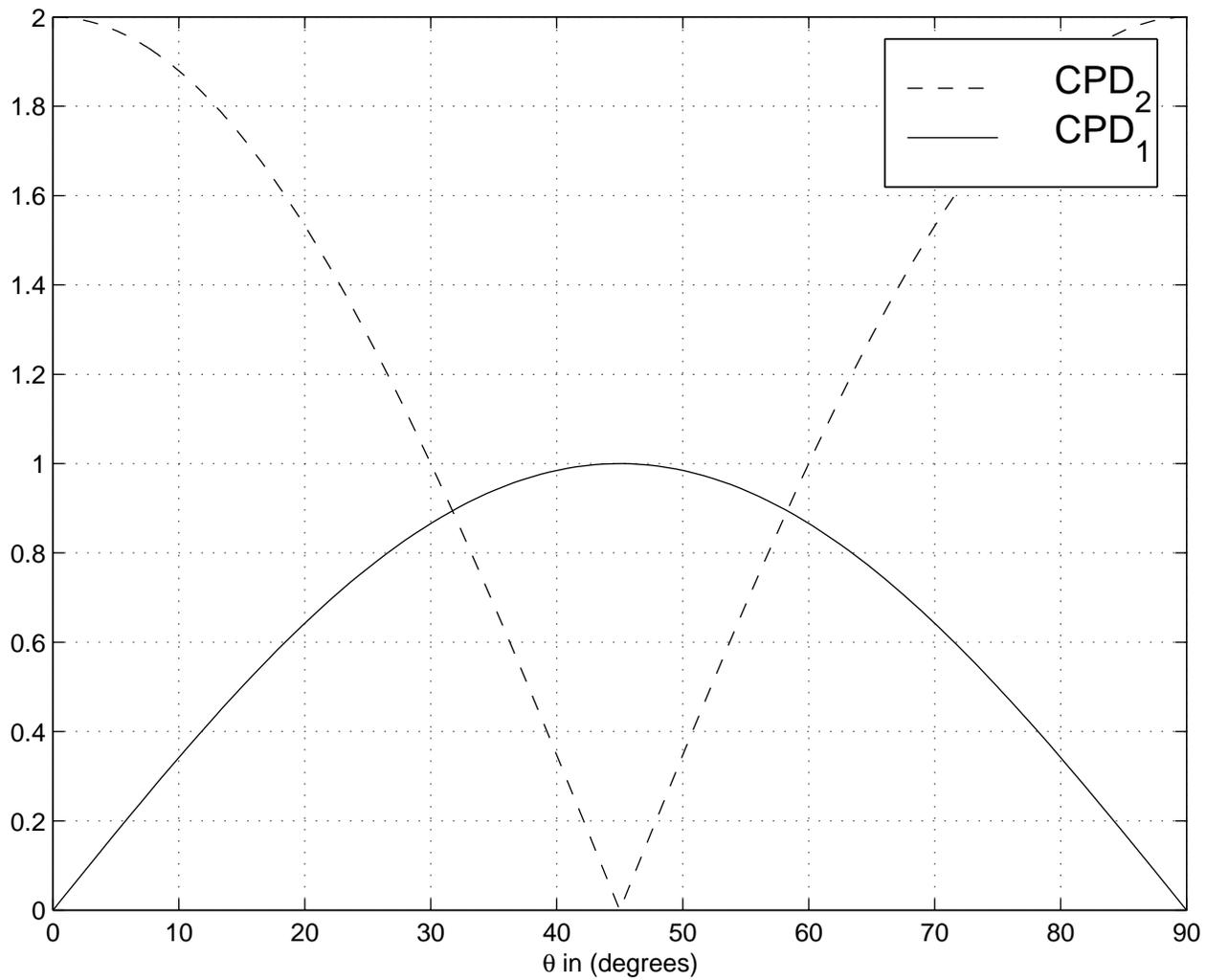}}%,height=4.2in,width=3.2in}
\caption{ The plots of $CPD_1,CPD_2$ for $\theta \in[0
~90^\circ]$. }\label{c2fig3}
\end{figure}

\begin{figure}[p]
\centerline{\psfig{figure=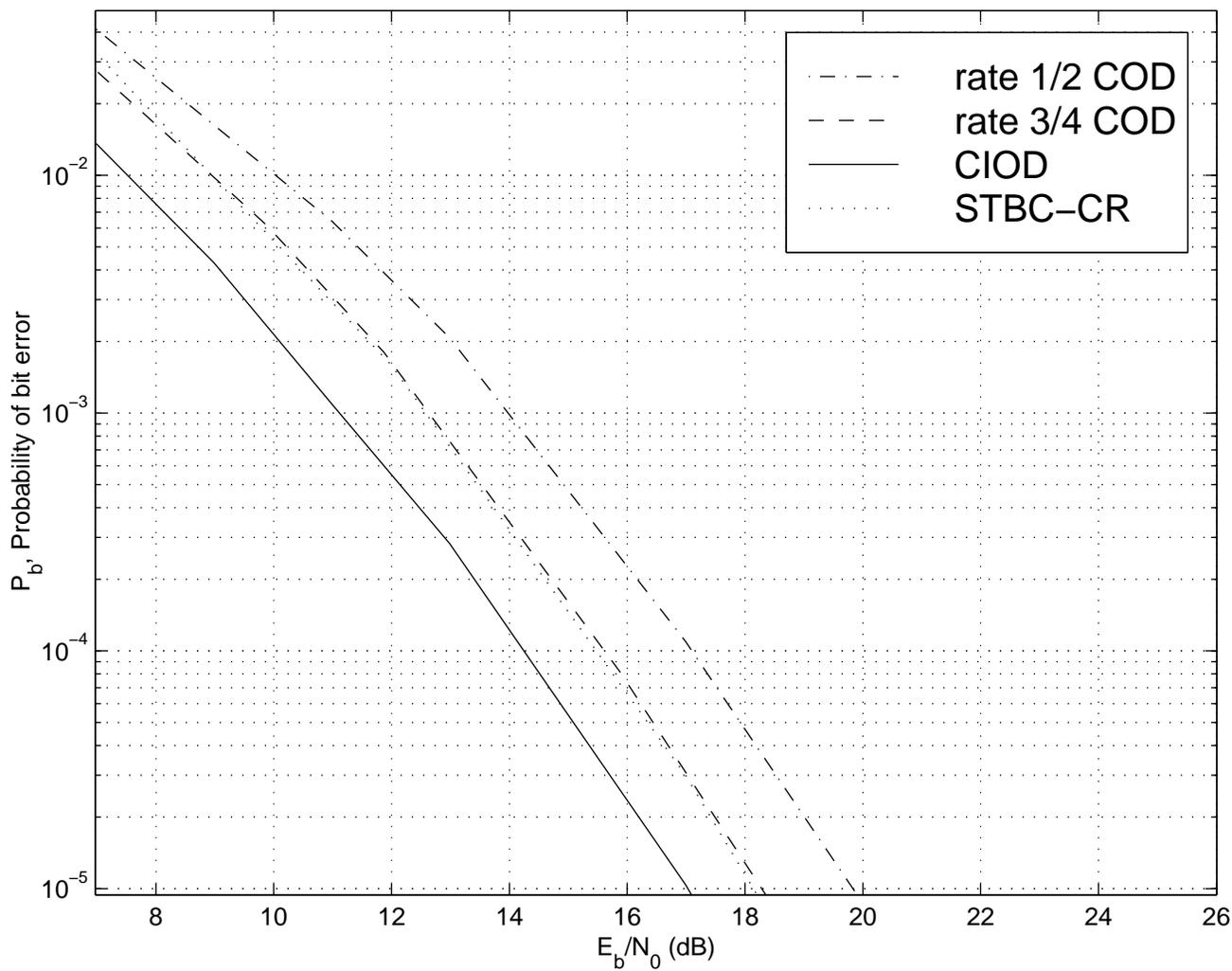}}
%,height=4.2in,width=3.2in}}
\caption{ The BER performance of coherent QPSK rotated by an angle
of $13.2825^\circ$ (Fig. \ref{c2fig2}) used by  the CIOD  scheme
for 4 transmit and 1 receive antenna compared with STBC-CR, rate
1/2 COD and rate 3/4 COD at a throughout of 2 bits/sec/Hz in
Rayleigh fading for the same number of transmit and receive
antennas.
 }\label{c2fig4}
\end{figure}

\begin{figure}[p]
\centerline{\psfig{figure=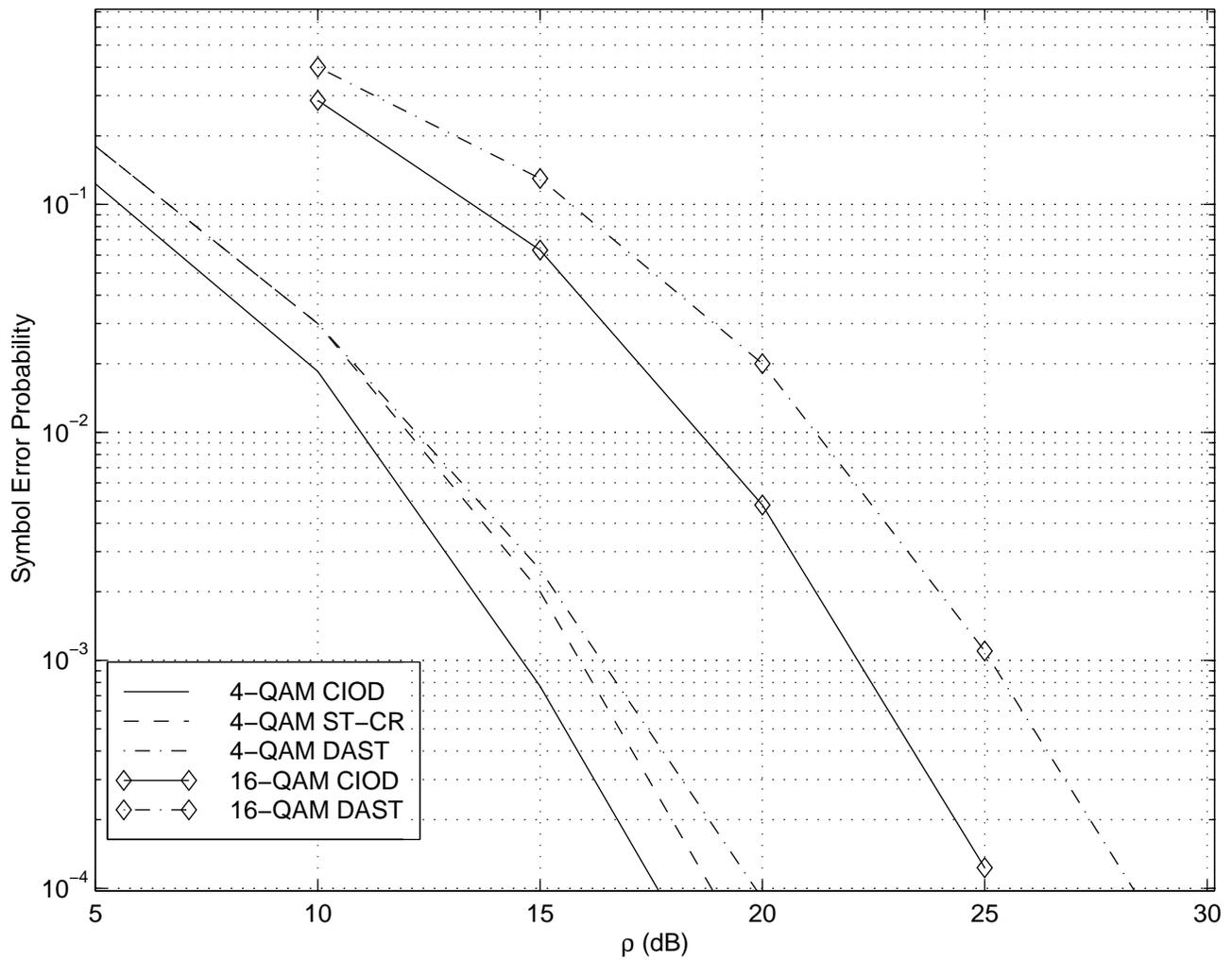}}
%,height=3.2in,width=4.2in}}
\caption{ The BER performance of the CIOD with 4- and 16-QAM
modulations and comparison with ST-CR and DAST schemes.}
\label{c2fig5}
\end{figure}

\begin{figure}
\centerline{\psfig{figure=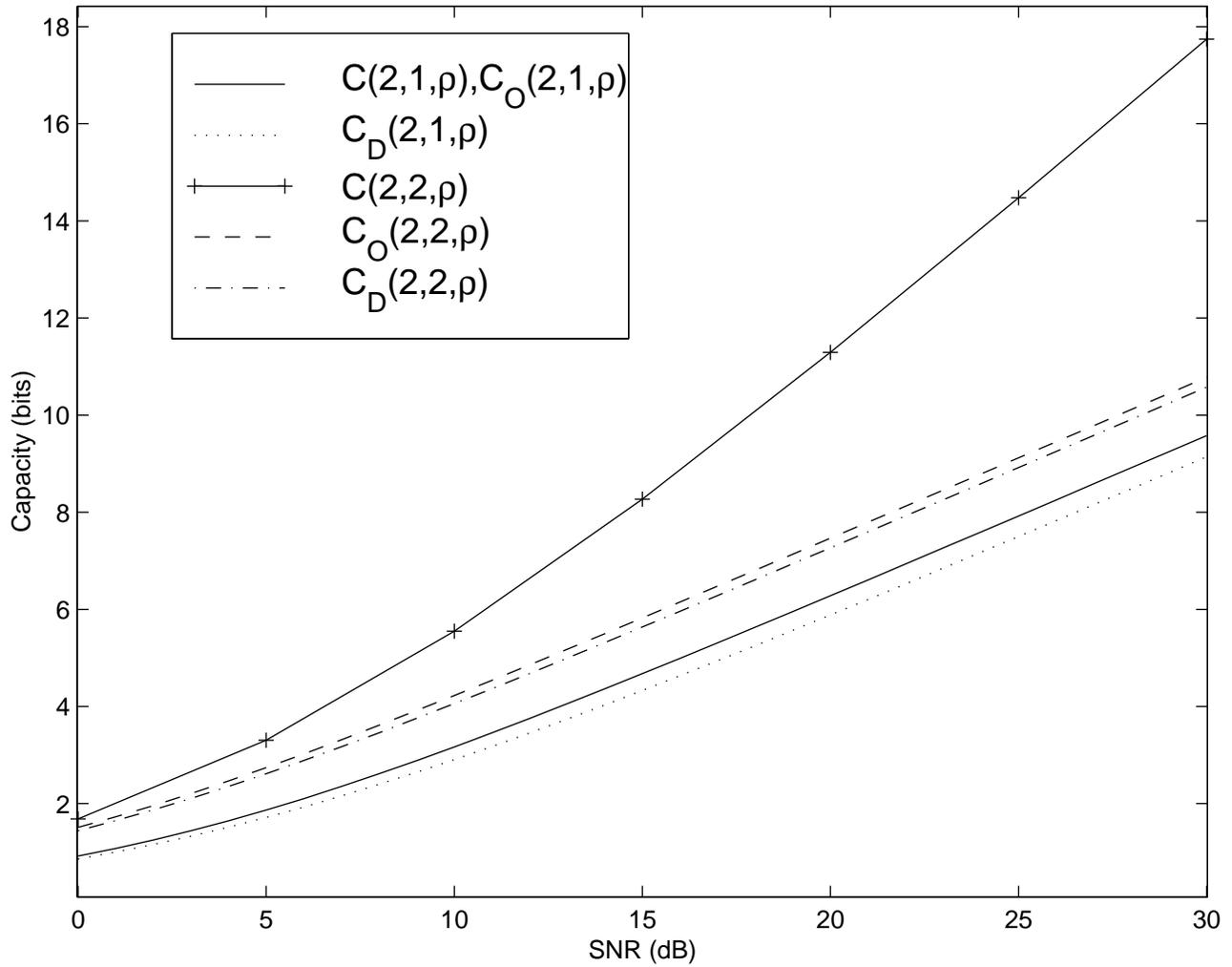}}%,height=3.in,width=3.2in}
\caption{ The  maximum mutual information (ergodic) of CIOD code
for two transmitters and one, two receivers compared with that of
complex orthogonal design (Alamouti scheme) and the actual channel
capacity. }
%for 2 Tx., 1 Rx.,2 Tx. 2 Rx. and 4 Tx., 1 Rx. antennas}
\label{fcap1}\end{figure}

\begin{figure}[p]
\centerline{\psfig{figure=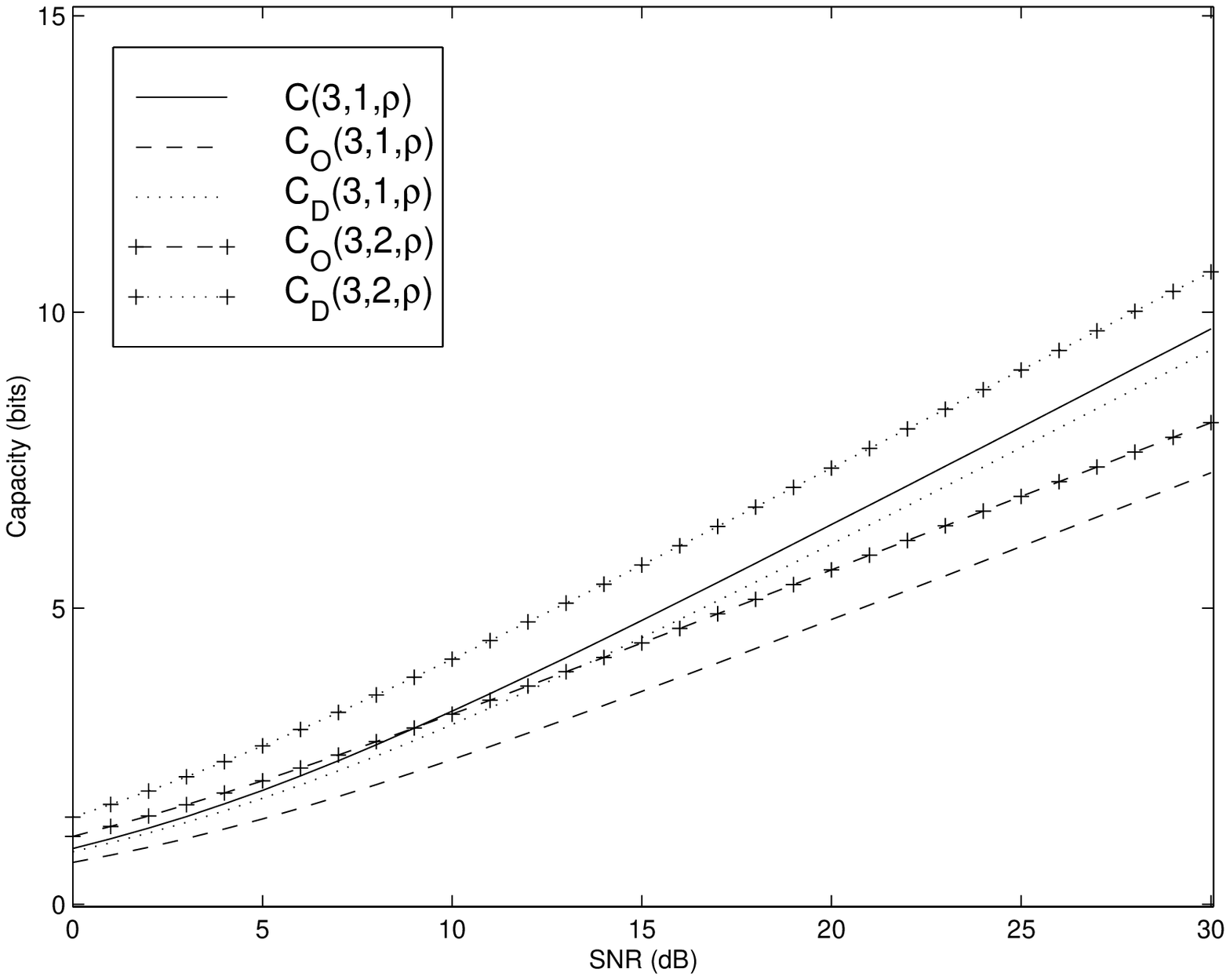}}%,height=3.in,width=3.2in}
\caption{ The maximum mutual information (ergodic) of GCIOD code
for three transmitters and one, two receivers compared with that
of code rate 3/4 complex orthogonal design  for three transmitters
and the actual channel capacity. }
%for 2 Tx., 1 Rx.,2 Tx. 2 Rx. and 4 Tx., 1 Rx. antennas}
\label{fcap3}
\end{figure}

\begin{figure}[p]
\centerline{\psfig{figure=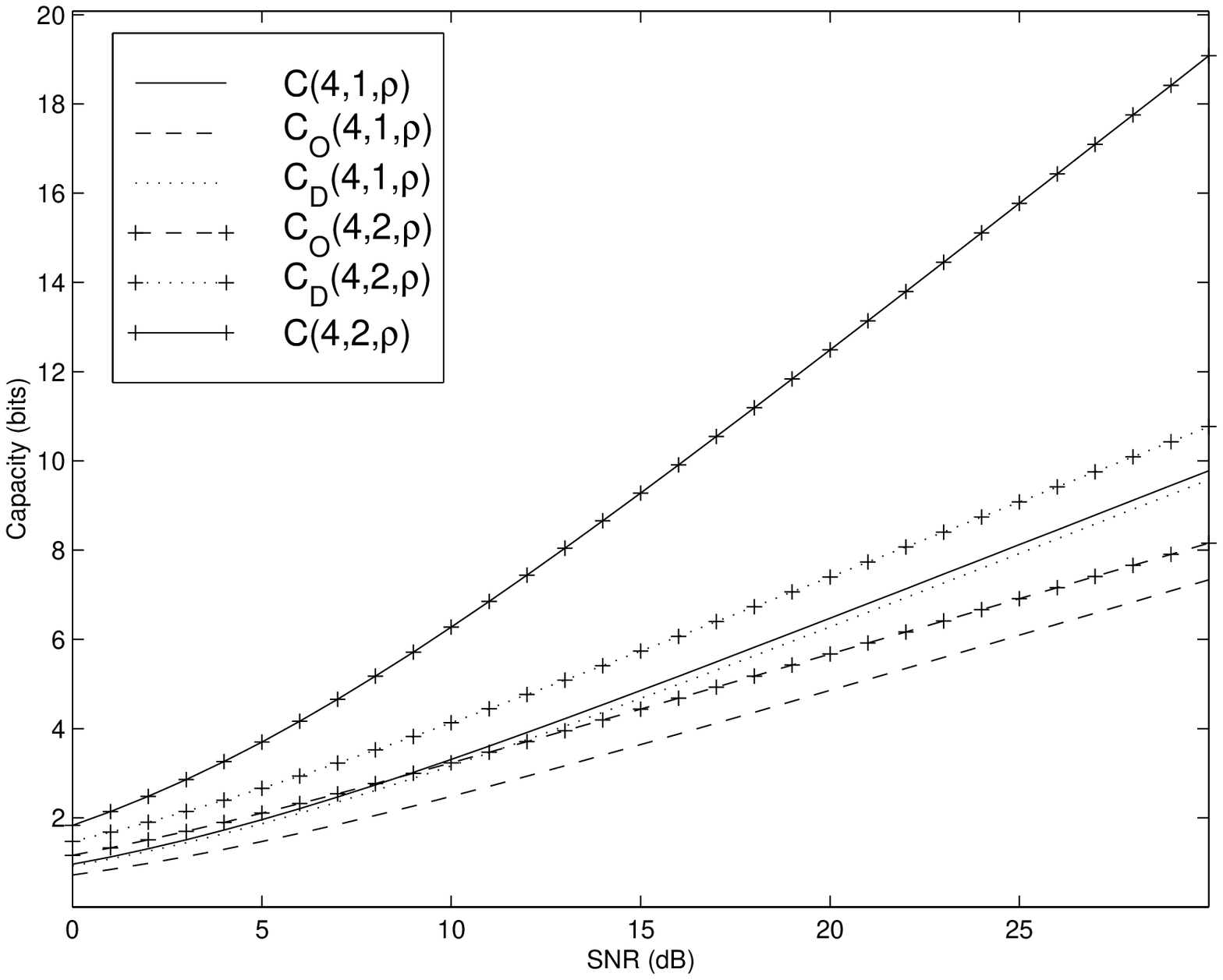}}%,height=3.in,width=3.2in}
\caption{ The maximum mutual information (ergodic) of CIOD code
for four transmitters and one, two receivers compared with that of
code rate 3/4 complex orthogonal design  for four transmitters and
the actual channel capacity. }
%for 2 Tx., 1 Rx.,2 Tx. 2 Rx. and 4 Tx., 1 Rx. antennas}
\label{fcap34}
\end{figure}

\begin{figure}[p]
\centerline{\psfig{figure=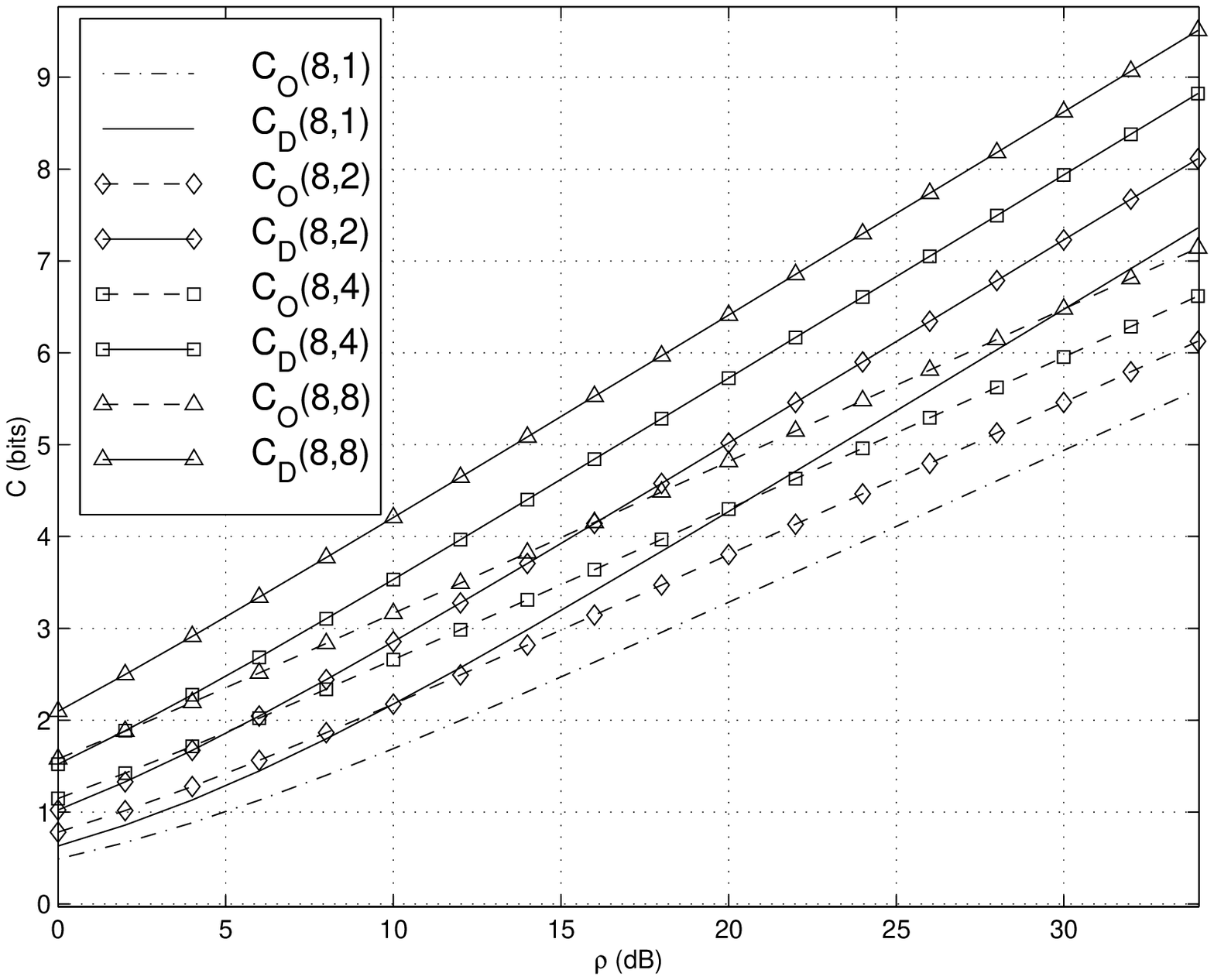}}%,height=3.in,width=3.2in}
\caption{ The maximum mutual information (average) of rate 2/3
GCIOD code for eight transmitters and one, two, four and eight
receivers compared with that of code rate 1/2 complex orthogonal
design  for eight transmitters over Rayleigh fading channels. }
%for 2 Tx., 1 Rx.,2 Tx. 2 Rx. and 4 Tx., 1 Rx. antennas}
\label{c2fig6}
\end{figure}

\begin{enumerate}
    \item Figure \ref{c1fig1}: The classes of
Full-rank Single-symbol Decodable Designs (FSDD).
\\
\item Figure \ref{c2fig2}: Expanded signal sets $\tilde {\cal A}$ for ${\cal A}=\{1, -1,
{\mathbf j}, -{\mathbf j} \}$ and a rotated version of it.
\\
\item Figure \ref{c2fig3}:  The plots of $CPD_1,CPD_2$ for $\theta \in[0
~90^\circ]$.
\\
\item Figure \ref{c2fig4}:  The BER performance of coherent QPSK rotated by an angle
of $13.2825^\circ$ (Fig. \ref{c2fig2}) used by  the CIOD  scheme
for 4 transmit and 1 receive antenna compared with STBC-CR, rate
1/2 COD and rate 3/4 COD at a throughout of 2 bits/sec/Hz in
Rayleigh fading for the same number of transmit and receive
antennas.
\\
\item Figure \ref{c2fig5}: The BER performance of the CIOD with 4- and 16-QAM
modulations and comparison with ST-CR and DAST schemes.
\\
\item Figure \ref{fcap1}: The  maximum mutual information (ergodic) of CIOD code
for two transmitters and one, two receivers compared with that of
complex orthogonal design (Alamouti scheme) and the actual channel
capacity.
\\
\item Figure \ref{fcap3}: The maximum mutual information (ergodic) of GCIOD code
for three transmitters and one, two receivers compared with that
of code rate 3/4 complex orthogonal design  for three transmitters
and the actual channel capacity.
\\
\item Figure \ref{fcap34}: The maximum mutual information (ergodic) of CIOD code
for four transmitters and one, two receivers compared with that of
code rate 3/4 complex orthogonal design  for four transmitters and
the actual channel capacity.
\\
\item Figure \ref{c2fig6}: The maximum mutual information (average) of rate 2/3
GCIOD code for eight transmitters and one, two, four and eight
receivers compared with that of code rate 1/2 complex orthogonal
design  for eight transmitters over Rayleigh fading channels.
\end{enumerate}
%%%%%%%% End Figure Captions%%%%%%%%%%%%%%%%%%%%%%%%%%%%%%%%%%%%%%%%%%%%%%%%%
%%%%%%%% Begin Tables %%%%%%%%%%%%%%%%%%%%%%%%%%%%%%%%%%%%%%%%%%%%%%%%
\begin{table}[p]
\caption{The Encoding And Transmission Sequence For $N=$2, Rate
1/2 CIOD} \vspace{0.1in}
\begin{center}\begin{tabular}{|c|c|c|}\hline\label{c2tab1}
 & antenna 0 & antenna 1 \\\hline
time $t$ & $x_{0I}$ & $0$
\\\hline time $t+T$ & $0$ & $x_{0Q}$ \\\hline
\end{tabular}
\end{center}
 \end{table}

\begin{table}[p]
\caption{The Encoding And Transmission Sequence For $N=$2,
Rate 1 CIOD} \vspace{0.1in}
\begin{center}\begin{tabular}{|c|c|c|}\hline\label{c2tab2}
 & antenna 0 & antenna 1 \\\hline
time $t$ & $x_{0I}+\J x_{1Q}$ & 0
\\\hline time $t+T$ & $0$ &$x_{1I}+ \J x_{0Q}$ \\\hline
\end{tabular}
\end{center}
 \end{table}

\begin{table}[p]
\caption{Comparison of rates of known   GLCODs and GCIODs for all
$N$} \label{c2tab3}
\begin{center}
\begin{tabular}{|c|c|c|c|} \hline
         Tx. Antennas                            & GLCODs  & GCIOD (rate-efficient) & CIOD (delay-efficient)    \\ \hline
N=2     &  1        &    1      & 1   \\ \hline
N=3,4   & 3/4       &    1& 1   \\ \hline
 N=5    & 2/3      &    6/7      & 3/4   \\ \hline
 N=6    &  2/3      &    6/7      &  3/4 \\ \hline
 N=7    &   5/8     & 4/5      &  3/4 \\ \hline
N=8    &   5/8      & 4/5      &  3/4 \\ \hline
 N=2m-3, odd   &   (m)/2(m-1)     &  2(m+1)/(3m+1)   &  7/11 \\ \hline
 N=2m-2, even   &   (m)/2(m-1)    &   2(m+1)/(3m+1)     &  7/11 \\ \hline
 \end{tabular}
\end{center}
\end{table}
\begin{table}[p]
\caption{Comparison of delays of known  GLCODs and GCIODs  $N \le
8 $} \label{c2tab4}
\begin{center}
\begin{tabular}{|c|c|c|c|} \hline
         Tx. Antennas                            & GLCODs  & GCIOD (rate-efficient) & GCIOD (delay-efficient)    \\ \hline
N=2     &  2        &    2     & 2   \\
\hline N=3,4   & 4 & 4&4   \\ \hline
 N=5    & 15      &    14      & 8   \\ \hline
 N=6    &  30      &    14      &  8 \\ \hline
 N=7    &   56     &    35      &  8 \\ \hline
N=8    &   112      &   50      &  8 \\ \hline
 \end{tabular}
\end{center}
\end{table}

\begin{table}[p]
\caption{The optimal angle of rotation for QPSK and normalized
$GCPD_{N_1,N_2}$ for various values of $N=N_1+N_2$.} \label{c2tab5}
\begin{center}
\begin{tabular}{|c|c|c|c|c|c|} \hline
$N$ & $N_1$ & $N_2$ & $x_0$  & $\theta_{opt}$ &
$GCPD_{N_1,N_2}$/4$d^2$ \\ \hline
 3 &  2 & 1  & 0.555 & $29^\circ$ & 0.3487 \\ \hline
5 &  $\begin{array}{c} 4 \\ 3 \end{array}$ & $\begin{array}{c} 1
\\ 2 \end{array}$  & $\begin{array}{c} 0.5246 \\ 0.5751 \end{array}$ & $\begin{array}{c}  27.76^\circ \\ 29.9^\circ \end{array}$ & $\begin{array}{c}  0.28 \\ 0.3869
\end{array}$ \\ \hline
6 &  $\begin{array}{c} 4 \\ 3 \end{array}$ & $\begin{array}{c} 2
\\ 3 \end{array}$  & $\begin{array}{c} 0.555 \\ 0.61 \end{array}$ & $\begin{array}{c}    29.9^\circ\\ 31.7175^\circ \end{array}$ & $\begin{array}{c}  0.3487 \\
0.4472
\end{array}$ \\ \hline
7 &  $\begin{array}{c} 5 \\ 4 \end{array}$ & $\begin{array}{c} 2
\\ 3 \end{array}$  & $\begin{array}{c} 0.543   \\ 0.5856 \end{array}$ & $\begin{array}{c}    28.51^\circ\\ 30.35^\circ \end{array}$ & $\begin{array}{c}  0.3229 \\
0.40
\end{array}$ \\ \hline
9 &  $\begin{array}{c} 7 \\ 5 \end{array}$ & $\begin{array}{c} 2
\\ 4 \end{array}$  & $\begin{array}{c} 0.53 \\ 0.591 \end{array}$ & $\begin{array}{c}    27.94^\circ\\ 30.622^\circ \end{array}$ & $\begin{array}{c}  0.29 \\
0.4135
\end{array}$ \\ \hline 10 &  $\begin{array}{c} 8 \\ 5
\end{array}$ & $\begin{array}{c} 2
\\ 5 \end{array}$  & $\begin{array}{c} 0.526 \\ 0.61 \end{array}$ & $\begin{array}{c}    27.76^\circ\\ 31.7175^\circ \end{array}$ & $\begin{array}{c}  0.3487 \\
0.4472
\end{array}$ \\ \hline
12 &  $\begin{array}{c} 10 \\ 6 \end{array}$ & $\begin{array}{c} 2
\\ 6 \end{array}$  & $\begin{array}{c} 0.52 \\ 0.61 \end{array}$ & $\begin{array}{c}    27.5^\circ\\ 31.7175^\circ \end{array}$ & $\begin{array}{c}  0.265 \\
0.4472
\end{array}$ \\ \hline
 $N$ &  $N-2$ & 2  & $> 0.5$ & $> 26.5656^\circ$ & $>0.2$ \\ \hline
 \end{tabular}
\end{center}
\end{table}

\begin{table}[h]
\caption{The coding gains of CIOD, STBC-CR, rate 3/4 COD and rate
1/2 COD for 4 tx. antennas and QAM constellations}\label{tab2}
\begin{center}
\begin{tabular}{|c|c|c|c|c|}
\hline
 R (bits/sec/Hz) & $\Lambda_{CIOD}$& $\Lambda_{STBC-CR}$&
  $\Lambda_{rate~ 3/4~ COD}$& $\Lambda_{rate ~1/2~ COD}$\\\hline
2 & 0.4478 & 0.5 & 0.333 & 0.2  \\\hline
 3 & 0.1491 & 0.165& 0.1333 & 0.0476  \\\hline
 4 & 0.0897 & 0.1& - & 0.0118  \\\hline
\end{tabular}
\end{center}
\end{table}
%%%%%%%%%%%%%% End Tables
\newpage
\twocolumn
\onecolumn
\newpage
%%%%%%%%%%%%%%%%%%% Figures 


%\newpage


%%%%%%%%%%%%%%%%End Figures %%%%%%%%%%%%%%%%%%%%%%%%%%%%%%%%%%%%%
%%%%%%%%%%%%%%%% Begin Tables %%%%%%%%%%%%%%%%%%%%%%%%%%%%%%%%%%%%%%%5

%\begin{biography}[{\includegraphics[width=1in,height=1.25in,clip,keepaspectratio]{mshell}}]{Michael Shell}
% where an .eps filename suffix will be assumed under latex, and a .pdf suffix
% will be assumed for pdflatex; or if you just want to reserve a space for
% a photo:

\begin{biographynophoto}{Md. Zafar Ali Khan}
(S'00-M'06) was born in Hyderabad, India. He received the B.E. in electronics and communication from Osmania University, M.Tech. in electrical engineering from IIT, Delhi and Ph.D. in electrical communication engineering from IISc, Bangalore in 1996, 1998 and 2003 respectively. He was a design engineer with Sasken, Bangalore in 1999, a senior design engineer at insilica semiconductors, Bangalore, India from 2003 to 2005 and Sr. member of tech. staff at Hellosoft, India in 2005. He is presently with IIIT, Hyderabad as an Assistant Professor.

His primary research interests
are in coded modulation, space-time coding and signal processing for wireless communications.
\end{biographynophoto}

\begin{biographynophoto}{B.~Sundar~Rajan}
(S'84-M'91-SM'98) was born in Tamil Nadu, India.
He received the B.Sc. degree in mathematics from
Madras University,
Madras, India, the B.Tech degree in electronics
from Madras Institute
of Technology, Madras, and the M.Tech and Ph.D.
degrees in electrical
engineering from the Indian Institute of
Technology, Kanpur, India,
in 1979, 1982, 1984, and 1989 respectively.

He was a faculty member with the Department of
Electrical Engineering at
the Indian Institute of Technology in Delhi,
India, from 1990 to 1997.
Since 1998, he has been a Professor in the
Department of
Electrical Communication Engineering at the
Indian Institute of Science,
Bangalore, India. His primary research interests
are in algebraic coding,
coded modulation and space-time coding.

Dr.~Rajan is a Member of the American
Mathematical Society.
\end{biographynophoto}

% insert where needed to balance the two columns on the last page
%\newpage


%\vfill



\begin{thebibliography}{1}

\bibitem{FoG}
J. G. J. Foschini and M. J. Gans, ``On limits of wireless
communication in a
  fading environment using multiple antennas,'' \emph{Wireless Personal
  Communication}, Vol. 6, no. 3, pp. 311-335, Mar. 1998.

\bibitem{Tel}
E. Teletar, ``Capacity of multi-antenna  Gaussian  channels,''
\emph{European
  Transactions on Telecommunications}, Vol. 10, no. 6, pp. 585-595, Nov. 1999.

%  \bibitem{Jak}
%J. W. C. Jakes, \emph{Microwave Mobile Communications}, New York:
%John Wiley and Sons, 1974.
%
%\bibitem{GSSSN}
%D. Gesbert, M. Shafi, D. S. Shiu, P. Smith, and A. Naquib, ``An
%overview of
%  {MIMO} space-time coded wireless systems,'' \emph{IEEE J. Select. Areas
%  Commun.}, Vol.21, no.3, pp.281--302, April. 2003.
%
%\bibitem{HTW}
%A. Hottinen, O. Tirkkonen and R. Wichman, \emph{Multi-antenna
%Transceiver Techniques for 3G and Beyond}, New York: John Wiley,
%2002.
%
%\bibitem{Pro}
%J. G. Proakis, \emph{Digital Communications}, New York:
%McGraw-Hill, 1989.



\bibitem{SeW}
N. Seshadri and J. H. Winters, ``Two signaling schemes for
improving the error
  performance of  FDD  transmission systems using transmitter antenna
  diversity,'' \emph{Int. Journal of Wireless Inform. Networks}, Vol. 1, pp.
  49-60, 1994.

\bibitem{GFBK}
J. Guey, M. P. Fitz, M. R. Bell, and W. Y. Kuo, ``Signal design
for transmitter
  diversity wireless communication systems over {Rayleigh} fading channels,''
  in \emph{Proc. of IEEE VTC 96}, 1996, pp. 136--140.

\bibitem{Wee}
V. Weerackody, ``Diversity of direct-sequence spread spectrum
system using
  multi transmit antennas,'' in \emph{Proc. of IEEE ICC 93}, 1993, pp.
  1775-1779.

\bibitem{Win1}
J. Winters, ``Switched diversity with feedback for DPSK mobile
radio
  systems,'' \emph{IEEE Trans. Veh. Technol.}, Vol. VT-32, pp. 134-150, Feb.
  1983.

\bibitem{Win2}
------, ``Diversity gain of transmit diversity in wireless systems with
  {Rayleigh} fading,'' in \emph{Proc. of IEEE ICC 94}, Vol. 2, 1994, pp.
  1121-1125.

\bibitem{TSC}
V. Tarokh, N. Seshadri, and A. R. Calderbank, ``Space-time codes
for high data
  rate wireless communication: performance criterion and code construction,''
  \emph{IEEE Trans. Inform. Theory}, Vol. 44, pp. 744-765, Mar. 1998.

\bibitem{TJC}
V. Tarokh, H. Jafarkhani, and A. R. Calderbank, ``Space-time block
codes from
  orthogonal designs,'' \emph{IEEE  Trans. Inform. Theory}, Vol. 45, pp.
  1456-1467, July 1999.

\bibitem{TJCc}
V. Tarokh, H. Jafarkhani and A. R. Calderbank , ``Correction to
``Space-time block codes from Orthogonal designs'',''{\em IEEE
Trans. on Inform. Theory}, Vol. 46, No.1, pp.314, Jan. 2000.

\bibitem{TiH1}
O. Tirkkonen and A. Hottinen, ``Square matrix embeddable STBC for
complex
  signal constellations,'' \emph{IEEE Trans. Inform. Theory}, Vol. 48, no. 2,
  pp. 384-395, Feb. 2002.

\bibitem{Ala}
S. M. Alamouti, ``A simple transmit diversity technique for
wireless
  communications,'' \emph{IEEE J. Select. Areas Commun.}, Vol. 16, no. 8, pp.
  1451-1458, Oct. 1998.

\bibitem{GaS1}
G. Ganesan and P. Stoica, ``Space-time diversity,'' in
\emph{Signal Processing
  Advances in Wireless and Mobile Communications}, 2000, Vol. 1, Ch. 2, pp.
  59-87.

\bibitem{GaS2}
------, ``Space-time block codes: a maximal SNR approach,'' \emph{IEEE Trans. Inform. Theory}, Vol. 47, no. 4,
  pp. 1650-1656, May 2001.

\bibitem{GaS3}
------, ``Space-time diversity using orthogonal and amicable orthogonal
  designs,'' in \emph{Proc. of ICASSP 2000}, Istanbul, Turkey, 2000, pp.
  2561-2564.

\bibitem{LiX}
X.-B. Liang and X.-G. Xia, ``On the Nonexistence of Rate-One
Generalized Complex Orthogonal Designs,
 ,'' \emph{IEEE Trans.
Inform. Theory},vol. 49, no. 11, Nov. 2003, pp.  2984-2988 .


\bibitem{Lia}
 X.-B. Liang, ``Orthogonal Designs With
Maximal Rates ,'' \emph{IEEE Trans. Inform. Theory}, Vol. 49, No.
10, pp.2468-2503, Oct. 2003.

\bibitem{SuX1}
W. Su and X.-G. Xia, ``Two generalized complex orthogonal
space-time block
  codes of rates 7/11 and 3/5 for 5 and 6 transmit antennas,'' \emph{IEEE
  Trans. Inform. Theory}, Vol. 49, pp. 313 -316, Jan. 2003.

\bibitem{SuX2}
------, ``On space-time block codes from complex orthogonal designs,''
  \emph{Wireless Personal Communications (Kluwer Academic Publishers)}, vol. 25, no. 1, pp.1-26, April 2003.

\bibitem{Gan}
G. Ganesan, ``Designing space-time codes using orthogonal
designs,'' Ph.D.
  dissertation, Uppsala University, Uppasal, Sweden, 2002.



\bibitem{Jaf}
H. Jafarkhani, ``A quasi-orthogonal space-time block code,''
\emph{IEEE
  Trans. Commun.}, Vol. 49, pp. 1-4, Jan. 2001.

\bibitem{TiH2}
O. Tirkkonen and A. Hottinen, ``Complex space-time block codes for
four {Tx}
  antennas,'' in \emph{Proc. of Globecom 2000},  pp. 1005-1009, Nov. 2000.

\bibitem{ShP1}
N. Sharma and C. B. Papadias, ``Improved quasi-orthogonal codes
through
  constellation rotation,'' \emph{IEEE Trans. Commun.}, Vol. 51, no. 3, pp.
  332--335, Mar. 2003.

\bibitem{ShP2}
------, ``Improved quasi-orthogonal codes,'' in \emph{Proc. of IEEE WCNC 2002},
  pp. 169-171, Mar. 2002.

\bibitem{PaF}
C. B. Papadias and G. J. Foschini, ``A space-time coding approach
for systems
  employing four transmit antennas,'' in \emph{Proc. of IEEE ICASSP 2002},
  Vol. 4, 2002, pp. 2481-2484.

\bibitem{SuX3}
Weifeng-Su and X. Xia, ``Quasi-orthogonal space-time block codes
with full-diversity,'' in \emph{Proc. of Globecom 2002}, Vol. 2,
Taipai, Taiwan, Nov.
  2002, pp. 1098-1102.

\bibitem{SuX4}
W. Su and X.-G. Xia, ``Signal constellations for quasi-orthogonal
space-time
  block codes with full-diversity,'' \emph{IEEE Trans. Inform. Theory},
  vol. 50, No. 10, Oct. 2004, pp. 2331-2347.

\bibitem{Jos}
T. Josefiak, ``Realization of {Hurwitz-Radon} matrices,''
\emph{Queen's papers
  on pure and applied mathmatics}, no. 36, pp. 346-351, 1976.

\bibitem{GeG}
A. V. Geramita and J. M. Geramita, ``Complex orthogonal designs,''
  \emph{Journal of Combin. Theory}, Vol. 25, pp. 211-225, 1978.

\bibitem{Her}
I. N. Herstein, \emph{Non-commutative rings}, ser. Carus
mathematical
  monograms.\hskip 1em plus 0.5em minus 0.4em\relax Washington DC, USA: Math.
  Assoc. Amer., 1968, Vol. 15.

\bibitem{GeS}
A. V. Geramita and J. Seberry, \emph{Orthogonal designs, quadratic
forms and
  {Hadamard} matrices}, ser. Lecture Notes in Pure and Applied
  Mathematics.\hskip 1em plus 0.5em minus 0.4em\relax Berlin, Germany:
  Springer, 1979, Vol. 43.

  \bibitem{XWG1}
Y. Xin, Z. Wang, and G. B. Giannakis, ``Space-time constellation
rotating codes
  maximizing diversity and coding gains,'' in \emph{Proc. IEEE Globecom
  2001}, Vol. 1, Vancouver, Canada, Nov. 2001, pp. 455-459.

\bibitem{XWG2}
------, ``Linear unitary precoders for maximum diversity gains with multiple
  transmit and receive antennas,'' in \emph{Proc. IEEE ASILOMAR 2000},
  Pacific Grove, USA, Nov. 2000, pp. 1553-1557.

\bibitem{DMB1}
M. O. Damen, K. Abed-Meraim, and J.-C. Belfiore, ``Diagonal
algebraic
  space-time block codes,'' \emph{IEEE Trans. Inform. Theory}, Vol. 48,
  no. 3, pp. 384-395, Mar. 2002.

\bibitem{DaB}
M. O. Damen and N. C. Beaulieu, ``On diagonal algebraic space-time
block
  codes,'' \emph{IEEE Trans. Commun.}, vol. 51, June 2003.

\bibitem{HaH}
B. Hassibi and B. Hochwald, ``High-rate codes that are linear in
space and
  time,'' \emph{IEEE Trans. Inform. Theory}, Vol. 48, no. 7, pp. 1804--1824,
  July 2002.

\bibitem{San}
S. Sandhu, ``Signal design for {MIMO} wireless: a unified
perspective,'' Ph.D.
  dissertation, Stanford University, Stanford, CA, Aug. 2002.

%\bibitem{DTB}
%M. O. Damen, A. Tewfik, and J. C. Belfiore, ``A construction of a
%space-time
%  code based on number theory,'' \emph{IEEE Trans. Inform. Theory}, Vol. 48,
%  no. 3, pp. 753--760, Mar. 2002.

\bibitem{DMB2}
M. O. Damen, K. Abed-Meraim, and J.-C. Belfiore, ``Lattice codes
decoder for
  space-time codes,'' \emph{IEEE Commun. Lett.}, Vol. 4, pp. 161--163, May
  2000.

\bibitem{DMB3}
------, ``Generalized sphere decoder for asymmetrical space-time communication
  architecture,'' \emph{Electronics Letters}, Vol. 36, no. 2, pp. 166--167,
  Jan. 2000.

\bibitem{Sli1}
S. B. Slimane, ``An improved {PSK} scheme for fading channels,''
\emph{IEEE
  Trans. Veh. Technol.}, Vol. 47, no. 2, pp. 703--710, May 1998.

\bibitem{Sli2}
------, ``An improved {PSK} scheme for fading channels,'' in \emph{Proc. IEEE
  GLOBECOM '96}, Nov. 1996, pp. 1276--1280.

\bibitem{JeR1}
B. D. Jelicic and S. Roy, ``Design of trellis coded {QAM} for flat
fading and
  {AWGN} channel,'' \emph{IEEE Trans. Veh. Technol.}, Vol. 44, pp. 192--201,
  Feb. 1995.

\bibitem{JeR2}
------, ``Cutoff rates for co-ordinate interleaved {QAM} over {Rayleigh} fading
  channel,'' \emph{IEEE Trans. Commun.}, Vol. 44, no. 10, pp. 1231--1233,
  Oct. 1996.

\bibitem{Goe}
D. Goeckel, ``Coded modulation with non-standard signal sets for
wireless
  {OFDM} systems,'' in \emph{Proc. IEEE ICC 1999}, Vol. 2, 1999, pp.
  791--795.

\bibitem{ChR}
A. Chindapol and J. Ritcey, ``Bit-interleaved coded modulation
with signal
  space diversity in {Rayleigh} fading,'' in \emph{Proc. IEEE ASILOMAR 1999},
  Vol. 2, Pacific Grove, USA, Nov. 1999, pp. 1003--1007.


\bibitem{KhR1}
Md. Zafar Ali Khan and B. Sundar Rajan, ``A co-ordinate
interleaved orthogonal design for
  four transmit antennas,'' {\em IISc-DRDO Report}, No. TR-PME-2002-17,
Department of Electrical Communication Engineering, Indian
Institute of Science, Bangalore, India, October 2002.
\bibitem{KhR2}
------, ``A co-ordinate interleaved orthogonal design for
  three transmit antennas,'' {\em  in Proc. of National Conference on Communications }, Mumbai, India, January 2002.

\bibitem{KhR3}
------, ``Space-time block codes from co-ordinate
interleaved orthogonal designs,''{\em in Proc. of ISIT 2002},  pp.
316, Lausanne, Switzerland, June 30 - July 5, 2002.

\bibitem{KhR4}
Md. Zafar Ali Khan, B. Sundar Rajan and M. H. Lee, ``On
single-symbol and double-symbol decodable STBCs,'' {\em Proc. of
ISIT 2003}, Yokohama, Japan, June 29-July 3, 2003, pp.127.


\bibitem{KhR5}
Md. Zafar Ali Khan and B. Sundar Rajan, ``Space-time block codes
from designs for fast-fading channels,'' {\em Proc. of  ISIT 2003},
Yokohama, Japan, June 29-July 4, 2003, pp.154.

\bibitem{KhR6}
------, ``A Generalization of some existence results on Orthogonal Designs for STBCs,''
\emph{to appear in IEEE Trans. Inform. Theory}, Vol. 50, No. 1,
Jan., 2004 pp.218-219.

\bibitem{KhR7}
------, ``Space-Time Block Codes from Designs for Fast-Fading Wireless Communication,''
 {\em IISc-DRDO Technical Report}, No: TR-PME-2003-07, Department of Electrical Communication Engineering, Indian Institute of Science, Bangalore, India, May 2003.

\bibitem{RKL}
B. Sundar Rajan, Md. Zafar Ali Khan and M. H. Lee, ``A co-ordinate
interleaved orthogonal design for
  eight transmit antennas,'' {\em IISc-DRDO Report}, No. TR-PME-2002-16,
Department of Electrical Communication Engineering, Indian
Institute of Science, Bangalore, India, October 2002.

\bibitem{KRL}
Zafar Ali Khan, B. Sundar Rajan and H.H.Lee ``Rectangular Coordinate
Interleaved Orthogonal Designs,'' Proceedings of IEEE GLOBECOM 2003,
Communication Theory Symposium, San Francisco, Dec. 2003, Vol.4,
pp.2004-2009.


\bibitem{KhR9}
Md. Zafar Ali Khan and B. Sundar Rajan, ``Bit and co-ordinate
interleaved coded
  modulation,'' in \emph{Proc. IEEE GLOBECOM 2000}, San Francisco, USA, Nov.
  2000, pp. 1595--1599.

\bibitem{Zafar}
Md. Zafar Ali Khan, ``Single-symbol and Double-symbol Decodable
STBCs for MIMO fading  channels,''  \emph{ Ph.D. Thesis, Indian
Institute of Science}, Bangalore, India, July 2003.


\bibitem{CYT}
Chau Yuen, Yong Liang Guan and T. T. Tjhung,``Construction of
Quasi-orhogonal STBC with Minimum Decoding Complexity from Amicable
Orthogonal Designs,'' {\em Proc. of ISIT 2004}, Chicago, USA, June
29-July 3, 2004.

\bibitem{CYT1}
Chau Yuen, Yong Liang Guan and T. T. Tjhung,``Quasi-Orthogonal STBC
with Minimum Decoding Complexity: Further Results,'' {\em Proc. of
WCNC 2005}, USA, pp. 483-488.

\bibitem{HWX}
Haiquan Wang, Dong Wang, and Xiang-Gen Xia, ``On Optimal
Quasi-Orthogonal Space-Time Block Codes with Minimum Decoding
Complexity,'' {\em Proc. of ISIT 2005}, pp. 1168-1172.

%\bibitem{KhR10}
%------, ``A new asymmetric {8-PSK TCM} scheme for {Rayleigh} fading channels,''
%  in \emph{Proc. SITA 1999}, Yuzawa, Niigata, Japan, pp. 749--752, Nov. 30- Dec. 4, 1999.
%
%\bibitem{KhR11}
%------, ``A 4-state asymmetric {8-PSK TCM} scheme for {Rayleigh} fading
%  channels optimum at high SNR's,'' in \emph{Proc. of IEEE VTC-2000-Spring},
%  Tokyo, Japan, pp. 685--689, Vol.1, May 2000.
%
%\bibitem{KhR12}
%------, ``A tight upper bound for {8-PSK BICM} schemes,'' in \emph{Proc. of
%  ISITA 2000}, Honolulu, Hawaii, USA, pp. 164-167, Nov. 2000.

\bibitem{BoB}
K. Boull$\acute{e}$ and J. C. Belfiore,``Modulation scheme
designed for Rayleigh fading channel'', presented at CISS'92,
Princeton, NJ, March 1992.

\bibitem{BVRB}
J.~Boutrous, E.~Viterbo, C. Rastello and J. C. Belfiore, ``Good
lattice constellationsfor both {Rayleigh} fading and Gaussian
channel,'' \emph{IEEE Trans. Inform. Theory}, Vol.42, pp.
502--518, March 1996.

\bibitem{BoV}
J. Boutrous and E. Viterbo, ``Signal space diversity: a power and
bandwidth
  efficient diversity technique for {Rayleigh} fading channel,'' \emph{IEEE
  Trans. Inform. Theory}, Vol. 44, no. 4, pp. 1453--1467, July 1998.
%
%\bibitem{SaP}
%S. Sandhu and A. Paulraj, ``Space-time block codes: a capacity
%perspective,''
%  \emph{IEEE Commun. Lett.}, Vol. 4, pp. 384--386, Dec. 2000.
%
%\bibitem{PaT}
%R. Patel and M. Toda, ``Trace inequalities involving {Hermitian}
%matrices,''
%  \emph{Linear algebra and its applications}, Vol. 23, pp. 13--20, 1979.

\bibitem{CYV}
Z. Chen, J. Yuan, and B. Vucetic, ``An improved space-time trellis
coded
  modulation scheme on slow {Rayleigh} fading channels,'' in \emph{Proc. IEEE
  International Conference on Communications 2001}, Vol. 4, Helsinki, Finland,
  June 2001, pp. 1110--1116.

\bibitem{Mohd}
Mohammed Ali Maddah-Ali and Amir K. Khandani,``A new
non-orthogonal space-time code with low decoding complexity,''
\emph{private communication}.
\bibitem{HoJ}
R.A.Horn and C.R.Johnson, {\it Matrix Analysis}, Cambridge
University Press, 1985.

\end{thebibliography}
\end{document}